\documentclass[preprintnumbers,tightenlines,notitlepage,nofootinbib,secnumarabic]{revtex4-1}

\usepackage{graphicx}
\usepackage{amsmath,amssymb,mathrsfs}
\usepackage{fancyhdr}
\usepackage{psfrag}
\usepackage{array,bbm}
\usepackage{url}
\usepackage{rotating}
\usepackage{multirow}
\usepackage{dsfont}
\usepackage{cancel}

\numberwithin{equation}{section}
\makeatletter       % Macros for arabic section numbering

\renewcommand{\p@subsection}{}

\makeatother

\newenvironment{Eqnarray}%
     {\arraycolsep 0.14em\begin{eqnarray}}{\end{eqnarray}}
\newcommand{\beqa}{\begin{Eqnarray}}
\newcommand{\eeqa}{\end{Eqnarray}}
\newcommand{\beq}{\begin{equation}}
\newcommand{\eeq}{\end{equation}}

\def\Fig#1{Fig.~\ref{#1}}
\def\eq#1{eq.~(\ref{#1})}
\def\eqs#1#2{eqs.~(\ref{#1}) and (\ref{#2})}

\def\eqst#1#2{eqs.~(\ref{#1})--(\ref{#2})}
\def\phm{\phantom{-}}
\def\half{\tfrac{1}{2}}
\def\centeron#1#2{{\setbox0=\hbox{#1}\setbox1=\hbox{#2}\ifdim
\wd1>\wd0\kern.5\wd1\kern-.5\wd0\fi
\copy0\kern-.5\wd0\kern-.5\wd1\copy1\ifdim\wd0>\wd1
\kern.5\wd0\kern-.5\wd1\fi}}
\def\ltap{\;\centeron{\raise.35ex\hbox{$<$}}{\lower.65ex\hbox{$\sim$}}\;}
\def\gtap{\;\centeron{\raise.35ex\hbox{$>$}}{\lower.65ex\hbox{$\sim$}}\;}
\def\gsim{\mathrel{\gtap}}
\def\lsim{\mathrel{\ltap}}
\def\wtil{\widetilde}
\def\anti{\overline}
\def\ur{U_R}
\def\dr{D_R}
\def\lt{\left}
\def\rt{\right}
\def\gam{\gamma}
\def\epem{e^+e^-}
\def\gev{~{\rm GeV}}
\def\tev{~{\rm TeV}}
\def\fbi{~{\rm fb}^{-1}}
\def\rts{\sqrt s}

\def\tanb{\tan\beta}
\def\tanbsq{\tan^2\beta}
\def\cotb{\cot\beta}
\def\sina{\sin\alpha}

\def\sinb{\sin\beta}
\def\sinbii{\sin^2\beta}
\def\sinbiv{\sin^4\beta}
\def\cosb{\cos\beta}
\def\cosbii{\cos^2\beta}
\def\cosbiv{\cos^4\beta}
\def\cbma{\cos(\beta-\alpha)}
\def\sbma{\sin(\beta-\alpha)}
\def\cbpa{\cos(\beta+\alpha)}
\def\sbpa{\sin(\beta+\alpha)}

\def\muflhc{\mu^h_f({\rm LHC})}
\def\mulhc#1{\mu^h_{#1}({\rm LHC})}
\def\muilc#1{\mu^h_{#1}({\rm ILC})}
\def\kgam{\kappa_\gam}
\def\kg{\kappa_g}

\def\lam{\lambda}
\def\hsm{h_{\rm SM}}
\def\beq{\begin{equation}}
\def\eeq{\end{equation}}
\def\hp{H^+}
\def\hm{H^-}
\def\ha{A}
\def\hl{h}
\def\hh{H}
\def\m#1{m_{#1}}
\def\mhh{\m{\hh}}
\def\mhl{\m{\hl}}
\def\mha{\m{\ha}}
\def\hpm{H^\pm}
\def\mhpm{\m{\hpm}}
\def\mw{m_W}
\def\call{{\cal L}}
\def\taui{\tau^{-1}}
\def\wp{W^+}
\def\wm{W^-}
\def\bec{\begin{center}}
\def\eec{\end{center}}
\def\kd{\kappa_D}
\def\ku{\kappa_U}

\begin{document}
\preprint{\vspace{0.1in}\large SCIPP 14/02\cr
\vspace{0.1in}\large UCD-2014-2\cr
\large March, 2014}
\vspace*{1.5cm}

\title{Probing wrong-sign Yukawa couplings at the LHC and a future linear collider}
\author{P.M.~Ferreira}
    \email[E-mail: ]{ferreira@cii.fc.ul.pt}
\affiliation{Instituto Superior de Engenharia de Lisboa-ISEL,
	1959-007 Lisboa, Portugal}
\affiliation{Centro de F\'{\i}sica Te\'{o}rica e Computacional,
    Faculdade de Ci\^{e}ncias,
    Universidade de Lisboa,
    Avenida\ Professor\ Gama Pinto 2,
    1649-003 Lisboa, Portugal}
    \author{John F. Gunion}
    \email[E-mail: ]{gunion@physics.ucdavis.edu}
    \affiliation{Davis Institute for High Energy Physics,
    University of California,
    Davis, California 95616, USA}
\author{Howard E.~Haber}
    \email[E-mail: ]{haber@scipp.ucsc.edu}
\affiliation{Santa Cruz Institute for Particle Physics,
    University of California,
    Santa Cruz, California 95064, USA}
\affiliation{Ernest Orlando Lawrence Berkeley National Laboratory,
University of California, Berkeley, California 94720, USA}
\author{Rui Santos}
    \email[E-mail: ]{rsantos@cii.fc.ul.pt}
\affiliation{Instituto Superior de Engenharia de Lisboa-ISEL,
	1959-007 Lisboa, Portugal}
\affiliation{Centro de F\'{\i}sica Te\'{o}rica e Computacional,
    Faculdade de Ci\^{e}ncias,
    Universidade de Lisboa,
    Avenida\ Professor\ Gama Pinto 2,
    1649-003 Lisboa, Portugal}
%\date{\today}

\begin{abstract}
We consider the two-Higgs-doublet model as a framework in which to evaluate the viability of scenarios in which the sign of the coupling of the observed Higgs boson to down-type fermions (in particular, $b$-quark pairs) is opposite to that of the Standard Model (SM), while at the same time all other tree-level couplings are close to the SM values.  We show that, whereas such a scenario is consistent with current LHC observations, both future running at the LHC and a future $\epem$ linear collider could determine the sign of the Higgs coupling to $b$-quark pairs.  Discrimination is possible for two reasons.  First, the interference between the $b$-quark and the $t$-quark loop contributions to the  $ggh$ coupling changes sign.  Second, the charged-Higgs loop contribution to the $\gam\gam h$ coupling is large and fairly constant up to the largest charged-Higgs mass allowed by tree-level unitarity bounds when the $b$-quark Yukawa coupling has the opposite sign from that of the SM (the change in sign of the interference terms  between the $b$-quark loop and the $W$ and $t$ loops having negligible impact).
\end{abstract}

\maketitle

%%%%%%%%%%%%%%%%%%%%%%%%%%%%%%%%%%%%%%%%%%%%%%%%%%%%%%%%%%%%%%%%%%%%%%%%
% Introduction
%
%%%%%%%%%%%%%%%%%%%%%%%%%%%%%%%%%%%%%%%%%%%%%%%%%%%%%%%%%%%%%%%%%%%%%%%
\section{Introduction}
\label{sec-intro}

Now that the existence of a Higgs boson is firmly established~\cite{ATLASHiggs, CMSHiggs},
the ATLAS and CMS Collaborations at the Large Hadron Collider (LHC)
have started probing the Higgs
couplings to the fermions and to the gauge bosons~\cite{Aad:2013wqa, Chatrchyan:2013mxa,pdg}.
With almost all data from the 8 TeV run analyzed,
it becomes increasingly clear that the Standard Model (SM)
predictions regarding the Higgs experimental rates are completely consistent with the current experimental data at the 95\% C.L., in some cases at the 68\% C.L. In the future, the LHC and an International Linear Collider (ILC) could  further reinforce this consistency with ever higher precision or could eventually reveal some discrepancies. At this moment in time, it is important to delineate  the portions of parameter space of models
where qualitative and quantitative differences of the couplings with respect to the SM  are consistent with current data but would be revealed by the upcoming LHC runs or at a future collider such as the ILC.

In this work, we will discuss the interesting possibility of a sign change in one of the Higgs Yukawa couplings, $h_D$ for down-type fermions  or $h_U$ for up-type fermions, relative to the Higgs coupling to $VV$ ($V=W^\pm$ or $Z$).
It is well known that the current LHC results cannot differentiate between scenarios where a sign change
occurs in the $h_D$ Yukawa couplings (see e.g.~Refs.~\cite{Espinosa:2012im, Falkowski:2013dza,Belanger:2013xza}) simply using the measured properties of the observed Higgs-like boson and assuming no particles beyond those of the SM. For example, in the most recent fit of Ref.~\cite{Belanger:2013xza}, it is found that while the coupling of the Higgs to top quarks must have the conventional positive sign relative to the Higgs coupling to $VV$,
the couplings of down-type quarks and leptons are only constrained to $|h_D/h_D^{\rm SM}|=1.0\pm 0.2$, where the sign ambiguity arises from the weak dependence of the $gg$ and $\gam\gam$ loops on the Higgs couplings to bottom-quark pairs.  The sign degeneracy in the determination of $h_D$\ at the LHC has also been stressed recently in Ref.~\cite{Celis:2013ixa}.

In this paper, we will show that the sign of the bottom Yukawa can be determined with sufficient LHC data or at an ILC.  The results of this paper will be established in the framework of
the softly-broken $\mathbb{Z}_2$ symmetric (CP-conserving) two-Higgs doublet model (2HDM).
The 2HDM possesses two limiting cases (called the decoupling and alignment limits introduced in Section~\ref{sec:decoup}), in which the Higgs couplings to $VV$, fermion pairs, and the cubic and quartic Higgs self-couplings approach their SM values.
But, the 2HDM is also sufficiently flexible as to allow for a SM-like limit for the Higgs couplings to $VV$, up-type quark pairs and Higgs self-couplings, but with a coupling to down-type fermions that is opposite in sign to that of the SM. We can thus explore what happens in the context of this specific model when the only tree-level difference relative to the SM is the sign of $h_D$.  The sign of $h_D$ impacts both the $ggh$ and $\gam\gam h$ couplings.
The $ggh$  coupling will change significantly when the sign of $h_D$ is changed due to the fact that the sign of the interference between the bottom-quark and top-quark  loops is reversed. The $h\to\gam \gam$ amplitude is altered primarily because the decoupling of the charged-Higgs loop contribution can be temporarily avoided until a rather large charged-Higgs mass, the boundary being set by the point at which the theory violates tree-level unitarity.  Indeed, the nondecoupling of the charged-Higgs loop dominates over the change in the sign of interference terms involving the $b$-quark loop (whose interference is unobservably small on its own), and leads to a potentially observable decrease in the magnitude of the $\gam\gam h$ effective coupling. While the change in the sign of  interference terms involving the bottom loop is a universal feature that can be used to
resolve the relative sign of  $h_D$ versus $h_U$, the charged-Higgs temporary nondecoupling need not be.  The latter proves essential in using the $\gam\gam$ final state of Higgs decay to determine the sign of $h_D$ relative to $h_U$, even allowing said discrimination at the next run of the LHC.  Using the $gg$ coupling is more generically useful  and allows the sign determination both at the LHC (albeit somewhat indirectly) and at a future linear collider.

As already implicit in the statements above, it is important to explore the $h_D$ sign issue in the context of a model in which both signs of $h_D$ are allowed and physically distinguishable.  The CP-conserving 2HDM provides one such context. Sensitivity to the sign of $h_D$ requires that the measurable collider event rates depend significantly on it.  The collider event rates are conveniently encoded in the cross section ratios
$\mu^h_f$ defined by
\begin{equation}
\mu^h_f \, = \, \frac{\sigma \, {\rm BR} (h \to
  f)}{\sigma^{\scriptscriptstyle {\rm SM}} \, {\rm BR} (\hsm \to f)}
\label{eg-rg}
\end{equation}
where $\sigma$ is the Higgs production cross section and ${\rm BR} (h \to f)$ is
the branching ratio of the decay to some given final state $f$;  $\sigma^{\scriptscriptstyle {\rm {SM}}}$
and ${\rm BR}(\hsm  \to f)$ are the expected values for the same quantities in the SM. 
The experimentally measured values of $\mu^h_f$ for a variety of final states $f$ at the LHC already provide interesting constraints on the 2HDM parameter space~\cite{manyrefs}.

In this paper, we do
not separate different LHC initial state production mechanisms ($gg\to h$, $VV\to h$, $b\bar{b}\to h$, $Vh$ associated production and $t\bar{t}h$ associated
production); that is, we sum over all production mechanisms in computing the cross section.  In our analysis of Higgs phenomena at the ILC, we consider only the $\epem\to Z h$ production
process. We employ the notation $\mu^h_f({\rm LHC}, {\rm ILC})$ when discussing these ratios for the LHC and ILC, respectively. In deciding whether or not a
given 2HDM parameter choice is excluded by LHC data for given values of $\muflhc$, all the currently well-measured final states $f=WW^*, ZZ^*, b\anti
b, \tau^+\tau^-,\gam\gam$ must be employed. In particular, we will find that $h_D<0$ is only consistent with current LHC Higgs data for a 2HDM of
type-II  if  the deviations in the $\gam \gam h$  and/or $gg h$ couplings will be detectable in the future with the LHC operating
at $\rts\sim 14\tev$, assuming an accumulation of luminosity $L\geq 300 \fbi$, and at a future ILC.

This paper is organized as follows. In Section 2, we describe the 2HDM and the constraints imposed
by theoretical and phenomenological considerations. In Section 3 we introduce the decoupling and alignment limits, and then define the wrong-sign Yukawa couplings scenario and discuss its properties.
In Section 4 we analyze the detailed phenomenology of the wrong-sign Yukawa coupling scenario, and in Section 5, we exhibit the results of our analysis.  Our conclusions are presented in Section 6.  Appendix \ref{app:higgsbasis} provides details regarding the Higgs basis scalar potential parameters of the 2HDM relevant for Section 3.   The Higgs sector of the minimal supersymmetric extension of the Standard Model (MSSM)~\cite{mssm} is a special case of the type-II 2HDM introduced in Section 2.   The possibility of an MSSM Higgs sector with an opposite-sign $hb\bar{b}$ coupling relative to the SM is addressed in Appendix \ref{app:mssm}.  Finally, Appendix \ref{apjfg} explains the nondecoupling behavior of the charged-Higgs loop contribution to the $h\to\gam\gam$ amplitude in a type-II 2HDM that is particularly relevant when $h_D$ has a sign opposite that of the SM.

\section{Models and constraints}
\label{sec-model}

%%%%%%%%%%%%%%%%%%%%%%%%%%%%%% minimal version of the model section

The 2HDM is an extension of the scalar sector of the SM with an extra hypercharge-one scalar doublet field,
first introduced in Ref.~\cite{Lee:1973iz} as a means to
explain matter-antimatter asymmetry (see Refs.~\cite{hhg, Branco:2011iw} for a detailed description
of the model).
The most general Yukawa Lagrangian, in terms of the quark mass-eigenstate fields, is:
\beq \label{yuk}
-\mathscr{L}_{\rm Y}=\anti U_L \wtil\Phi_{a}^0{\eta^U_a} \ur +\overline
D_L K^\dagger\wtil\Phi_{a}^- {\eta^U_a}\ur
+\overline U_L K\Phi_a^+{\eta^{D\,\dagger}_{a}} \dr
+\overline D_L\Phi_a^0 {\eta^{D\,\dagger}_{a}}\dr+{\rm h.c.}\,,
\eeq
where $\wtil\Phi_{a}\equiv (\wtil\Phi^0\,,\,\wtil\Phi^-)^T
=i\sigma_2\Phi_{a}^*$ and $K$ is the CKM mixing matrix.
In \eq{yuk} there is an implicit sum over the index $a=1,2$, and the $\eta^{U,D}$
are $3\times 3$ Yukawa coupling matrices.
In general, such models give rise to  couplings corresponding to tree-level Higgs-mediated flavor-changing neutral currents (FCNCs),
in clear disagreement with experimental data.

A natural way to avoid FCNC interactions is to impose
a $\mathbb{Z}_2$ symmetry on the dimension-four terms of the Higgs Lagrangian
in order to set two of the $\eta_a^Q$ equal to zero in~\eq{yuk}~\cite{GWP}.
This in turn implies that one of the two Higgs fields is odd under
the $\mathbb{Z}_2$ symmetry. The Higgs potential can thus be written as:
\beqa
 \mathcal{V} &=& m^2_{11}\Phi_1^\dagger\Phi_1+m^2_{22}\Phi_2^\dagger\Phi_2
   -\lt(m^2_{12}\Phi_1^\dagger\Phi_2+{\rm h.c.}\rt)
   +\half\lambda_1\lt(\Phi_1^\dagger\Phi_1\rt)^2
   +\half\lambda_2\lt(\Phi_2^\dagger\Phi_2\rt)^2 \nonumber \\
&&\qquad\qquad\qquad\qquad\qquad\qquad +\lambda_3\Phi_1^\dagger\Phi_1\Phi_2^\dagger\Phi_2
   +\lambda_4\Phi_1^\dagger\Phi_2\Phi_2^\dagger\Phi_1
   +\lt[\half\lambda_5\lt(\Phi_1^\dagger\Phi_2\rt)^2+{\rm h.c.}\rt]\,,\label{pot}
\eeqa
where $m_{12}^2$ softly breaks the $\mathbb{Z}_2$ symmetry.
In particular, we do not allow a hard breaking of the $\mathbb{Z}_2$ symmetry, which implies that the term of the form $\bigl(\Phi_1^\dagger\Phi_2\bigr)\bigl(\lambda_6\Phi_1^\dagger\Phi_1+\lambda_7\Phi_2^\dagger\Phi_2\bigr)+{\rm h.c.}$ is absent.  For simplicity we
will work with a CP-conserving scalar potential by choosing  $m_{12}^2$ and
$\lambda_5$ to be real.

The 2HDM parameters are chosen such that electric charge is conserved
while neutral Higgs fields acquire real vacuum expectation values,\footnote{A sufficient condition for
guaranteeing that the vacuum is CP invariant is $\lambda_5|v_1| |v_2|\leq |m_{12}^2|$
(see e.g.~Appendix B of Ref.~\cite{decoupling}).  Moreover, 
the existence of a tree-level scalar potential minimum that breaks the electroweak symmetry but preserves both the electric charge and CP symmetries, ensures that no additional tree-level potential minima that spontaneously break the electric charge and/or CP symmetry
can exist~\cite{vacstab}. As such, in our simulations we can be certain that $v_1$ and $v_2$ can be chosen real.}
$\langle\Phi_a^0\rangle=v_a/\sqrt{2}$ (for $a=1,2$),
where
\beq
v^2\equiv v_1^2+v_2^2=\frac{4m_W^2}{g^2}=(246~{\rm GeV})^2\,,
\quad { \rm and} \quad \tan\beta\equiv \frac{v_2}{v_1}\,.
\eeq
By convention, we take $0\leq\beta\leq\half\pi$ (after a suitable rephasing of the Higgs doublet fields).
From the 8 degrees of freedom we end up with
three Goldstone bosons, a charged-Higgs pair,
 two CP-even neutral Higgs states, $h$ and $H$ (defined such that
$m_{h}\leq m_{H}$), and one CP-odd
neutral Higgs boson $A$. The CP-even Higgs squared-mass matrix is diagonalized
by an angle $\alpha$, which is defined modulo $\pi$.
The coupling of $h$ to $VV$ is specified by
\beq
g_{hWW}=g\mw \sbma\,.
\label{hwwform}
\eeq

As noted above, Higgs-mediated tree-level FCNCs can be avoided by imposing a $\mathbb{Z}_2$ symmetry that is preserved by all dimension-four interactions of the Higgs Lagrangian.
Different choices for the transformation of the fermion fields under this $\mathbb{Z}_2$
lead to different Higgs-fermion interactions.
In this paper, we shall focus on two different choices,
which lead to models that are called the type-I~\cite{type1,hallwise} and
type-II~\cite{type2,hallwise} 2HDM.
In the type-I 2HDM, $\eta_1^U=\eta_1^D=0$ in
\eq{yuk}, whereas in the type-II 2HDM, $\eta_1^U=\eta_2^D=0$.
In the former all fermions couple exclusively to $\Phi_2$ while
in the latter the up-type quarks couple
exclusively to $\Phi_2$ and the down-type quarks and charged leptons
couple exclusively to
$\Phi_1$.
In both the type-I and type-II 2HDM, the Higgs-fermion couplings are
flavor diagonal and depend on the
two angles $\alpha$ and $\beta$ as shown in
Table~\ref{tab:couplings}.  The tree-level MSSM Higgs sector
is a special case of the type-II 2HDM~\cite{mssm}.

\begin{table}[t!]
\begin{center}
\begin{tabular}{|c|ccccccc|ccccccc|}
\hline
 & & \multicolumn{5}{c}{Type-I} &  \hspace{3ex}& &\multicolumn{5}{c}{Type-II} &\\
 & & $h$  & & $A$ & & $H$ & \hspace{3ex}& & $h$ & & $A$ & & $H$& \\
\hline
Up-type quarks  & &
$\cos{\alpha}/\sin{{\beta}}$   & &
$\phm\cot{\beta}$  & &
$\sin{\alpha}/\sin{{\beta}}$   &  \hspace{3ex} & &
$\phm\cos{\alpha}/\sin{{\beta}}$   & &
$\cot{\beta}$ & &
$\sin{\alpha}/\sin{{\beta}}$ &\\
Down-type quarks and charged leptons & &
$\cos{\alpha}/\sin{{\beta}}$   & &
$- \cot{\beta}$  & &
$\sin{\alpha}/\sin{{\beta}}$  &  \hspace{3ex} & &
$- \sin{\alpha}/\cos{{\beta}}$   & &
$\tan{\beta}$ & &
$\cos{\alpha}/\cos{{\beta}}$ & \\
\hline
\end{tabular}
\end{center}
\caption{Couplings of the fermions to the lighter and heavier
CP-even scalars ($h$ and $H$),
and the CP-odd scalar ($A$), relative to the corresponding SM value of $m_f/v$.
\label{tab:couplings}}
\end{table}

%%%%%%%%%%%%%%%%%%%%%%%%%%%%%% end of minimal version of the model section

%%%%%%%%%%%%%%%%%%%%%%%%%%%%%% constraints

The most relevant constraints on the 2HDM are briefly discussed in Ref.~\cite{Arhrib:2013oia}. Here, we will just enumerate
the constraints imposed on the parameters of the CP-conserving 2HDM.

\renewcommand{\theenumi}{\roman{enumi}}%
\begin{enumerate}

\item The Higgs potential is bounded from below~\cite{vac1};
\item Tree-level unitarity is imposed
on the quartic Higgs couplings~\cite{unitarity};
\item It complies with $S$ and $T$ parameters~\cite{Peskin:1991sw,STHiggs}
as derived from electroweak precision
observables~\cite{lepewwg,gfitter1,gfitter2};
\item  The global minimum of the Higgs potential is unique~\cite{Barroso:2013awa} and no spontaneous charge or CP-breaking occurs~\cite{vacstab};
\item Indirect constraints on the ($m_{H^\pm}$,$\tan\beta$) plane
stem from loop processes involving charged Higgs bosons. They originate
mainly from  $B$ physics observables~\cite{BB,BB2,gfitter1} and from the
$R_b\equiv\Gamma(Z\to b\bar{b})/\Gamma(Z\to{\rm hadrons})$
~\cite{Freitas:2012sy, Denner:1991ie,Boulware:1991vp,Grant:1994ak,Haber:1999zh, gfitter1} measurement.  In particular,  for the type-II 2HDM, $\mhpm\gsim 340\gev$ is required.
\item LEP searches based on $e^+ e^- \to H^+ H^-$~\cite{Abbiendi:2013hk} and recent LHC results~\cite{ATLASICHEP, CMSICHEP} based on
$pp \to \bar t \, t (\to H^+ \bar b ) $ constrain the mass of the charged Higgs to be above $O(100)$ GeV, depending on the model Type.

\end{enumerate}

Finally we should note that there is an unexplained discrepancy
between the  value of $\overline{B}\to D^{(*)}\tau^-\overline{\nu}_\tau$  measured by
the {B\sc{a}B\sc{ar}} collaboration~\cite{Lees:2012xj} and the corresponding SM prediction. The observed deviation is of the order  3.4~$\sigma$.   If confirmed, this observation would exclude both the SM and the version of the 2HDM considered in this work.

\section{Decoupling, alignment, delayed decoupling and the wrong-sign Yukawa couplings}
\label{sec:decoup}

In light of the fact that the LHC Higgs data is consistent with the predictions of the Standard Model with one complex hypercharge-one Higgs doublet, it is of interest to consider the limit of the 2HDM in which the properties of the lightest CP-even Higgs boson $h$ approach those of the SM Higgs boson.  It is convenient in this section to adopt a sign convention in which $\sin(\beta-\alpha)$ is non-negative,\footnote{The implications of an alternative convention $|\alpha|\leq\half\pi$, employed in the 2HDM parameter scans of Sections~\ref{sec-results} and \ref{sec-results2}, will be addressed later in this section.}
i.e.~ $0\leq\beta-\alpha\leq\pi$.
Since
\beq \label{VV}
\frac{g_{hVV}}{g_{h_{\rm SM}VV}}=\sbma\,,\quad \text{where $V=W^\pm$ or $Z$}\,,
\eeq
it follows that $h$ is SM-like in the limit of $\cbma\to 0$.
%\clearpage

It is convenient to rewrite the Higgs potential of \eq{pot} in terms of new scalar doublet fields
defined in the Higgs basis~\cite{Donoghue:1978cj,Georgi,silva,lavoura,lavoura2,branco,Davidson:2005cw}.
The coefficients of the quartic terms of the scalar potential in the Higgs basis are denoted by
$Z_i$ (where $i=1,2,\ldots,7$).  Expressions for the $Z_i$ in terms of the $\lambda_i$ defined by \eqst{z1}{z7} are given in Appendix~\ref{app:higgsbasis}.
In particular, using \eqs{z1}{z6}, it follows that,
\beqa
\cos^2(\beta-\alpha)&=&\frac{Z_1 v^2-m_h^2}{m_H^2-m_h^2}\,,\label{c2exact}\\
\sin(\beta-\alpha)\cos(\beta-\alpha)&=&-\frac{Z_6v^2}{m_H^2-m_h^2}
\label{scexact}\,.
\eeqa

By assumption, the sizes of the scalar potential parameters (in any basis) are limited by tree-level unitarity constraints.   This means that $Z_1/(4\pi)\lsim\mathcal{O}(1)$ and $Z_6/(4\pi)\lsim\mathcal{O}(1)$.
It follows that if $m_H\gg v$ then $|\cbma|\ll 1$ in which case $h$ has SM-like couplings to $VV$.
This is the \textit{decoupling limit}~\cite{Haber:1989xc,decoupling}, where $m_{H^\pm}^2-m_A^2\sim\mathcal{O}(v^2)$ and
$m_{H}^2-m_A^2\sim\mathcal{O}(v^2)$ [i.e.~$m_H\sim m_A\sim m_{H^{\pm}}\gg m_h$],
and the $hVV$ couplings approach those of the Standard Model.   That is, below the common scale of the heavy Higgs states, the effective field theory that describes Higgs physics is the Standard Model with a single hypercharge-one Higgs doublet.  However,
note that if $h$ is SM-like, it does not necessarily follow that the masses of $H$, $A$ and $H^\pm$ are large.  Indeed, \eq{scexact} implies that it is possible to achieve $|\cbma|\ll 1$ by taking $|Z_6|\ll 1$~\cite{Carena:2001bg,decoupling}.
The limit where $Z_6\to 0$ is called the \textit{alignment limit}~\cite{Craig:2013hca,Asner:2013psa,Carena:2013ooa,Haberinprep}, since in this limit the mixing of the two-Higgs-doublet fields in the Higgs basis is suppressed.\footnote{In the alignment limit where $Z_6\to 0$, it is possible to have $\sbma\to 0$, in which case we would identify the heavier CP-even state $H$ as the SM-like Higgs boson.  We will not consider this possibility further in this paper.}

In both the decoupling and alignment limits, the couplings of $h$ to the fermions should also approach their SM values.  To see how this happens, consider the $h\overline{f}f$ couplings in the case of the type-II 2HDM.  Using the results displayed in Table~\ref{tab:couplings}, the $h\overline{f}f$ couplings relative to those of the SM
(for $f=U,D$) are given by:
\beqa
h\overline{D}D:&& \qquad -\frac{\sin\alpha}{\cos\beta}=\sbma-\cbma\tan\beta\,,\label{hDD}\\
h\overline{U}U:&& \qquad
\phantom{-}\frac{\cos\alpha}{\sin\beta}=\sbma+\cbma\cot\beta\,.\label{hUU}
\label{hddhuu}
\eeqa

In the case of $\cbma=0$, the $h\overline{f}f$ couplings reduce precisely to the corresponding SM values.   However, for values of $\cbma$ that are small but nonzero, the decoupling limit can be ``delayed'' if either $\tanb$ or $\cotb$ is large.   On the other hand, it is desirable to have $(m_t/v)\cot\beta\!\lsim\!1$ and $(m_b/v)\tan\beta\!\lsim\!1$,
in order to avoid nonperturbative behavior in the couplings of $H$, $A$ and $H^\pm$ to the
third generation at scales far below the Planck scale.  In addition, phenomenological constraints arising from $B$ physics observables and $R_b$ mentioned above rule out regions of $\tan\beta\!\lsim \!1$ for large regions of the 2HDM parameter space~\cite{BB2}.  Consequently, we shall focus on the parameter region where
\beq \label{range}
1\lsim\tanb\lsim 50\,.
\eeq
In this case, decoupling is not delayed for the coupling of $h$ to up-type fermions.  On the other hand, for $\tanb$ in the range of interest, it is certainly possible to have $\sbma$ close to 1 and yet have significant departures from decoupling in the coupling of $h$ to down-type fermions.  That is, it is possible to have $\sbma$ close to 1 and yet have $\cbma\tanb\sim\mathcal{O}(1)$.  Since $\cbma$ behaves as $v^2/m_H^2$ in the decoupling limit [cf.~\eq {scexact}], we see that the $h\overline{D}D$ coupling approaches its SM value if
\beq \label{true}
m_H^2\gg v^2\tanb\,.
\eeq
Thus, if $\tanb\gg 1$ we say that we have \textit{delayed decoupling}~\cite{Haber:2000kq},
since a much larger value of the heavy Higgs mass scale is required to achieve decoupling of the heavy Higgs states (i.e.~$m_H\gg v$ is not sufficient).\footnote{Likewise, the alignment limit is also delayed, since the condition $|Z_6|\ll 1$ is now replaced by $|Z_6|\tan\beta\ll 1$.}

The \textit{wrong-sign Yukawa coupling} regime is defined as the region of 2HDM parameter space in which at least one of the couplings of $h$ to down-type and up-type fermion pairs is \textit{opposite} in sign to the corresponding coupling of $h$ to $VV$.  This is to be contrasted with the Standard Model, where the couplings of $h$ to $\overline{f}f$ and $VV$ are of the same sign.  Note that in the convention where $\sbma\geq 0$, the $hVV$ couplings in the 2HDM are always non-negative.
To analyze the wrong-sign coupling regime, it is more convenient to rewrite the type-II Higgs-fermion Yukawa couplings, given by \eqs{hDD}{hUU}, in the following form:
\beqa
h\overline{D}D: &&\qquad -\frac{\sin\alpha}{\cos\beta}=-\sbpa+\cbpa\tan\beta\,,\label{hDDalt}\\
h\overline{U}U:&& \qquad
\phantom{-}\frac{\cos\alpha}{\sin\beta}=\phantom{-}\sbpa+\cbpa\cot\beta\,.\label{hUUalt}
\eeqa

In the case of $\sbpa=1$, the $h\overline{D}D$ coupling normalized to its SM value is equal to $-1$ (whereas the normalized $h\overline{U}U$ coupling is $+1$).
Note that in this limiting case, $\sbma=-\cos 2\beta$, which implies that
the wrong-sign $h\overline{D}D$ Yukawa coupling can only be achieved for values of $\tan\beta > 1$.
Likewise, in the case of $\sbpa=-1$, the $h\overline{U}U$ coupling normalized to its SM value is equal to $-1$ (whereas the normalized $h\overline{D}D$ coupling is $+1$).  In this limiting case, $\sbma=\cos 2\beta$, which implies that the wrong-sign $h\overline{U}U$ couplings can only be achieved for $\tan\beta<1$.  In the type-I 2HDM, both the $h\overline{D}D$ and $h\overline{U}U$ couplings are given by \eq{hUU} [or equivalently by \eq{hUUalt}].  Thus, for $\sbma=-1$, both the normalized $h\overline{D}D$ and $h\overline{U}U$ couplings are equal to $-1$, which is only possible if $\tanb<1$.  In light of \eq{range},
only the wrong-sign $h\overline{D}D$ coupling regime of the type-II 2HDM can be realistically achieved.

It should be emphasized that the above conclusions do not depend on the convention adopted for the range of the angle $\alpha$.  In the convention used in Sections~\ref{sec-results} and \ref{sec-results2} of this paper, we scan over $|\alpha|\leq\pi/2$, which allows for the possibility of negative $\sbma$.  However, the definition of the wrong-sign Yukawa coupling is not changed as it refers to the \textit{relative} sign of the $h\bar{f}f$ and $hVV$ couplings.  To translate between both conventions, one simply must shift $\alpha\to\alpha\pm\pi$ (the sign chosen so that $\alpha$ is in its desired range).    In practice, the scans of Section~\ref{sec-results} and \ref{sec-results2} focus on the wrong-sign $h\overline{D}D$ coupling regime where $\tanb>1$, in which case $\sbma>0$ and the distinction between the two conventions becomes moot.

In the above discussion of the wrong-sign Yukawa coupling regime, we have not yet imposed the requirement that $h$ is SM-like.  In particular, for a fixed value of $\tan\beta$, the limit of $\sbpa\to 1$ is \textit{not} the decoupling limit (indeed the $hVV$ couplings do not approach their SM values except in the limit of $\alpha\to 0$ and $\beta\to\half\pi$).   This implies that for $|\cbma|\ll 1$ we must have $\tanb\gg 1$,   Likewise, the limit of $\sbpa=-1$ is not the decoupling limit unless $\beta\to 0$ and $\alpha\to -\half\pi$, i.e.~$\cot\beta\gg 1$.  Again, we see that for values of $\tan\beta>1$, among all possible wrong-sign Yukawa coupling scenarios only the wrong-sign $h\overline{D}D$ coupling in the type-II 2HDM is phenomenologically viable.

Therefore, in this paper, we shall explore the possibility that the $h\overline{D}D$ coupling normalized to its SM value is close to $-1$ in the type-II 2HDM.\footnote{The possibility that a parameter regime of the MSSM Higgs sector exists with a wrong-sign $h\overline{D}D$ coupling is addressed in Appendix~\ref{app:mssm}.}  
This scenario was first examined in Ref.~\cite{GKO} and then later clarified in Ref.~\cite{decoupling}.
Current LHC Higgs observations are not sufficiently precise as to allow one to distinguish this case from that of the SM Higgs boson.
To study this case, we first define a parameter $\epsilon$ by defining the normalized $h\overline{D}D$ coupling to be given by
\beq \label{edef}
-\frac{\sin\alpha}{\cos\beta}=-1+\epsilon\,.
\eeq
Multiplying \eq{edef} by $-2\cos^2\beta$, and employing the trigonometric identity,
$2\cos\beta\sin\alpha=\sbpa-\sbma$, it follows that\footnote{Although we are interested in the 2HDM parameter regime where $\epsilon$ is small, \eq{epsexact} is valid for all
values of $\epsilon$.   In particular, for $\epsilon=2$ we have $\sbma=1$ and $\sbpa=-\cos 2\beta$, which is consistent with the result of \eq{epsexact}.}
\beq \label{epsexact}
\sbpa-\sbma=2(1-\epsilon)\cos^2\beta\,.
\eeq
By employing the trigonometric identity $\sbma=\sin 2\beta\cbpa-\cos 2\beta\sbpa$ and taking $0\leq\beta\leq\half\pi$, one can also derive
\beqa
\sbpa&=&\phantom{-}(1-\epsilon)\cos^2\beta+\sin\beta\sqrt{1-(1-\epsilon)^2\cos^2\beta}\,,\label{sp}\\
\cbpa&=&-(1-\epsilon)\sin\beta\cos\beta+\cos\beta\sqrt{1-(1-\epsilon)^2\cos^2\beta}\,.\label{cp}
\eeqa

Using \eqs{hDD}{edef}, it follows that
\beq \label{edef2}
\tanb=\frac{1+\sbma-\epsilon}{\cbma}\,.
\eeq
Since the $hVV$ couplings are assumed to be close to the SM, we still must impose the constraint that $|\cbma|\ll 1$.   Thus, in the case of the wrong-sign $h\overline{D}D$ Yukawa coupling, we must have
$\tan\beta\gg 1$, which is the region of delayed decoupling defined below \eq{true}.

For completeness, we also also examine the case of a wrong-sign $h\overline{U}U$ coupling in the type-II 2HDM (or the case of the wrong-sign $h\overline{U}U$ and $h\overline{D}D$ couplings in the type-I 2HDM) by taking $\sbpa$ close to $-1$ [cf.~\eq{hUUalt}].  To study this case, we first define a parameter $\epsilon^\prime$ via
\beq \label{epdef}
\frac{\cos\alpha}{\sinb}=-1+\epsilon^\prime\,,
\eeq
which yields an $h\overline{U}U$ coupling normalized to its SM value given by $-1+\epsilon^\prime$.
An analysis similar to the one used in the case of the wrong-sign $h\overline{D}D$ Yukawa coupling yields
\beq \label{epspexact}
\sbpa+\sbma=-2(1-\epsilon^\prime)\sin^2\beta\,.
\eeq
and
\beqa
\sbpa&=&-(1-\epsilon^\prime)\sin^2\beta-\cos\beta\sqrt{1-(1-\epsilon^\prime)^2\sin^2\beta}\,,\label{sp2}\\
\cbpa&=&-(1-\epsilon^\prime)\sin\beta\cos\beta+\sin\beta\sqrt{1-(1-\epsilon^\prime)^2\sin^2\beta}\,.\label{cp2}
\eeqa
Using \eqs{hUU}{edef2}, it follows that
\beq \label{epdef2}
\cot\beta=\frac{-\sbma-1+\epsilon^\prime}{\cbma}\,,
\eeq
For values of $|\cbma|\ll 1$,  \eq{epdef2} can only be satisfied if $\cot\beta\gg 1$, which lies outside the range of $\tan\beta$ under consideration [cf.~\eq{range}], as previously noted.

To complete the analysis of the tree-level Higgs couplings, we briefly look at the $h$ self-coupling.  In the 2HDM, the $hhh$ coupling\footnote{A similar analysis can be given for the $hhhh$ coupling using the results given in Ref.~\cite{decoupling}.  However, this coupling cannot be realistically probed by the LHC and ILC, so we will not provide the explicit expressions here.}  
is given by~\cite{decoupling}:
\beq
G_{hhh}=-3v\bigl[Z_1\sin^3(\beta-\alpha)+3Z_6\cbma\sin^2(\beta-\alpha)+(Z_3+Z_4+Z_5)\sbma\cos^2(\beta-\alpha)+Z_7\cos^3(\beta-\alpha)\bigr]\,,
\eeq
where in the softly broken $\mathbb{Z}_2$ symmetric 2HDM, the $Z_i$ are given in 
\eqst{z1}{z7}.
Rewriting the $Z_i$ in terms of the $\lambda_i$ yields
\beq
G_{hhh}=3v\bigl[-\cos\beta\sin^2\alpha \lambda_1+\sin\beta\cos^3\alpha\lambda_2
-\sin\alpha\cos\alpha\cbpa(\lambda_3+\lambda_4+\lambda_5)\bigr]\,,
\eeq
which reproduces the result given in Ref.~\cite{Ginzburg}.  Using the results of Appendix D of Ref.~\cite{decoupling}, we can rewrite the $hhh$ coupling in a more convenient form,
\beq
G_{hhh}=\frac{-3}{v\sin^2\beta\cos^2\beta}\biggl[\sin\beta\cos\beta(\cos\beta\cos^3\alpha-\sin\beta\sin^3\alpha)m_h^2-\cos^2(\beta-\alpha)\cbpa m_{12}^2\biggr]\,,
\eeq
which reproduces the result given in Ref.~\cite{Arhrib} (after correcting a missing factor of 2).

In the decoupling/alignment limit where $\sbma=1$, we have $\cos\alpha=\sin\beta$ and $\sin\alpha=-\cos\beta$.  Then, the $hhh$ coupling reduces to the SM value,
\beq
G_{hhh}\to G_{hhh}^{\rm SM}=-{3m_h^2\over v}\,.
\eeq
In the wrong-sign Yukawa coupling limit for type-II Higgs couplings to down-type [up-type] fermions, respectively, where $\sbpa=+1$ [$-1$], we have $\cos\alpha=+[-]\sin\beta$ and $\sin\alpha=+[-]\cos\beta$, so that
\beq
G_{hhh}\to -[+]G_{hhh}^{\rm SM}\cos 2\beta\,,
\eeq
which reduces to the SM value only when $\beta\to \half\pi$ [$\beta\to 0$] for type-II Higgs couplings to down-type [up-type] fermions, respectively.
It is quite remarkable that this matches the behavior of the $hVV$ coupling in the same limit.  In particular, for $\sbpa=\pm1$, we have $\sbma=\mp\cos 2\beta$, as previously noted.  Hence in the wrong-sign Yukawa coupling limit, \eq{VV} yields
\beq
G_{hVV}\to -[+]G_{hVV}^{\rm SM}\cos 2\beta\,.
\eeq
Of course, the corresponding first order corrections to the $hhh$ and $hVV$ couplings will differ as one moves away from the strict limiting case treated above.

In the decoupling and alignment  limits discussed at the beginning of this subsection, the tree level couplings of $h$ approach the corresponding values of the SM Higgs boson.  The behavior of the decoupling and alignment limits differ when one-loop effects are taken into account.  In the decoupling limit, the properties of $h$ continue to mimic those of the SM-like Higgs boson since the effects of the $H$, $A$ and $H^\pm$ loops decouple in the limit of heavy scalar masses.  In contrast, the alignment limit only requires that $|Z_6|\ll 1$, so that in principle the masses of $H$, $A$ and $H^\pm$ could be relatively close to the electroweak scale.  In this case, the loop effects mediated by $H$, $A$ and $H^\pm$ can compete with other electroweak radiative effects and thus distinguish between $h$ and the SM Higgs boson.

In processes in which the one-loop effects are small corrections to tree level results, very precise measurements will be required to distinguish between $h$ and the SM Higgs boson in the alignment limit.  Indeed, a much more fruitful experimental approach in this case is to search directly for the $H$, $A$ and $H^\pm$ scalars!  However, in Higgs processes that are absent at tree level but arise at one-loop, the loop effects mediated by  $H$, $A$ or $H^\pm$ can compete directly with deviations that arise due to small departures from the alignment limit.  The most prominent example is the decay rate for $h\to\gamma\gamma$.  Departures from the SM decay rate for $h\to\gamma\gamma$ can arise either from deviations in the $hW^+ W^-$, $h\bar{t}t$ and/or $h\bar{b}b$ couplings from their SM values, or from the contributions of the charged-Higgs boson loop (which is not present in the SM).    To compute the latter, we need to compute the $hH^+H^-$ coupling.  Using the results of Ref.~\cite{decoupling}, 
\beq \label{hHH}
G_{hH^+H^-}=-v\bigl[Z_3\sbma+Z_7\cbma\bigr]\,,
\eeq
where $Z_3$ and $Z_7$ are defined in terms of the $\lambda_i$ in \eqs{z345}{z7}.
It is convenient to reexpress the $hH^+H^-$ coupling in terms of the Higgs masses and $\lambda_5$.
Using the expressions given in
Appendix D of Ref.~\cite{decoupling}, we obtain
\beq \label{alt}
G_{hH^+H^-}=\frac{1}{v}\left[\bigl(2m_A^2-2m_{H^\pm}^2-m_h^2+2\lambda_5 v^2\bigr)\sbma
+\bigl(m_A^2-m_h^2+\lambda_5 v^2\bigr)(\cot\beta-\tan\beta)\cbma\right]\,,
\eeq
where
\beq \label{mdiff}
m_A^2-m_{H^\pm}^2=\half v^2(\lambda_4-\lambda_5)\,.
\eeq
In the alignment limit where the masses of $H$, $A$ and $H^\pm$ are of order the electroweak scale,
$G_{hH^+H^-}\sim\mathcal{O}(v)$ and the charged-Higgs loops can compete with the SM loops that contribute to the $h\to\gamma\gamma$ one-loop amplitude.  In the normal decoupling limit where $m_{H^\pm}^2\sim m_A^2\gg \mathcal{O}(v^2)$ and $|\cbma|\sim v^2/\mha^2$, $G_{hH^+H^-}\sim\mathcal{O}(v)$  as expected, in which case the charged-Higgs loop contribution to the
$h\to\gamma\gamma$ amplitude is suppressed by a factor of $v^2/m^2_{H^\pm}$. Note that this factor is of the same order as $\cbma$.   The contribution of the fermion loops also deviates from the SM by a factor of $\mathcal{O}\bigl(\cbma\bigr)$  due to the modified tree level $h\overline{f}f$
couplings [cf.~\eqs{hDD}{hUU}].\footnote{The contribution of the $W^\pm$ loop deviates from the SM by a factor of $\mathcal{O}\bigl(\cos^2(\beta-\alpha)\bigr)$ in light of \eq{VV}.}  However, in the decoupling limit the contribution of the bottom-quark loop is suppressed by a factor of $m_b^2/v^2$ and can thus be ignored.  We conclude that the deviation from the SM in the decoupling limit is due primarily to the top-quark loop and the charged-Higgs loop, whose contributions to the $h\to\gamma\gamma$ decay amplitude are of the same order of magnitude.

The form of the $hH^+H^-$ coupling given in \eq{alt} suggests the existence of 2HDM parameter regimes in which $G_{hH^+H^-}\gg \mathcal{O}(v)$, even under the assumption that $\cbma\sim\mathcal{O}(v^2/\mhh^2)\ll 1$.  For example,
if we allow $\lambda_4-\lambda_5$ to be large and if 
$\mhpm$, $\mha$, $\mhh\gg v$, then it possible to have $m_A^2-\mhpm^2\sim\mathcal{O}(\mhpm^2)$.  It would then follow that the contribution of
the charged-Higgs loop contribution to the $h\to \gamma\gamma$ amplitude, which scales as
$G_{hH^+H^-}/m_{H^\pm}^2$, approaches a constant in the region of $m_{H^\pm}\gg m_h$.
This nondecoupling behavior was first emphasized in Ref.~\cite{Arhrib:2003ph} and subsequently reexamined in Ref.~\cite{Bhattacharyya:2013rya}.  Indeed, the behavior of the charged-Higgs loop in the nondecoupling regime is similar to the contribution of a heavy fermion loop to the $h\to \gamma\gamma$ amplitude, which scales as $G_{h\bar{f}f}/m_f$ and approaches a constant for $m_f\gg m_h$.  In particular, if $m_f$ is too large, then $G_{h\bar{f}f}=m_f/v\gg 1$  and tree level unitarity is violated.  However, there is an intermediate range of heavy fermion masses above $m_h$ but below the mass scale at which tree level unitarity is violated, in which the fermion loop contribution to the $h\to \gamma\gamma$ amplitude is approximately constant.   Likewise, if $G_{hH^+H^-}/v\gg 1$ is too large then one would need to take $Z_3$ above its unitarity bound [in light of \eq{hHH}].  Again, there is an intermediate region of heavy Higgs masses (where tree level unitarity is still maintained) in which the charged-Higgs loop contribution to the $h\to \gamma\gamma$ amplitude is approximately constant.
Thus, we expect regions of 2HDM parameter space in which a SM-like Higgs can exhibit a non-negligible deviation in $\Gamma(h\to\gamma\gamma)$ from SM 
expectations.

Alternatively, the second term on the right-hand side of \eq{alt} can be enhanced in the delayed decoupling regime where $\tan\beta\gg 1$ and $|\cbma|\tan\beta\sim\mathcal{O}(1)$.
In this case, $G_{hH^+H^-}\sim\mathcal{O}(m_A^2/v)
\sim\mathcal{O}(m_{H^\pm}^2/v)$ [under the assumption that $\lambda_4$, $\lambda_5\lsim\mathcal{O}(1)$].   However, this behavior is also associated with growing quartic couplings that can potentially violate tree level unitarity.
Indeed, by comparing with \eq{hHH}, we see that $Z_7$ is being enhanced.  More directly,
it is straightforward to obtain
\beq
\bigl(m_A^2-m_h^2+\lambda_5 v^2\bigr)(\cot\beta-\tan\beta)\cbma=-v^2 \cos 2\beta\bigl[
(\lambda_1-\lambda_3-\lambda_4-\lambda_5)\cosb\sin\alpha+
(\lambda_2-\lambda_3-\lambda_4-\lambda_5)\sinb\cos\alpha\bigr]\,,
\eeq
which again implies that some of the Higgs potential parameters must be enhanced by a factor
of $\mathcal{O}(m_A^2/v^2)$ if $\mha^2\gg v^2$ and $|\cbma|\tan\beta\sim\mathcal{O}(1)$.
Thus, if $m_A$ becomes too large, the unitarity constraints on the Higgs potential quartic coupling parameters will be violated.   Nevertheless, there exists an intermediate range of charged-Higgs masses in which tree level unitarity is maintained while  the charged-Higgs loop contribution to the $h\to \gamma\gamma$ amplitude is approximately constant.   That is, there exists a
region of 2HDM parameter space, in which $|\cbma|$ is small and the $h\overline{D}D$ coupling is opposite in sign to that of the SM Higgs boson, where
a deviation in the $h\to\gamma\gamma$ decay rate from the predicted SM rate due to the contribution of the charged-Higgs loop can be detected.

Details on the nondecoupling of the $H^\pm$ loop contribution to the $h\to\gam\gam$ amplitude can be found in Appendix~\ref{apjfg}, where it is shown that such nondecoupling is inevitable for the wrong-sign $h\overline{D}D$ coupling scenario.  
The resulting magnitude of the effect yields deviations from the SM that will ultimately be observable at the LHC and a future linear collider, as discussed in the following sections.

%%%%%%%%%%%%%%%%%%%%%%%%%%%%%%%%%%%%%%%%%%%%%%%%%%%%%%%%%%%%%%%%%%%%%%%%
% Results
%
%%%%%%%%%%%%%%%%%%%%%%%%%%%%%%%%%%%%%%%%%%%%%%%%%%%%%%%%%%%%%%%%%%%%%%%
\section{Phenomenology of the Wrong-sign Yukawa couplings}
\label{sec-results}

It is convenient to define the ratio of the $h\to f$ coupling to the corresponding SM value as
\beq
\kappa_f= {g_{hf}\over g_{\hsm f}}\,,
\label{kfrat}
\eeq
where we will be considering $f=b\anti b$, $c\anti c$, $\tau^+\tau^-$, $WW^*$, $ZZ^*$.
As for the coupling to photons, $\kappa_\gamma$ is defined as
\begin{equation}
\kappa_\gamma^2\,=\,\frac{\Gamma^{\rm 2HDM}(h\to \gamma\gamma)}{\Gamma(\hsm\to \gamma\gamma)}\,,
\end{equation}
with an analogous definition for $\kappa_g$. Note that $\kappa_\gamma$ and $\kappa_g$
are strictly positive, whereas the remaining $\kappa_f$ could be either positive or negative. These definitions
for the couplings $\kappa$ coincide with the definitions used by the experimental
groups at the LHC~\cite{rec}, at leading order.  We shall also make the simplifying assumption (which holds in the SM and in the 2HDM under consideration) that all down-type [up-type] fermion final states are governed by the same $\kd$ [$\ku$].
It is convenient to begin with a simplified discussion of the impact of changing the sign of $\kd$ in order to set the stage. In this section, we employ the convention of $|\alpha|\leq \pi/2$ for which $\ku>0$ in both type-I and type-II models. For this choice, the $hVV$  coupling of \eq{hwwform} can, in principle, be either positive or negative.  However, for $\ku>0$, the phenomenology of the $\gam\gam$ final state requires that the $hVV$ coupling be positive, which means that acceptable regions of parameter space must have $\beta-\alpha>0$.

Consider first the amplitude of the process $h \to gg$. In an appropriate normalization, the top- and bottom-quark loops contribute $4.1289$ and $-0.2513 + 0.3601i$, respectively,
when $\ku=1$ and $\kd=1$. This implies a large fractional change in the $ggh$ coupling with a change of sign of $\kd$. One finds a
shift in  $\kg$ of  $+13\%$  in going from positive $\kd=+1$ to  $\kd=-1$. Naively, one would suppose that this large shift would
be easily observed. However, this is a difficult task at the LHC due to the challenge in
identifying gluons (even if indirectly) in the final state.
In addition, the primary $gg$ fusion production cross section has  some systematic errors associated with higher order corrections.
Nonetheless, Table~1-20 of Ref.~\cite{Dawson:2013bba} gives expected errors for $\kg$ of $6$--$8\%$ for $L=300\fbi$ and $3$--$5\%$ for $L=3000\fbi$, based on
fitting all the rates rather than directly observing the $gg$ final state.  At the ILC, the primary production mechanism of $\epem\to Z^*\to Zh$
is very well determined in terms of the $ZZh$ coupling and isolation of the $gg$ final state is easier. The error on $\kg$ estimated in
Ref.~\cite{Dawson:2013bba} is $2\%$ for a combination of $L=250\fbi$ at $\rts=250\gev$ and $L=500\fbi$ at $\rts=500\gev$. Other error estimates are to
be found in Ref.~\cite{Ono:2012ah, Asner:2013psa} where it is concluded that $\kg$ can be measured at the ILC with an accuracy of 8.5\% at a
center-of-mass (CM) energy of
250 GeV and 7.3\% at a CM energy of 350 GeV with an integrated luminosity of 250 fb$^{-1}$ and beam polarization
of $-$80\% (electron) and 30\% (positron)~\cite{Ono:2012ah}.  The error estimate for $500\gev$ with $L=500\fbi$ decreases to $\sim 2.3\%$ (see
Tables 6.1 and 6.2 of Ref.~\cite{Asner:2013psa}), consistent with the estimate from Ref.~\cite{Dawson:2013bba}.
Thus, in the end, we can anticipate that  both the LHC and ILC will be able to determine whether or not $h_D$ is positive using the indirect fit
and direct measurement of $\kg$, respectively.

 In $h \to \gamma \gamma$, the presence of the large $W$-boson loop contribution means that considerable precision is required to identify  the
 interference effects.  In more detail, the contributions to the amplitude of this process assuming SM couplings are as follows [these are the $I^h$'s defined in \eq{sform}];  $W$ boson,
 $-8.3233$; top-quark loop (with $\ku>0$), $1.8351$; bottom loop for $\kd>0$, $-0.0279 + 0.0400i$.  As a result, switching the sign of $\kd$ would
 change $\kgam$ from $1$ to $0.991$, i.e.~a $\lsim 1\%$ shift.  The accuracy with which $\kgam$ can be measured at
 the 14 TeV LHC is given in Table~1-20 of Ref.~\cite{Dawson:2013bba} as $5\%$--$7\%$ for integrated luminosity of $L=300\fbi$ and $2\%$--$5\%$ for $L=3000\fbi$. The ranges correspond to
 optimistic/pessimistic estimates regarding systematic and theoretical errors. Thus, if the change in $\kgam$ was only of order $1\%$ this could
 not be detected at the LHC. Nonetheless, we claim that with high enough integrated luminosity one can distinguish $\kd<0$ from $\kd>0$ in
 the context of the type-II 2HDM using the high precision $\gam\gam$ final state due to the fact that the $\gam\gam h$ coupling is inevitably suppressed
 in the $\kd=-1$ case as a result of a large nondecoupling charged-Higgs loop contribution, {\it i.e.} one that approaches a constant at large
 charged-Higgs mass (up to the limit at which the $\lam_i$ couplings violate the tree level unitarity bounds).  There is also a $\kd>0$ region of parameter space for which the charged-Higgs loop does not decouple, and this region would be ruled out in very similar fashion to the $\kd<0$ region we focus on. In fact, the modification to the $\gam\gam$ coupling is the {\it only} way of revealing this $\kd>0$ nondecoupling region. However, for $\kd>0$ there is also the standard decoupling region for which the charged-Higgs loop does decouple and that would allow arbitrarily precise agreement with the SM predictions.   A detailed discussion of nondecoupling and perturbativity/unitarity bounds is given in Appendix \ref{apjfg}. Typically, $\kgam$ is inevitably suppressed relative to the SM prediction by
 about 5\% for $\kd<0$, which  should be measurable at the $\rts=14\tev$ LHC run for $L=3000\fbi$. At the ILC, for a combination of $L=250\fbi$ at
 $\rts=250\gev$ and $L=500\fbi$ at $\rts=500\gev$ the expected error on $\kgam$ is $\sim 8.3\%$ (including a $0.5\%$ theory uncertainty) based on
 measuring $\epem\to Zh$ with $h\to\gam\gam$, implying that the sign of $\kd$ cannot be directly determined at the ILC using the $\gam\gam$ final
 state.

To explore in more detail, we scan over the 2HDM parameter space subject to all
the constraints described in section~\ref{sec-model}.
We begin by fixing $m_h = 125\, \textrm{GeV}$.
The charged-Higgs mass is varied between 100 and 900 GeV in type-I
and 340 and 900 GeV in type-II. The heavier CP-even scalar
mass was kept between 125 and 900 GeV while the
pseudo-scalar mass range is between 90 and 900 GeV.
The soft-breaking parameter was varied in the range
$- (900~{\rm GeV})^2<m_{12}^2 <(900~{\rm GeV})^2$
while $1 < \tan \beta < 30$.
After passing all constraints, the points were used to calculate the various
$\muflhc$, which is the ratio of the number of events predicted by the model
for the process $pp \to h \to f$
to the SM prediction for the same final state,
\begin{equation}
\mu^h_f ({\rm LHC})\,=\, \frac{\sigma^{\rm 2HDM} (pp\rightarrow h) \,{\rm BR}^{\rm 2HDM}(h \rightarrow f)}
{\sigma^{\rm SM} (pp\rightarrow \hsm) \,{\rm BR}(\hsm \rightarrow f)} \, .
\label{eq:Rh}
\end{equation}
In computing the $pp\to h$ cross section, we have summed over
all light Higgs production mechanisms ($gg$ fusion, $VV$ fusion, $Vh$ associated production, $b\bar{b}$ fusion, and $t\bar{t}h$ associated production),\footnote{Further discrimination among models and parameter choices within a model can be obtained by separately considering each individual initial state $\times$ final state combination, as is now done by the LHC experimental collaborations and considered in Ref.~\cite{Belanger:2013xza}.} 
employing a mass of $m_h=125\gev$.
The processes that involve only Higgs couplings to gauge bosons can be obtained
by simply rescaling the SM cross sections. Hence, for
Higgs Strahlung and vector boson fusion we have used the results of Ref.~\cite{LHCHiggs} (the same applies to the final state $h t \bar t$).
We have included QCD corrections but not the SM electroweak corrections
because they can be quite different for the 2HDMs. Cross sections for
$b \bar{b} ~\to h$ are included at next-to-next-to-leading order (NNLO)~\cite{Harlander:2003ai}
and the gluon fusion process $gg \to h$ was calculated using
HIGLU~\cite{Spira:1995mt}.  (When the Higgs boson is SM-like, the $b\bar b \to h$ cross section is much smaller than the $gg\to h$ cross section.)
 Our baseline will be to require that the $\muflhc$ for final states $f=WW$, $ZZ$, $b\bar b$, $\gamma \gamma$ and $\tau^+ \tau^ -$  are each consistent with unity within 20\%, which is a rough approximation to the precision of current data. We then examine the consequences of requiring that all the $\muflhc$ be within 10\% or 5\% of the SM prediction. This enables us to understand how an increase in precision affects the scenario we will now describe in detail.

%%%%%%%%%%%%%%%%%%%%%%%%%%%%%%%%%%%%%%%%%%%%%%%%%%%%%%%%%%%%%%%%%%%%%

%
\begin{figure}[t!]
\bec
\includegraphics[width=3.5in,angle=0]{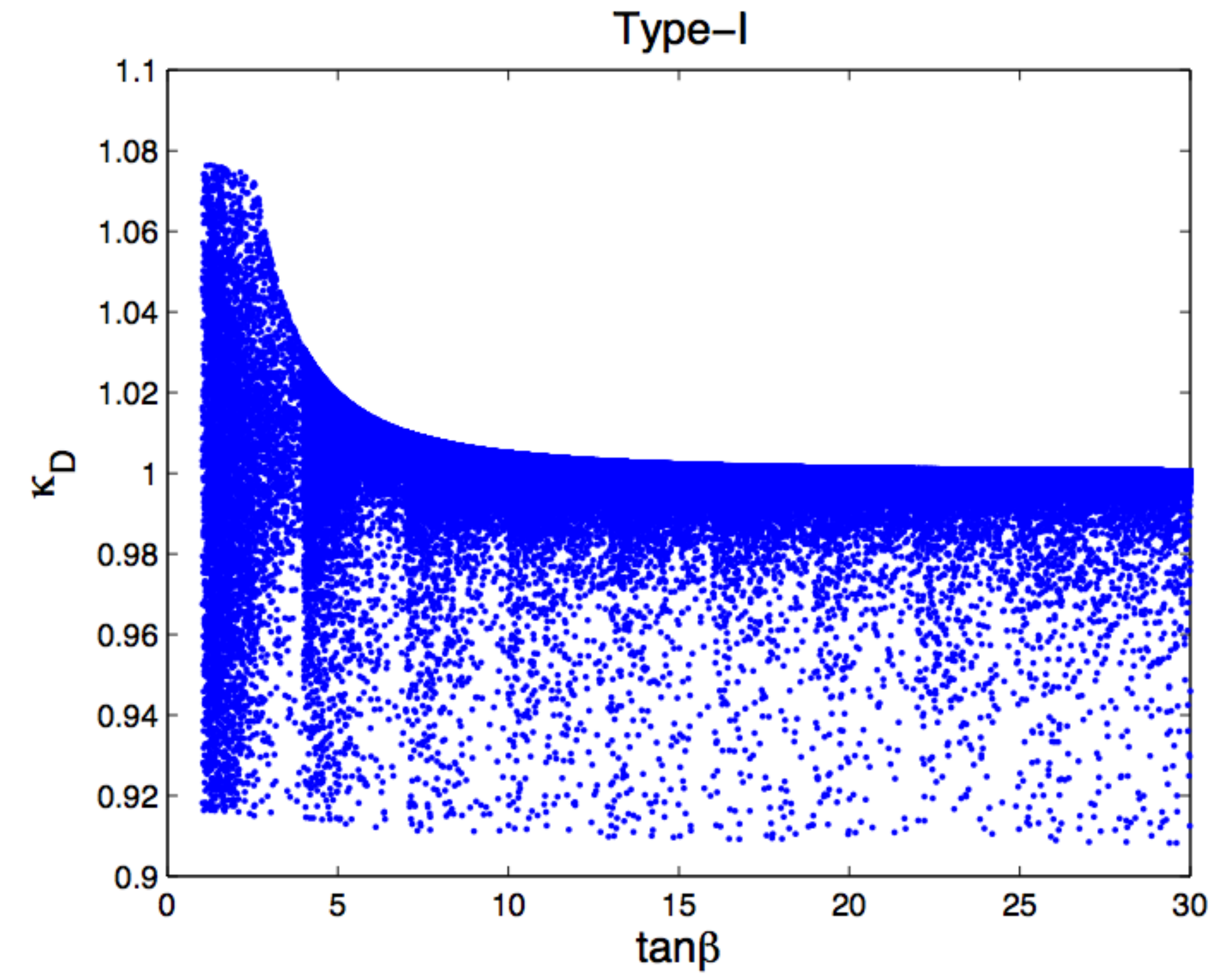}
\hspace{-.3cm}
\includegraphics[width=3.5in,angle=0]{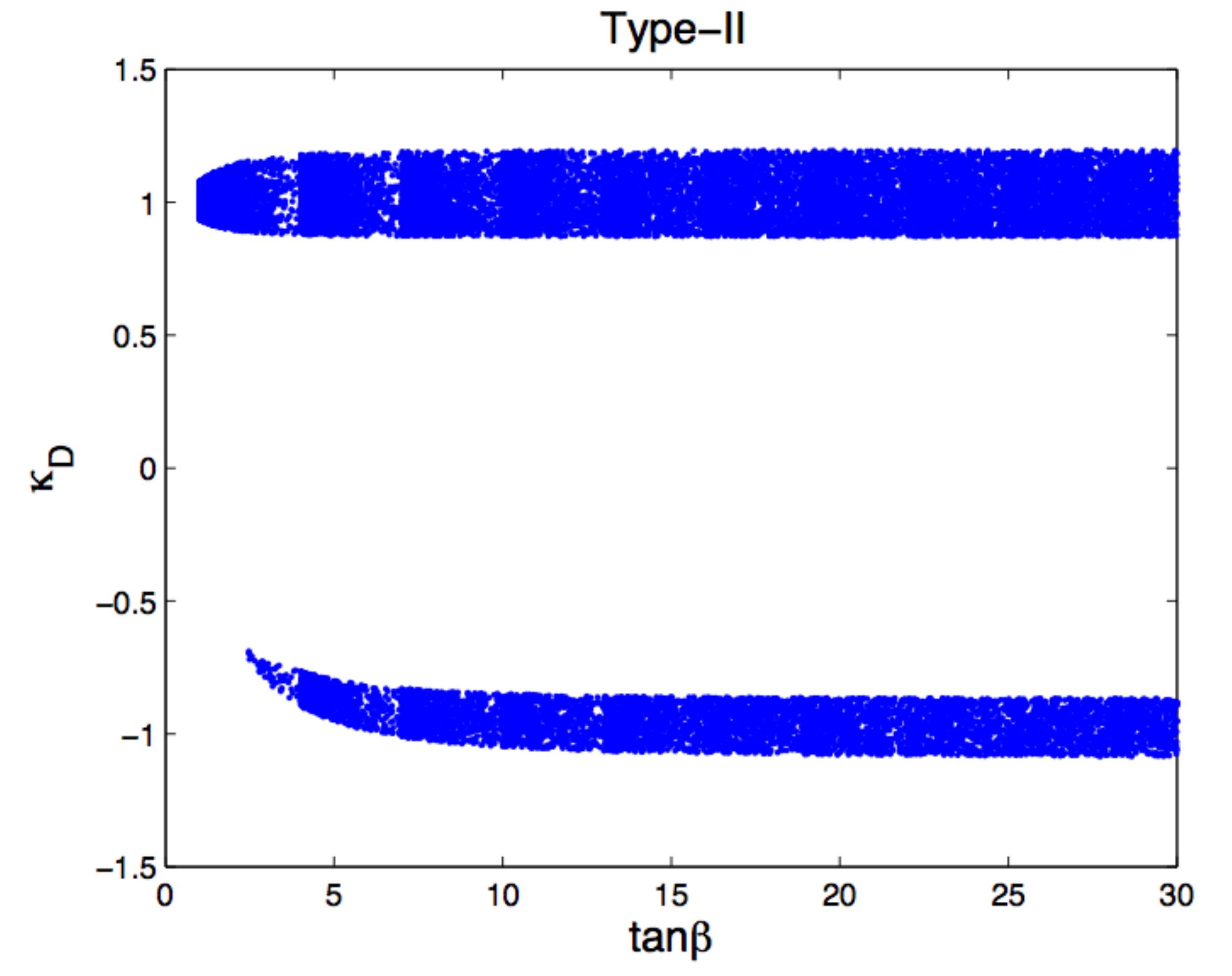}
\eec
\caption{Ratio of the lightest Higgs couplings to down quarks in the 2HDM relative to the SM as a function of $\tan \beta$.
Left: Type-I and right: type-II. All  $\muflhc$ are within 20\% of the SM value. }
\label{fig:TI}
\end{figure}
As discussed in Sections \ref{sec-model} and \ref{sec:decoup},
the relative $hVV$ coupling is given by $\kappa_V=g^{\rm 2HDM}_{hVV}/g^{\rm SM}_{hVV}= \sin ( \beta - \alpha)$.  The relative Higgs-fermion couplings are
$\ku=\kd=\cos \alpha / \sin \beta$ in the type-I 2HDM, whereas in the type-II 2HDM,
\begin{equation}
\ku= \frac{\cos \alpha}{\sin \beta} \,,\qquad\qquad
\kd = - \frac{\sin \alpha}{\cos \beta}  \, .
\end{equation}
In Fig.~\ref{fig:TI} we show $\kd$ in  type-I and type-II models as a function of $\tan \beta$ for those
parameter space points that pass all theoretical and experimental constraints and have all $\muflhc$ within 20\% of the SM prediction of 1.   In all cases, $\kappa_V>0$, which implies that a wrong-sign Yukawa coupling would correspond to a negative value of $\kappa_D$ or $\kappa_U$.  Noting that $\kappa_U\geq 0$ in the convention of $|\alpha|\leq\pi/2$, it follows that only regions with $\kappa_D<0$ correspond to a wrong-sign Yukawa coupling scenario.

As expected, the
left panel of Fig.~\ref{fig:TI} shows that all points are very close to $\kd=1$ for a type-I 2HDM, while the right panel shows that in the case of the type-II 2HDM all points fall within two
main regions: one where $\kd \approx 1$ and the other one where $\kd \approx -1$.
In short, although the LHC results have clearly shown that the Higgs rates to fermions and gauge bosons are very consistent with the SM predictions, it is clear that the roughly $20\%$ precision with which LHC rates are currently measured allows for a second non-SM-like region with the opposite sign of $h_D$ that can fit within the context of the type-II 2HDM.

In Section~\ref{sec:decoup}, we showed that
in the type-II 2HDM,  the $\kd\sim +1$ region corresponds to the limit $\sin (\beta - \alpha) \approx 1$ whereas the $\kd\sim -1$
region is attained in the limit $\sin (\beta + \alpha) \approx 1$,
\begin{equation}
\kd^{II}\to 1 \qquad (\sin (\beta- \alpha) \to 1) \, ; \qquad
\kd^{II}  \to -1 \qquad (\sin (\beta+ \alpha) \to 1)  \,,
\end{equation}
corresponding to negative and positive values of $\sin \alpha$, respectively (in a convention where $0\leq\beta\leq\half\pi$).
On the other hand, the relative $hVV$ and $hhh$ couplings satisfy
\begin{equation}
\kappa_V, \frac{G^{\scriptscriptstyle {\rm 2HDM}}_{hhh}}{G^{\scriptscriptstyle {\rm SM}}_{hhh}} \to 1 \quad (\sbma\to 1);\qquad \kappa_V, \frac{G^{\scriptscriptstyle {\rm 2HDM}}_{hhh}}{G_{hhh}^{\scriptscriptstyle {\rm SM}}}\to \frac{\tan^2\beta-1}{\tan^2\beta+1} \quad (\sbpa\to 1).\label{eq-tan}
\end{equation}

\begin{figure}[t!]
\bec
\includegraphics[width=3.5in,angle=0]{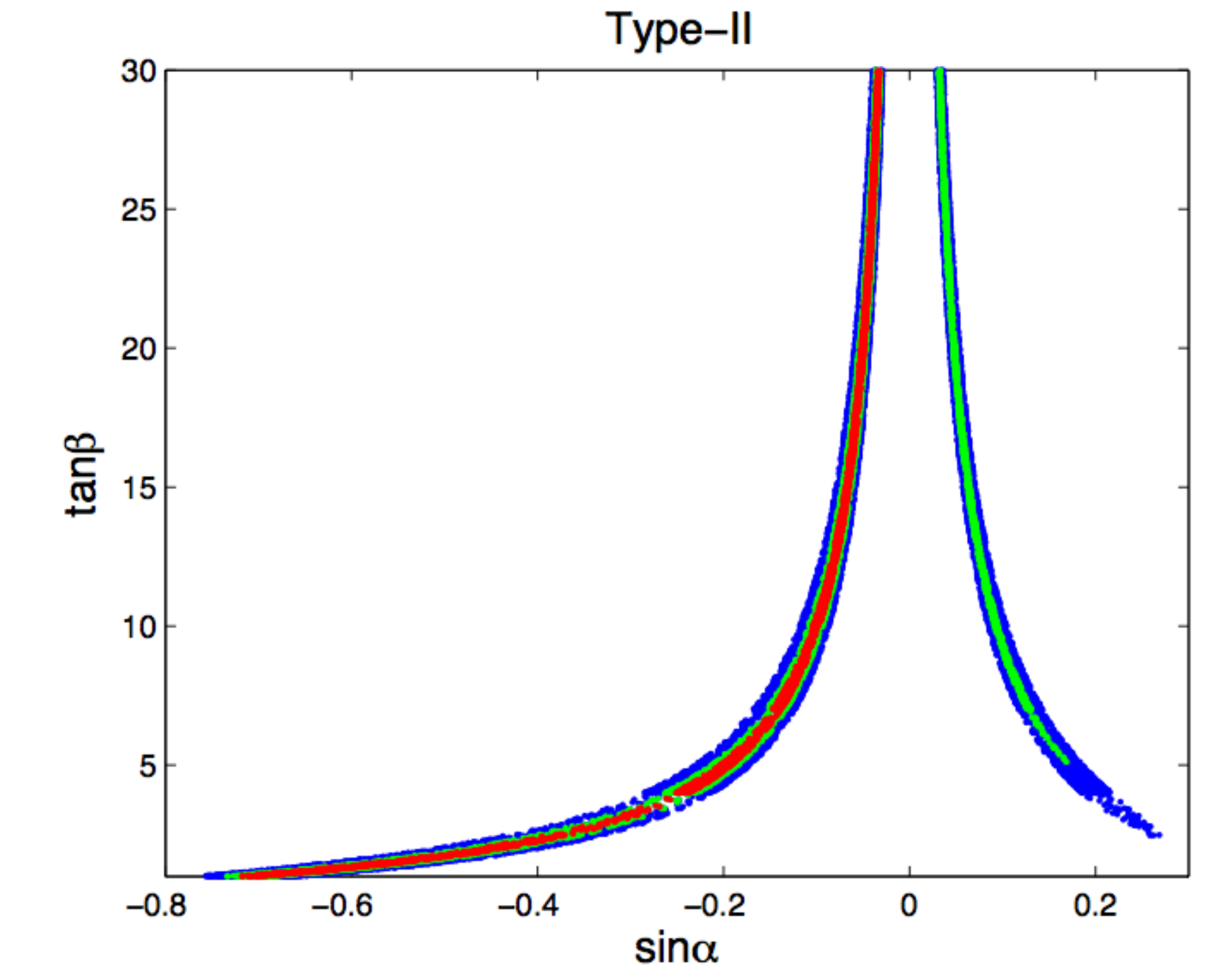}
\includegraphics[width=3.5in,angle=0]{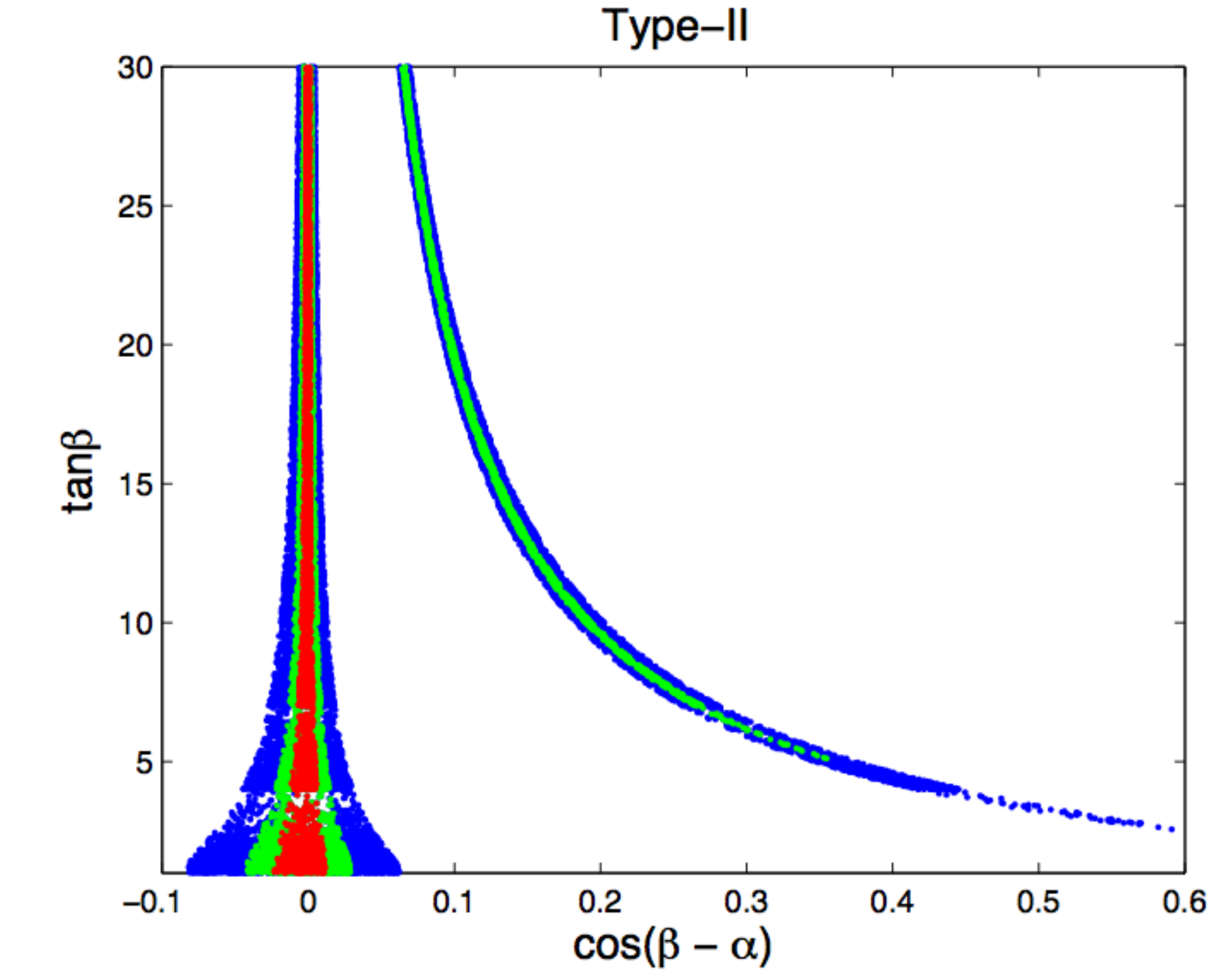}\eec
\caption{Allowed regions for 2HDM type-II with all $\muflhc$ within 20\% (blue/black),
 10\% (green/light grey) and 5\% (red/dark grey) of the SM value of unity. Left: In the $\tan \beta$ vs. $\sin \alpha$ plane. Right: In $\tanb$ vs. $\cos(\beta-\alpha)$ space, }
\label{fig:T2}
\end{figure}

In Fig.~\ref{fig:T2}, the left panel shows $\tan \beta$ as a function of $\sin \alpha$ with all $\muflhc$ within 20\% (blue/black),
 10\% (green/light grey) and 5\% (red/dark grey) of their SM values. We clearly see two branches---one with $\sin \alpha <0$
corresponding to the SM limit and one with $\sin \alpha  > 0$ corresponding to the wrong-sign Yukawa coupling scenario.
In the left branch, the points are all such that  $\sin (\beta - \alpha) \sim 1$; the points in the right branch
all have $\sin (\beta + \alpha) \sim 1$. The right panel shows that as $\tanb$ increases the $\kd<0$ branch corresponds to parameters with small $\cbma$, i.e.~$\sbma\sim 1$.
Note that the second branch is excluded if we demand that all the $\muflhc$ fall within 5\% of unity. %excludes the second branch.
%, a point to which we return shortly.

It is instructive to consider why
  $\sin (\beta + \alpha) \approx 1$ with $\kd\sim -1$ is still allowed by current data.
Note that \eq{epsexact} implies that at very large $\tan\beta$ where $\beta\to\half\pi$,
\begin{equation}
\sin (\beta+ \alpha) - \sin (\beta - \alpha)=\frac{2(1-\epsilon)}{1+\tan^2\beta}\ll 1  \qquad (\text{for $\tan \beta \gg 1$})  \, .
\label{eq-sba1}
\end{equation}
In particular, when $\epsilon<1$ we see that $\sin (\beta - \alpha)$ is always below $\sin (\beta + \alpha)$.
Fig.~\ref{fig:T2} reflects the behavior shown in \eq{eq-sba1}
in that  the larger $\tan \beta$ is, the closer the negative and positive $\sin \alpha$ regions are.
Furthermore, as $\epsilon$ decreases the region where the low values of $\tan \beta$ are allowed decreases.
Therefore, when $\tan \beta$ is very large we see that $|\cbma|\ll 1$, and we recover
the SM $VV$ and $hhh$ couplings.  Furthermore, as discussed earlier there is limited sensitivity to the sign of the Yukawa couplings for the one-loop induced $\gam\gam$ and $gg$ couplings.
Thus, due to the limited accuracy with which the $\gam\gam$ and $gg$ couplings are (indirectly) measured, the region of wrong-sign Yukawa couplings (where $\sin (\beta + \alpha) \approx 1$
and $\sin \alpha > 0$) in the type-II 2HDM is still allowed by the current LHC data.

\section{Results and discussion}
\label{sec-results2}
\begin{figure}[h!]
\bec
\includegraphics[width=3.5in,angle=0]{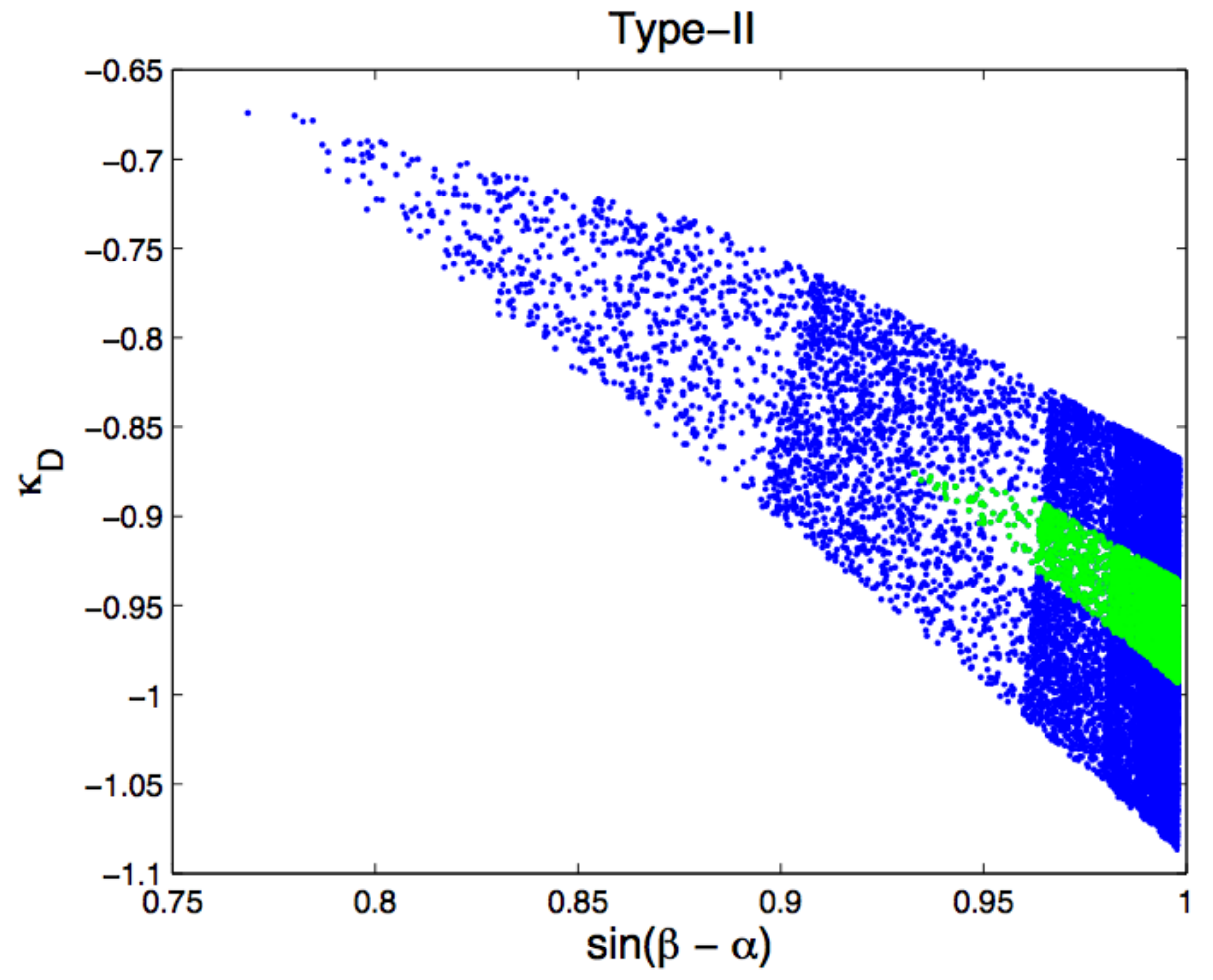}
\hspace{-.3cm}
\includegraphics[width=3.5in,angle=0]{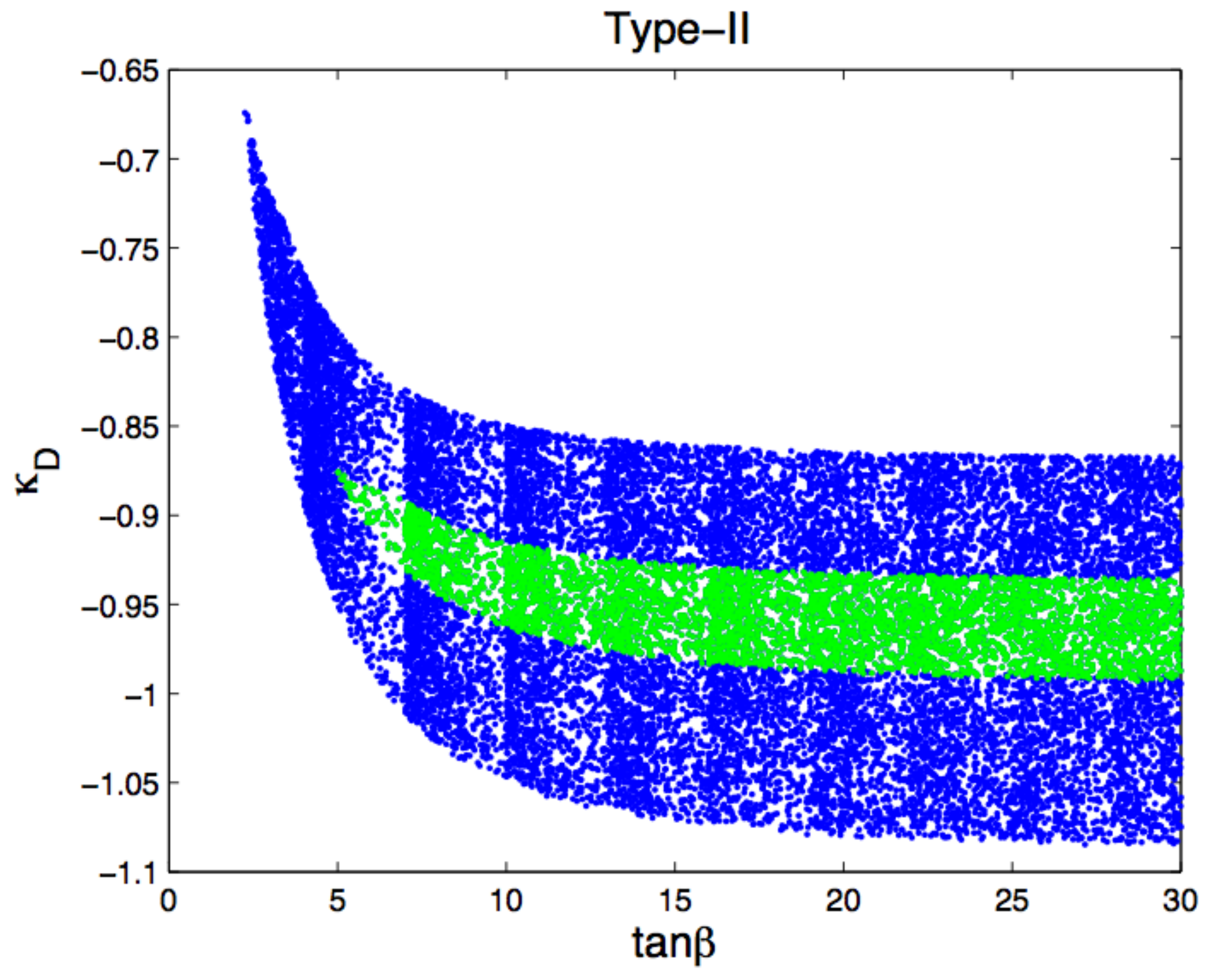}
\eec
\caption{Left panel: The Yukawa coupling ratio $\kd=h_D^{\rm 2HDM}/h_D^{\rm SM}$ as a function of $\sin (\beta - \alpha)$
in the type-II 2HDM, with all $\muflhc$ within 20\% (blue/black) and
 10\% (green/light grey) of their SM values.
Right panel:  Same ratio as a function of $\tan \beta$. If one demands consistency at the $5\%$ level, no points survive.}
\label{fig:T3}
\end{figure}

In order to study in more detail the wrong-sign region of the type-II 2HDM, we have generated a new set of points where
we have further imposed that $\sin \alpha >0$.
In the left panel of Fig.~\ref{fig:T3} we present  $\kd $ as a function of
$\sin (\beta - \alpha)$, with all $\muflhc$ within 20\% of the SM values in blue (black) and
10\% of the SM values in green (light grey). As expected, the values are very close to the region where  $\kd=-1$ while simultaneously $\sin (\beta - \alpha)$ approaches 1.
As the $\muflhc$ values are required to agree more precisely with the SM value of 1, the points move closer to the above limit.
In the right panel we show the same ratio as a function of
$\tan \beta$. As $\tan \beta$ grows, $\sin (\beta + \alpha)$ is forced to be closer to $\sin (\beta - \alpha)$
as indicated in \eq{eq-sba1} and is forced to be closer to 1 due to the LHC constraints.
As indicated by Fig.~\ref{fig:T2}, increasing the precision of the Higgs measurements would allow
exclusion of the low $\tan \beta$ region if all $\muflhc$ are within 10\% of unity. Moreover, the entire $\kd<0$ region is eliminated if all $\muflhc$ are within 5\% of unity.

\begin{figure}[t!]
\bec
\includegraphics[width=3.5in,angle=0]{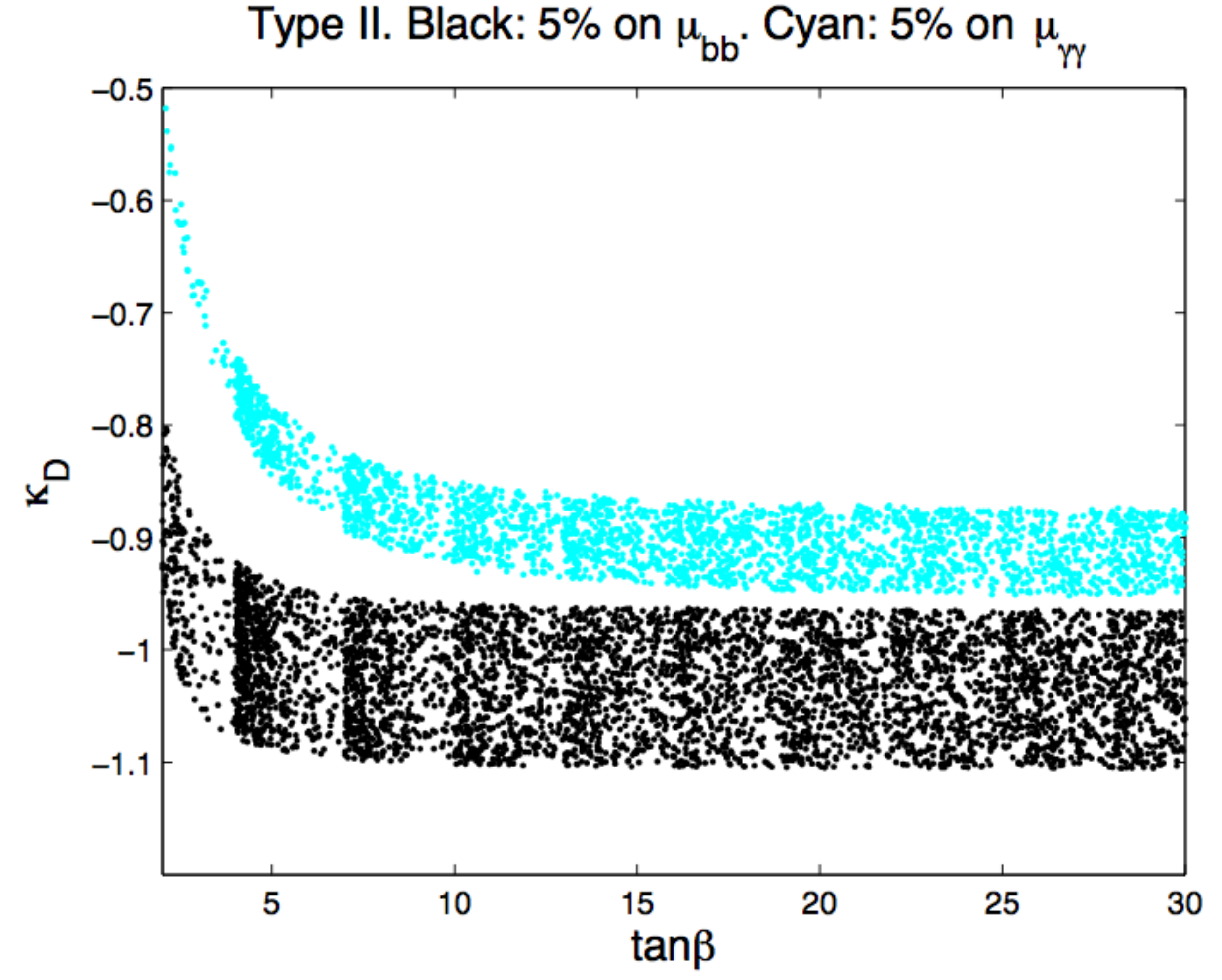}
\eec
\caption{For the 2HDM type-II,
 we show regions in $\kd$ vs. $\tanb$ space having $\sina>0$ that are allowed when $\mulhc{\gam\gam}$ (cyan/grey) and
$\mulhc{b b}$ (black) are within 5\% of the SM prediction of unity.
}
\label{contra}
\end{figure}
\begin{figure}[ht!]
\includegraphics[width=3.5in,angle=0]{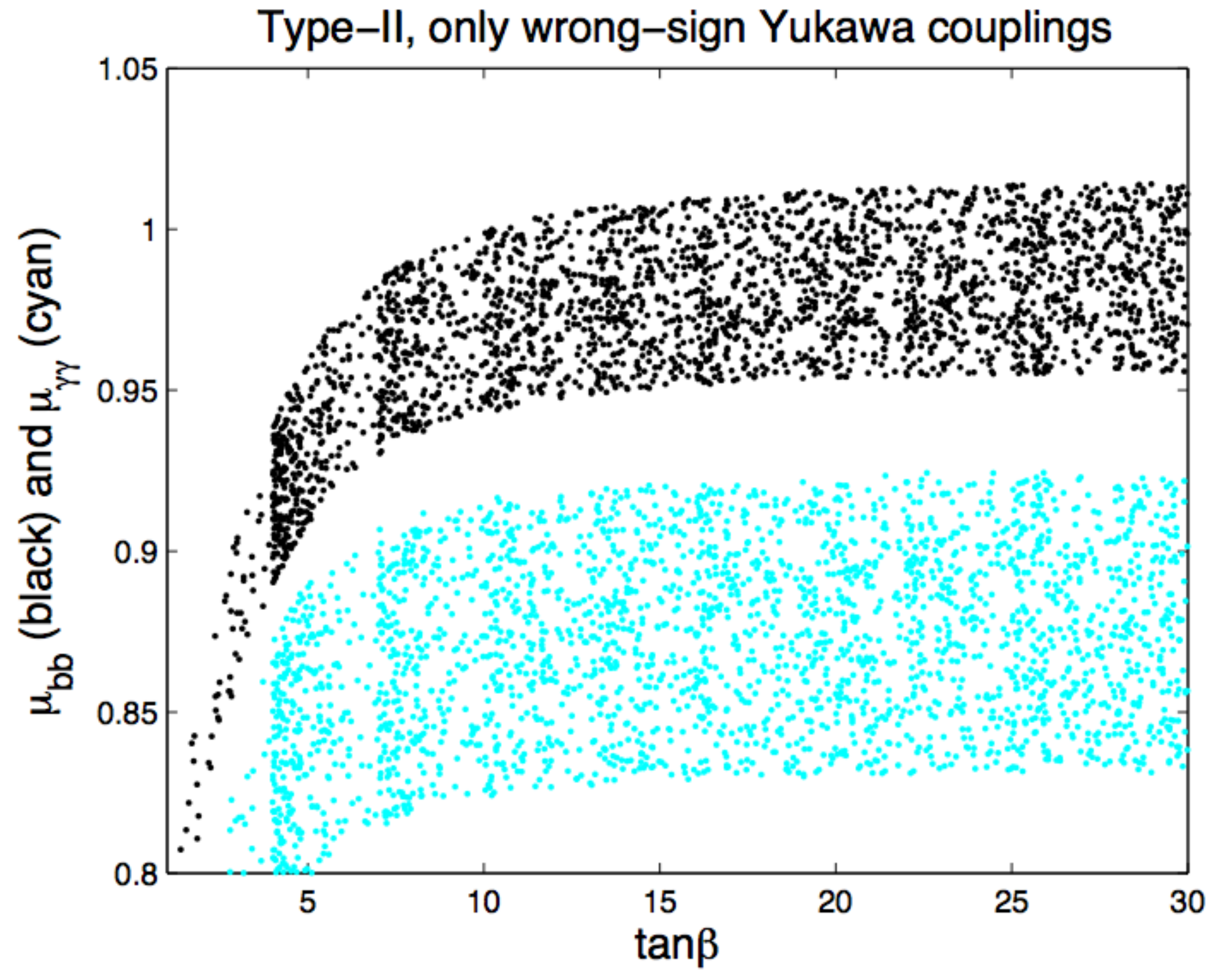}\includegraphics[width=3.5in,angle=0]
{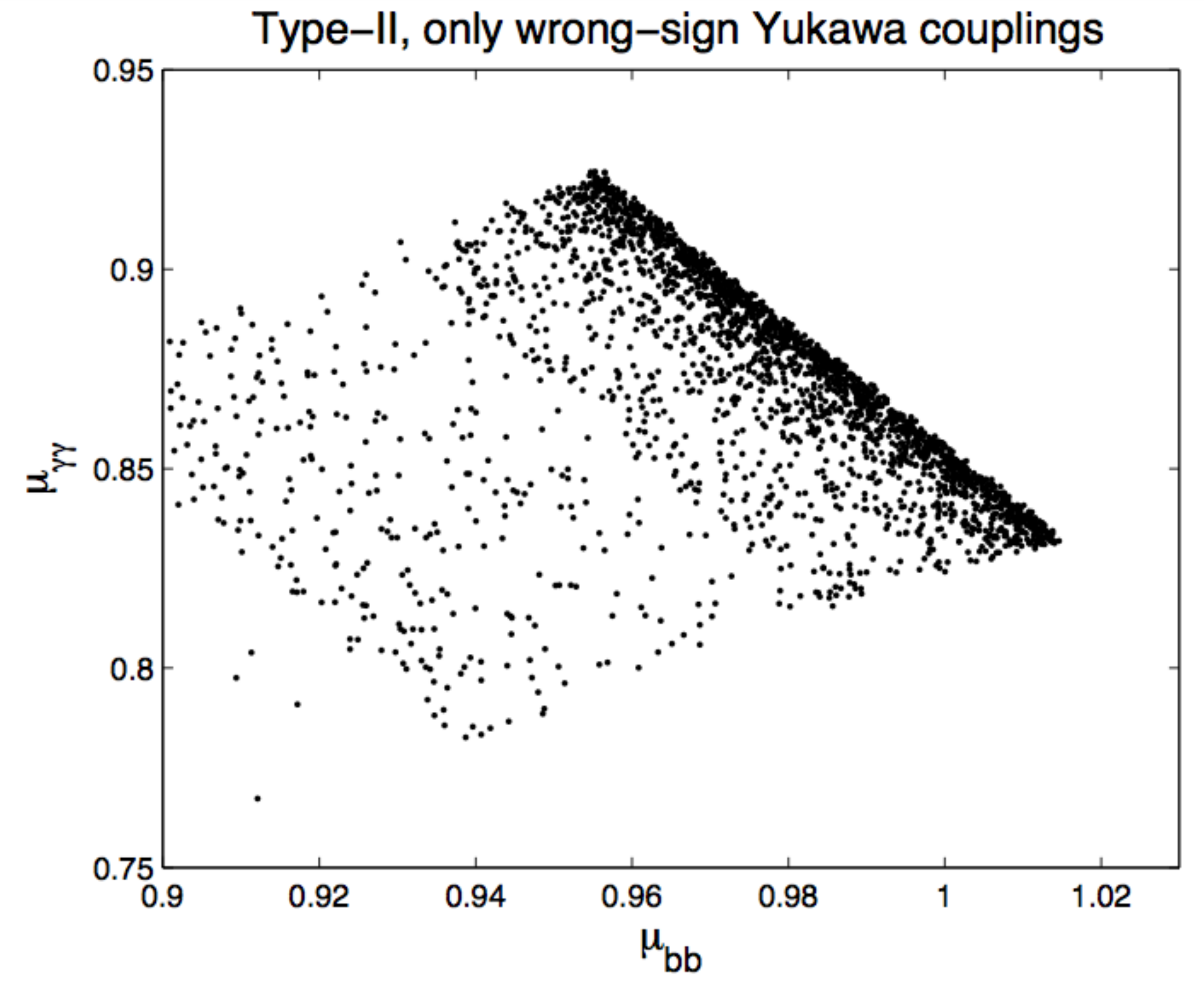}
\caption{Assuming that the $WW,ZZ$ rates are measured to be within 5\% of the SM prediction, we plot  $\mulhc{\gam\gam}$ and $\mulhc{b b}$ vs. $\tanb$ (left) and  $\mulhc{\gam\gam}$ vs. $\mulhc{b b}$ (right).}
\label{muplots}
\end{figure}

In fact, we will see that it is
$\mulhc{\gam\gam}$ that makes overall consistency with SM rates at the
5\% level impossible in the $\sin\alpha>0$ branch.  This is due to the
fact that for all the $\kd<0$ points we are in the nondecoupling
regime for which the charged-Higgs boson loop contribution to the $h\to\gamma\gamma$ amplitude
is approximately constant as a function of
$\mhpm$ (up until the tree level unitarity upper limit of $\mhpm\sim 650\gev$, beyond which $\kd<0$ is not a perturbatively consistent possibility).  The charged-Higgs loop gives about a $10\%$ reduction in $\Gamma(\hl\to \gam\gam)$ that is inconsistent with $\mulhc{\gam\gam}$ being within $5\%$ of unity.  The details of the nondecoupling regime are discussed at length in Appendix~\ref{apjfg}.

%
%
%LHC discussion
Another perspective is obtained by examining \Fig{contra}.  There, we have shown regions in  $\kd$  vs. $\tanb$ space where either
$\mulhc{\gam\gam}$ (cyan/grey) or $\mulhc{b b}$
(black) are within $5\%$ of unity for points in the $\sina>0$ branch.\footnote{Note that $\mulhc{\tau^+\tau^-}=\mulhc{b b}$ in the 2HDM, implying that
measurements in the $\tau^+\tau^-$ channel are equally useful.  Further, at the LHC, the $\tau^+\tau^-$ final state will be more precisely measured than for the $b \bar{b}$ final state.}
We observe that the two branches represented do not intersect, and as such it is impossible to achieve
5\% agreement with the SM in both of these channels. This explains why there are no red points in the right branch of the plots
in Fig.~\ref{fig:T2}.

Further insight is gained from Fig.~\ref{muplots}, which considers points for which the $\mulhc{WW,ZZ}$ are within 5\% of the SM value of 1. On
the left, we exhibit the values of $\mulhc{\gam\gam}$ and $\mulhc{b b}$ vs $\tanb$.  This shows that while $\mulhc{bb}$ can be within 5\%
of unity, $\mulhc{\gam\gam}$ cannot --- it is always more than 7-8\% below unity, implying that 5\% accuracy on this channel would exclude the
$\kd<0$ branch.  On the right, we plot $\mulhc{\gam\gam}$ vs. $\mulhc{b b}$.  The largest value of $\mulhc{\gam\gam}$ that can be achieved is
$\sim 0.925$, and this only if $\mulhc{bb}\lsim 0.96$. Thus, it is the suppression of the $\gam\gam$ final state at the LHC that is key to
ruling out the $\kd<0$ possibility for $\rts=14\tev$ operation at high luminosity.  This same conclusion is found in the work of Ref.~\cite{dgjk2hdm}.
There, different initial states are separated from one another and one finds that the $VV\to h\to \gam\gam$ rate is the most suppressed relative
to other processes --- because of the 6\% enhancement of $\sigma(gg\to h)$ when $\kd<0$ the $gg\to h\to \gam\gam$ rate is not as suppressed
relative to the remaining processes but still contributes to the overall inconsistency for $\kd<0$ between these $\gam\gam$ final state channels
with other final states such as $ZZ$, $WW$ and $\tau\tau$ when all are measured with 5\% accuracy.

\begin{figure}[t!]
 \includegraphics[width=3.5in,angle=0]{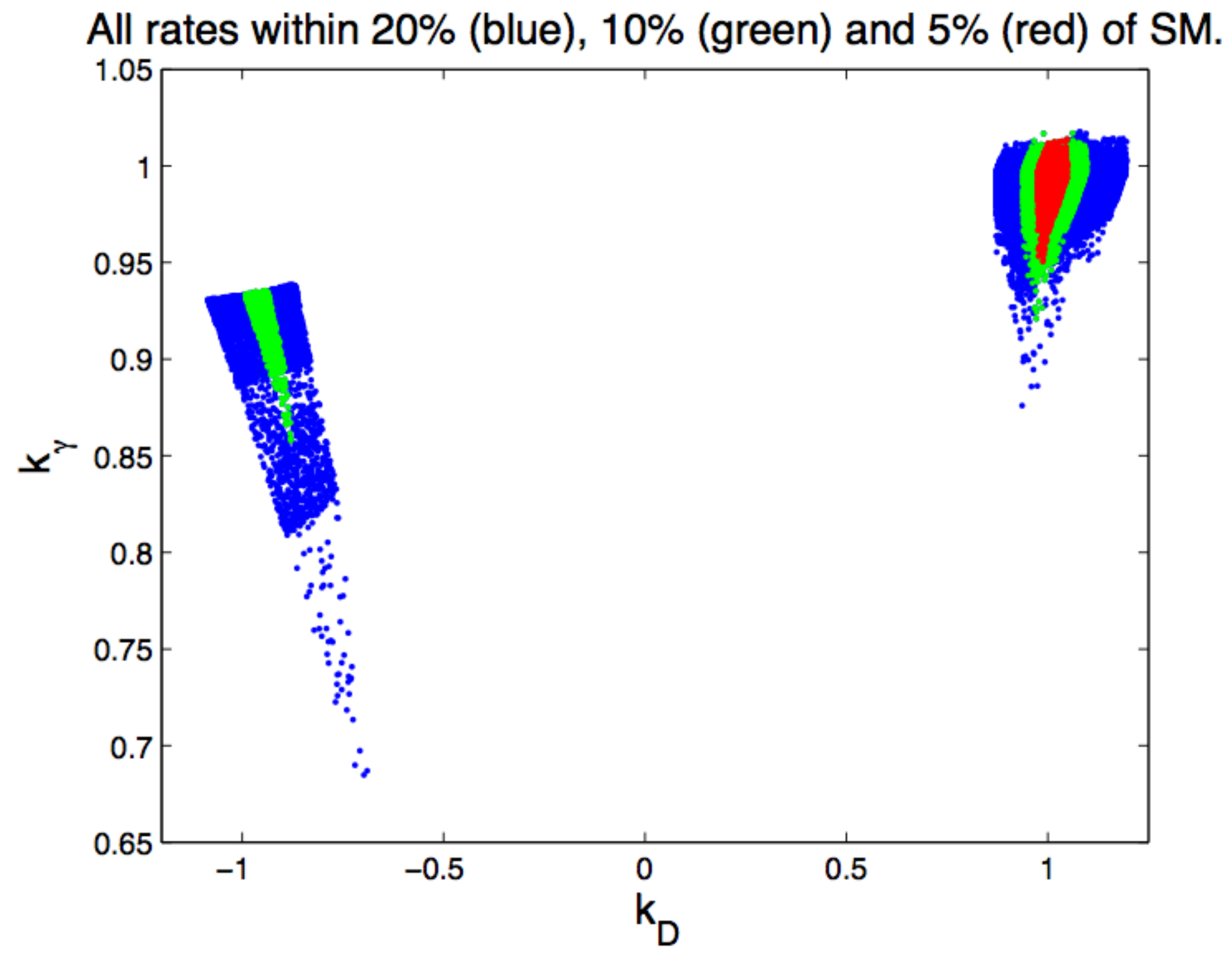} \includegraphics[width=3.5in,angle=0]
{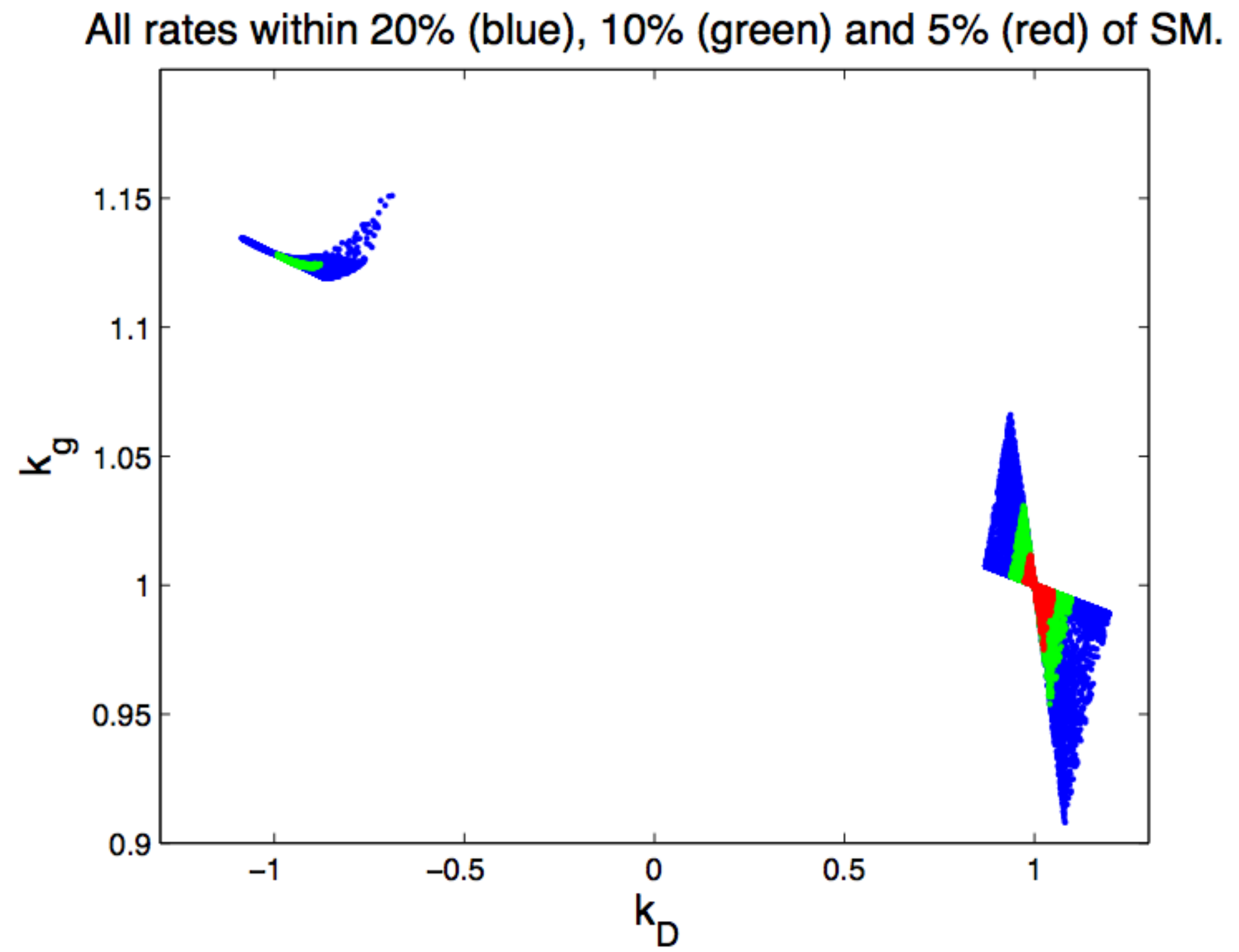}
\caption{Allowed regions for 2HDM type-II with all $\muflhc$ within 20\% (blue/black),
 10\% (green/light grey) and 5\% (red/dark grey) of their SM values in $\kgam$ vs. $\kd$ space (left); $\kg$ vs. $\kd$ space (right). }
\label{kgamplot}
\end{figure}

In \Fig{kgamplot}, we show in $\kgam$ or $\kg$ vs. $\kd$ space the points that are allowed if the $\muflhc$'s are each within $20\%$ (blue), $10\%$ (green), or $5\%$ (red) of unity (the SM limit).  We observe from the left hand plot that $\kgam$ is always at least 5\% below unity in the $\kd<0$ region and that $5\%$ accuracy on the $\muflhc$'s will eliminate this region entirely.  In fact, as we saw in Fig.~\ref{muplots} it is $\mulhc{\gam\gam}$ that necessarily has a greater than $5\%$ deviation from unity. The right hand plot shows that in the $\kd<0$ region, $\kg$ is always bigger than 1.13.  However, since currently the LHC is unable to determine $\kg$ with the necessary accuracy this does not help to exclude the $\kd<0$ region.  But, as summarized earlier, with $L=300\fbi$ at $\rts=14\tev$, $\kg$ can be determined to about $8\%$ accuracy and such a deviation will certainly be observable.

\begin{figure}[t!]
\centering
\includegraphics[width=3.5in,angle=0]{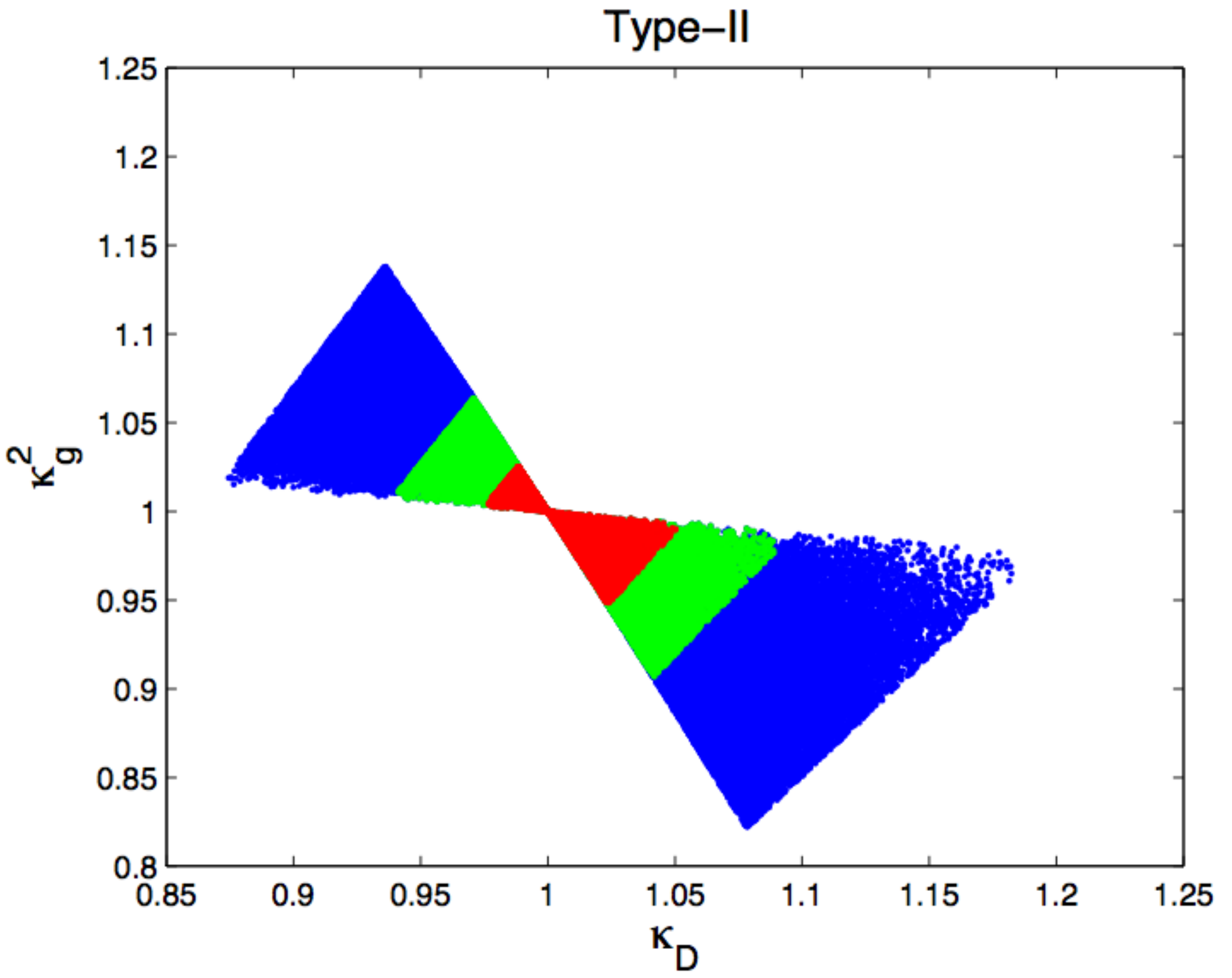}
\hspace{-.3cm}
\includegraphics[width=3.5in,angle=0]{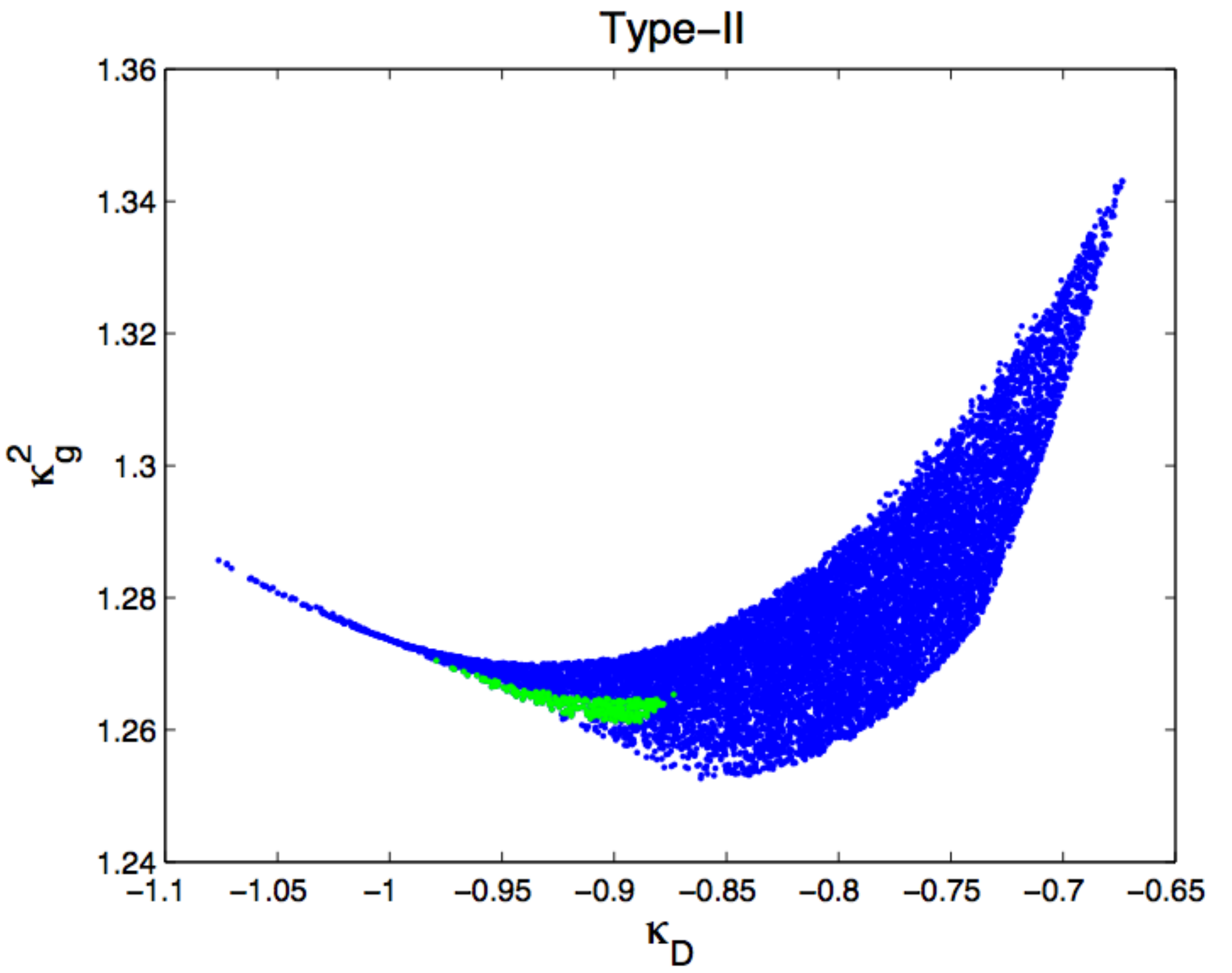}
\caption{$\kg^2=\Gamma (h \to gg)^{\rm 2HDM}/\Gamma (\hsm \to gg)$ as a function of $\kd=h_D^{\rm 2HDM}/h_D^{\rm SM}$ in type-II, with all $\muflhc$ within 20\%
(blue/black), 10\% (green/light grey) and 5\% (red/dark grey) of their SM values.
Left panel: $\sin \alpha <0$. Right panel: $\sin \alpha >0$.  }
\label{fig:T4}
\end{figure}

As noted earlier, at the ILC the $gg$ final state becomes a powerful  tool for determining the sign of $\kd$. Thus, we shall explore the $gg$ final state issues in more detail.
In \Fig{fig:T4}  we exhibit $\kg^2=\Gamma (h \to gg)^{{\rm 2HDM}}/\Gamma (\hsm \to gg)$ as a function of $\kd$
for $\sin \alpha <0$ (left) and $\sin \alpha >0$ (right)  with all $\muflhc$ within 20\% of the SM values in blue (dark grey)
and 10\% of the SM values in green (light grey). Contrary to the SM-like scenario, when $\sin \alpha >0$ (wrong-sign Yukawa coupling) the value of the ratio
of the widths is always above 1.25. \Fig{fig:T4} shows that the minimum value of $\kg^2$ becomes larger when smaller deviations of the $\muflhc$s from unity are required. In particular, when the $h_D$ coupling
changes sign but all tree level couplings have  SM magnitude, the ratio between the two widths is exactly
\begin{equation}
\frac{\Gamma (h \to gg)^{\scriptscriptstyle {\rm 2HDM}}} {\Gamma (\hsm \to gg)}  = 1.27  \qquad  (\sin (\beta + \alpha) = 1)\,,
\label{eg-ga}
\end{equation}
which is in agreement with \texttt{HDECAY}~\cite{Djouadi:1997yw, Harlander:2013qxa} and \texttt{2HDMC}\cite{Eriksson:2009ws, Harlander:2013qxa}.
Note that this interference effect, which is almost 30\% relative to the SM, does not manifest itself in the
production process $gg \to h$ that is important for the LHC and might therefore have been quite easily detectable.  In contrast to the leading order (LO) result,
\begin{equation}
\frac{\sigma (gg \to h)^{\scriptscriptstyle {\rm 2HDM}}_{{\rm LO}}} {\sigma (gg \to \hsm)_{{\rm LO}}}  \approx
\frac{\Gamma (h \to gg)^{\scriptscriptstyle {\rm 2HDM}}_{{\rm LO}}} {\Gamma (\hsm \to gg)_{{\rm LO}}}  \approx 1.27  \qquad  (\sin (\beta + \alpha) = 1)\,,
\label{eg-ga2}
\end{equation}
at NNLO in the limit of $\sbpa=1$,
$\sigma (gg \to h)^{\scriptscriptstyle {\rm 2HDM}}_{{\rm NNLO}} /  \sigma (gg \to \hsm)_{{\rm NNLO}}  \approx 1.06$~\cite{Spira:1995mt}
while the ratio of the partial widths of $h \to gg$ does not suffer any significant change in going from LO to NNLO. Therefore, the present LHC data cannot discriminate between the two scenarios
based on interference effects at the production level; it is only through a luminosity $L\geq 300\fbi$ of data accumulated at $\rts=14\tev$ and a combined fit of the rates for all final states that one can manage to determine the underlying $\kg$ with adequate precision.

Of course the ILC can probe ${\rm BR}(h\to gg)$ more easily and directly using the process $e^+ e^- \to Z h \to Z gg$. We define 
%the quantity $\muilc{gg}$ as
\begin{equation}
\muilc{gg} \, = \, \frac{\sigma \, {\rm BR} (h \to
  gg)}{\sigma^{\scriptscriptstyle {\rm SM}} \, {\rm BR}(\hsm \to gg)}
\label{eg-la}
\end{equation}
where $\sigma$ is the measured $\epem\to Z^*\to Z h$ Higgs production cross section at the ILC and $\sigma^{\scriptscriptstyle {\rm SM}}$
and ${\rm BR}(\hsm \to gg)$ are the SM values of the production cross section at the ILC and branching
ratio of a Higgs decaying to a pair of gluons. The ratio of the cross sections in the process
 $e^+ e^- \to Z h $ is just $\sin^2 (\beta - \alpha)$. Likewise we can define similar ratios for the processes
$e^+ e^- \to Z h \to Z b  \bar{b}$ and $e^+ e^- \to Z h \to Z c  \bar{c}$ which we will call $\muilc{bb}$ and  $\muilc{cc}$, respectively.

\begin{figure}[t!]
\centering
\includegraphics[width=3.5in,angle=0]{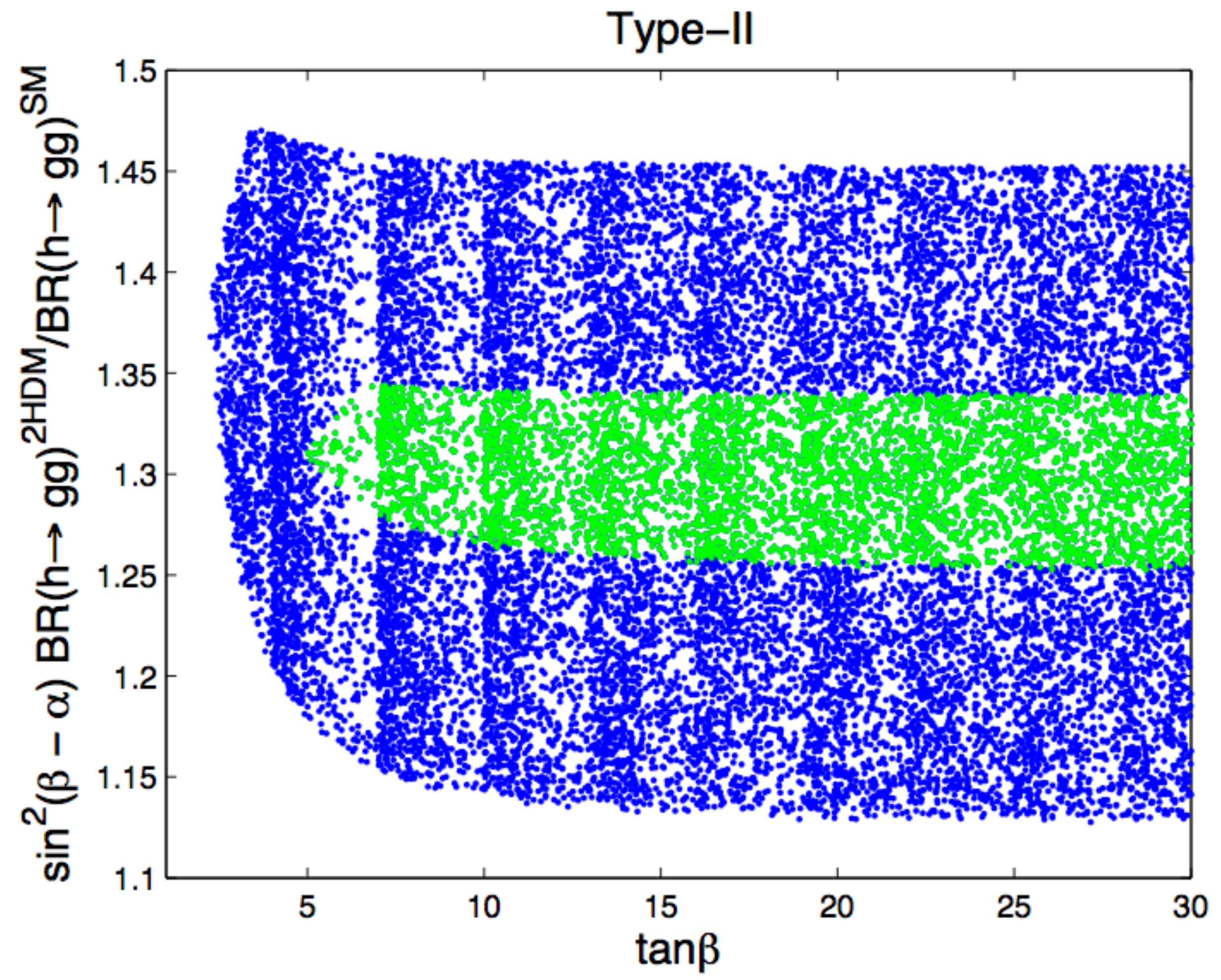}
\hspace{-.3cm}
\includegraphics[width=3.5in,angle=0]{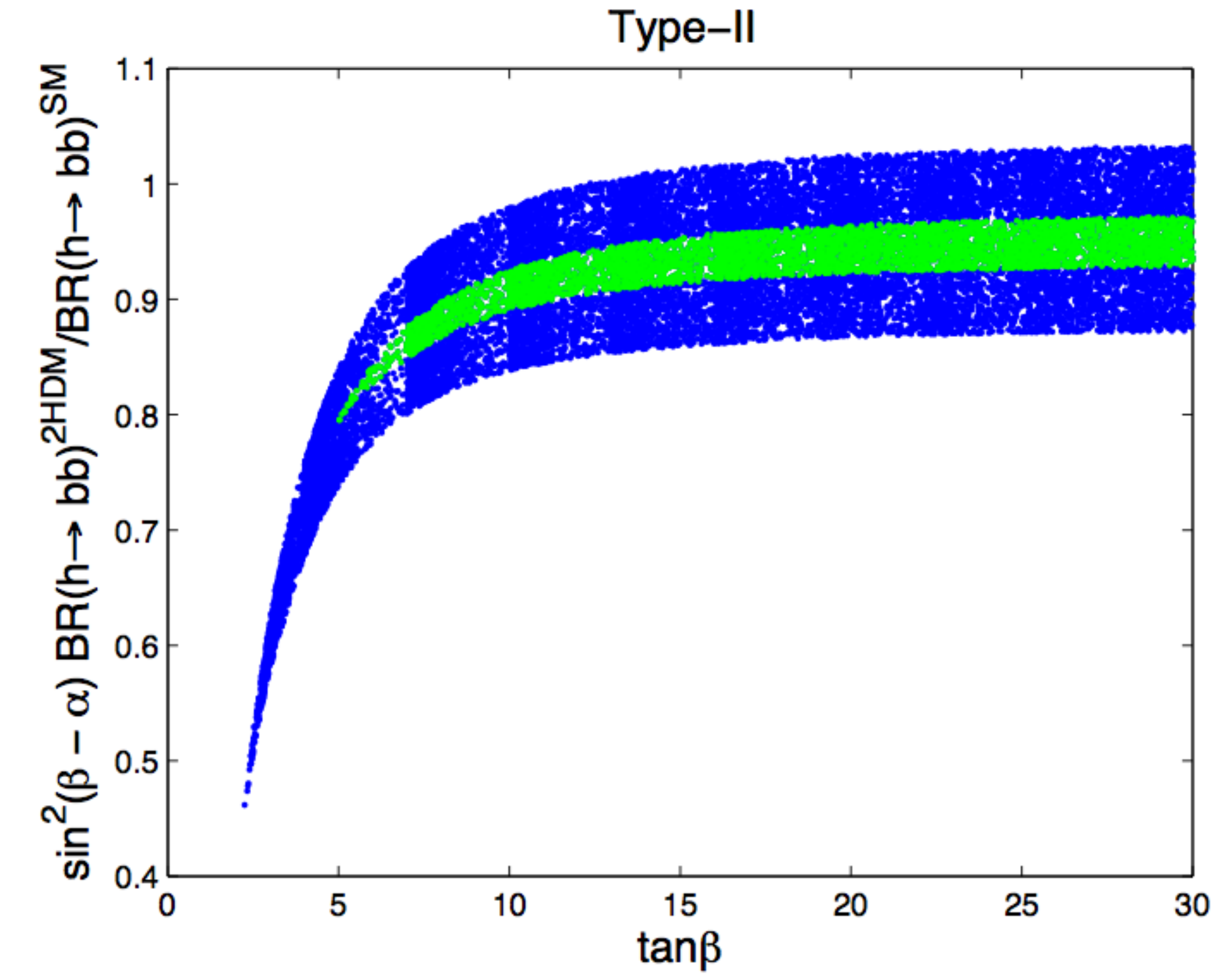}
\caption{Left panel: $\muilc{gg}$ as a function $\tan \beta$.
Right panel:  $\muilc{bb}$ as a function of $\tan \beta$. The model is type-II, requiring $\sina>0$, with all $\muflhc$ within 20\% of the SM values in blue (black)
and 10\% of the SM values in green (light grey). }
\label{fig:T5}
\end{figure}

 In the left panel of \Fig{fig:T5} we show
the quantity $\muilc{gg}$ as a function of $\tan \beta$. When all $\muflhc$'s measured at the LHC are forced to be within 20\% of the SM values (blue/black)
all points are above 1.12. If the precision is increased to 10\%, the bound is increased to 1.25. Recently, it was shown that
$\muilc{gg}$ can be measured at the ILC with  an accuracy of 8.5\% at a CM energy of
250 GeV and 7.3\% at a CM energy of 350 GeV with an integrated luminosity of 250 fb$^{-1}$ and beam polarization
of $-$80\% (electron) and 30\% (positron)~\cite{Ono:2012ah, Asner:2013psa}. The 95\% C.L. predicted measurement for $\sqrt s = 350$ GeV and
250 fb$^{-1}$ luminosity is 1.02 $\pm$ 0.07~\cite{Ono:2012ah}, assuming SM expectations. Therefore, this measurement could
 exclude all points in the left panel of \Fig{fig:T5}.
In the right panel we present $\muilc{bb}$ as a function of $\tan \beta$. The corresponding SM predicted measurement for the ILC is 1.00 $\pm$ 0.01. Clearly, a better than about $5\%$ measurement of $\muilc{bb}$  can also help probe the wrong-sign coupling provided enough precision is attained at the LHC in the measurements of the
Higgs couplings to fermions and gauge bosons. The values of $\muilc{bb}$ are slightly
below 1 because, as can be seen from \eq{eq-sba1}, when $\sin (\beta + \alpha) \to 1$, $\sin (\beta - \alpha)$ is slightly
below 1 as the right-hand side of the equation is positive. Note that the ratio of the branching ratios in $\muilc{bb}$ is very close to 1 in the limit we are considering
and as such $\muilc{bb} \approx \sin^2 (\beta - \alpha)$.
Similar results would be obtained for $\muilc{cc}$, where the final state is $c \bar c$, but
the precision in the $\muilc{cc}$ measurement is not as good as for  $\muilc{bb}$.

\section{Conclusions}

The couplings of the Higgs boson recently discovered at the LHC
to the fermions and gauge bosons are starting to be measured with some precision.
It is important to understand the implications of these results in the context of specific Higgs sector models. In this paper, we considered type-I and type-II
 $\mathbb{Z}_2$-symmetric and CP-conserving 2HDMs. Our focus was on the fact that the sign of the Yukawa coupling to the down-type fermions could be opposite to that of the SM.
Using scans over type-I and type-II parameter spaces, subject to basic theoretical and experimental constraints
as described in the main text, we found that
a sign change in the down-quark Yukawa couplings can be accommodated in
the context of the current LHC data set at 95\% C.L., but only in the case of the type-II 2HDM when  $\sin(\beta+\alpha)\sim 1$.
The situation is different in the type-I 2HDM --- because only one doublet couples to all fermions
the sign change would result in deviations from the SM predictions that are incompatible  with the current Higgs data set.
In this paper, we address the possibility of probing
the wrong-sign Yukawa coupling of the Higgs to down-type quarks with future measurements of Higgs properties at the $\rts=14\tev$ LHC and at
the International Linear Collider.

In particular, we performed a scan dedicated to the part of type-II 2HDM parameter space where the wrong-sign down-type quark coupling is currently acceptable.
We filtered parameter space points requiring that the
values of $\muflhc$, the production rate of a given final state $f$ relative to the SM, are within either 20\%, 10\% or 5\% of
the SM predictions for the LHC. Of greatest immediate interest is the fact that projected precisions for the determination of the magnitude of the $\gam\gam h $ coupling relative to its SM value, $\kgam$ (using $pp\to h\to \gam\gam$ in particular) imply that the LHC with $\rts=14\tev$ and $L\geq 300\fbi$ will either rule out or confirm the wrong-sign scenario. Of particular importance for this conclusion is the fact that the charged-Higgs loop contribution to the $\gam\gam h$ couplings does not decouple for the $\sin(\beta+\alpha)\to 1$ scenario, leading to a $\sim 10\%$ decrease in $\Gamma(h\to \gam\gam)$.  This statement applies for any charged-Higgs mass below the bound of about $650\gev$ for which the Higgs coupling parameters satisfy tree level unitarity bounds. In the context of the model, a finding that the $h_D$ Yukawa has a negative sign {\it and also} detecting a charged-Higgs with mass above $650\gev$ would imply that the theory is in a realm where perturbative calculations become suspect.

In addition, we have shown that
the predictions for the measurements of $\muilc{gg}$ and $\muilc{bb}$
at the ILC would allow us to probe the wrong-sign Yukawa
coupling of a type-II 2HDM. Therefore, at both collider facilities, either a measurement
or a definite 95\% exclusion limit could be set on the wrong-sign Yukawa coupling scenario.

\acknowledgments{
The works of P.M.F. and R.S. are supported in part by the Portuguese
\textit{Funda\c{c}\~{a}o para a Ci\^{e}ncia e a Tecnologia} (FCT)
under contract PTDC/FIS/117951/2010, by FP7 Reintegration Grant, number PERG08-GA-2010-277025,
and by PEst-OE/FIS/UI0618/2011.
The work of J.F.G. and H.E.H. is supported
in part by the U.S. Department of Energy, under grant
numbers DE-SC-000999 and DE-FG02-04ER41268, respectively.  J.F.G. and H.E.H. also thank the Aspen Center for Physics, supported by the National Science Foundation under grant number PHYS-1066293, for hospitality and a great working atmosphere. J.F.G. acknowledges conversations and related collaboration with B. Dumont, Y. Jiang and S. Kraml.  R.S. acknowledges discussions with M.~M{\"u}hlleitner, M.~Spira
and J.P. Silva. Finally, H.E.H. is grateful for probing questions by JoAnne Hewett, which inspired the inclusion of Appendix B.
}

%%%%%%%%%%%%%%%%%%%%%%%%%%%%%%%%%%%%%%%%%%%%%%%%%%%%%%%%%%%%%%%%%%%%%%%%
%%%%%%%%%%%%
%%%%%%%%%%%%%%%

\appendix

\section{The Higgs basis of the softly broken $\mathbb{Z}_2$ symmetric CP-conserving 2HDM}
\label{app:higgsbasis}

It is convenient to reexpress the Higgs potential given by \eq{pot} in the
Higgs basis~\cite{Donoghue:1978cj,Georgi,silva,lavoura,lavoura2,branco,Davidson:2005cw}.
By assumption, we have assumed that all the scalar potential parameters and the two vacuum expectation values $v_1$ and $v_2$ are real, which implies that the scalar potential and the vacuum are CP invariant.
By a suitable transformation on the two-Higgs-doublet fields $\Phi_a$ ($a=1,2$), one can define two new linearly independent Higgs doublet fields $H_1$ and $H_2$ such that $\langle H_1^0\rangle=v/\sqrt{2}$ and $\langle H_2^0\rangle=0$.  This is accomplished by defining
\beq
H_1=\begin{pmatrix}H_1^+\\ H_1^0\end{pmatrix}\equiv \frac{v_1 \Phi_1+v_2 \Phi_2}{v}\,,
\qquad\quad H_2=\begin{pmatrix} H_2^+\\ H_2^0\end{pmatrix}\equiv\frac{-v_2 \Phi_1+v_1 \Phi_2}{v}\,.
\eeq
The Higgs basis is uniquely defined
up to an overall sign of the $H_2$ scalar doublet field.  In the Higgs basis, the scalar potential is given by
\beqa
\mathcal{V}&=& Y_1 H_1^\dagger H_1+ Y_2 H_2^\dagger H_2 +[Y_3
H_1^\dagger H_2+{\rm h.c.}]
+\half Z_1(H_1^\dagger H_1)^2+\half Z_2(H_2^\dagger H_2)^2
+Z_3(H_1^\dagger H_1)(H_2^\dagger H_2)\nonumber\\
&&\qquad +Z_4( H_1^\dagger H_2)(H_2^\dagger H_1)
+\left\{\half Z_5 (H_1^\dagger H_2)^2 +\big[Z_6 (H_1^\dagger
H_1) +Z_7 (H_2^\dagger H_2)\big] H_1^\dagger H_2+{\rm
h.c.}\right\}\,,\label{hbasispot}
\eeqa
where the squared-mass terms are given by:
\beqa
Y_1&=& m_{11}^2\cosbii+m_{22}^2\sinbii-2m_{12}^2\sinb\cosb\,,\\
Y_2&=& m_{11}^2\sinbii+m_{22}^2\cosbii+2m_{12}^2\sinb\cosb\,,\\
Y_3&=& (m_{22}^2-m_{11}^2)\sinb\cosb-m_{12}^2\cos 2\beta\,,
\eeqa
and the Higgs basis quartic couplings are given by
\beqa
Z_1 & \equiv & \lambda_1\cosbiv+\lambda_2\sinbiv+2(\lambda_3+\lambda_4+\lambda_5)\sinbii\cosbii\,,\label{z1}\\
Z_2 & \equiv & \lambda_1\sinbiv+\lambda_2\cosbiv+2(\lambda_3+\lambda_4+\lambda_5)\sinbii\cosbii\,,\label{z2}\\
Z_i&\equiv&(\lambda_1+\lambda_2-2\lambda_3-2\lambda_4-2\lambda_5)\sinbii\cosbii+\lambda_i\,,\qquad\text{for $i=3,4,5$}\,,\label{z345}\\
Z_6 & \equiv & -\sinb\cosb\bigl[\lambda_1\cosbii-\lambda_2\sinbii-(\lambda_3+\lambda_4+\lambda_5)\cos 2\beta\bigr]\,,\label{z6}\\
Z_7&\equiv&-\sinb\cosb\bigl[\lambda_1\sinbii-\lambda_2\cosbii+(\lambda_3+\lambda_4+\lambda_5)\cos 2\beta\bigr]\label{z7}\,.
\eeqa
Note that one is free to redefine $Y_3$, $Z_6$ and $Z_7$ by an overall sign in light of the sign ambiguity in defining the Higgs basis.
The potential minimum conditions are especially simple in the Higgs basis,
\beq \label{minconds}
Y_1=-\tfrac{1}{2}Z_1 v^2\,,\qquad\quad
Y_3=-\tfrac{1}{2} Z_6 v^2\,,
\eeq
leaving $Y_2$ as the only free squared-mass parameter of the model.

Finally, we note some useful relations that relate the Higgs basis parameters to the Higgs masses~\cite{decoupling,Haber:2006ue}:
\beqa
Z_1 v^2&=&m_h^2 \sin^2(\beta-\alpha)+m_H^2\cos^2(\beta-\alpha)\,,\\
Z_3 v^2&=& 2(m_{H^\pm}^2-Y_2)\,,\\
Z_4 v^2&=&m_h^2 \cos^2(\beta-\alpha)+m_H^2\sin^2(\beta-\alpha)+m_A^2-2m_{H^\pm}^2\,,\\
Z_5 v^2&=&m_h^2\cos^2(\beta-\alpha)+m_H^2\sin^2(\beta-\alpha)-m_A^2\,,\\
Z_6 v^2&=&-(m_H^2-m_h^2)\sbma\cbma\,.
\eeqa
The Higgs masses and $\cbma$ do not depend on the parameters $Z_2$ and $Z_7$.

\section{The wrong-sign $\boldsymbol{h\overline{D}D}$ coupling and the MSSM Higgs sector}
\label{app:mssm}

The tree level scalar potential of the MSSM Higgs sector is given by \eq{pot}, with~\cite{mssm}
\beq \label{lamsusy}
\lambda_1=\lambda_2=\tfrac{1}{4}(g^2+g^{\prime\,2})\,,\qquad\quad
\lambda_3=\tfrac{1}{4}(g^2-g^{\prime\,2})\,,\qquad\quad \lambda_4=-\half g^2\,,\qquad\quad 
\lambda_5=0\,.
\eeq
In particular $\lambda_6=\lambda_7=0$ [defined below \eq{pot}].  Inserting \eq{lamsusy} into
\eqst{z1}{z7} yields~\cite{Haber:2006ue} 
\beqa
Z_1&=&Z_2=\tfrac{1}{4}(g^2+g^{\prime\,2})\cos^2 2\beta\,,\qquad Z_3=Z_5+\tfrac{1}{4}(g^2-g^{\prime\,2})\,,\qquad
Z_4=Z_5-\half g^2\,,\nonumber \\
Z_5&=&\tfrac{1}{4}(g^2+g^{\prime\,2})\sin^2 2\beta\,,\qquad\qquad\,\,\,\,
Z_7=-Z_6=\tfrac{1}{4}(g^2+g^{\prime\,2})\sin 2\beta\cos 2\beta\,.\label{zsusy}
\eeqa
Using the result for $Z_6$ given above in \eqs{c2exact}{scexact} yields the tree level expressions,
\beqa
\cos^2(\beta-\alpha)&=&\frac{m_Z^2\cos^2 2\beta-m_h^2}{m_H^2-m_h^2}\,,\label{susyc2exact}\\
\sin(\beta-\alpha)\cos(\beta-\alpha)&=&\frac{m_Z^2\sin 2\beta\cos 2\beta}{m_H^2-m_h^2}
\label{susyscexact}\,,
\eeqa
where $m^2_{h,H}$ are the MSSM \textit{tree level} CP-even Higgs squared masses,
\beq \label{mssmhm}
  m^2_{H,h} = \half \left( \mha^2 + m^2_Z \pm
                  \sqrt{(\mha^2+m^2_Z)^2 - 4m^2_Z \mha^2 \cos^2 2\beta}
                  \; \right)\,.
\eeq
In addition, the MSSM tree level Higgs-fermion Yukawa couplings possess a type-II structure due to supersymmetry.

In the decoupling limit where $m_{H}\sim m_A\gg m_Z$, \eq{susyc2exact} implies that $\sbma\simeq 1$ (in the convention where
$0\leq\beta-\alpha\leq\pi$).  Using \eq{susyscexact},
\beq
\cbma\tanb\simeq\frac{2m_Z^2\sin^2\beta\cos 2\beta}{m_A^2}\ll 1\,,
\eeq
for \textit{all} values of $\tan\beta$.  In particular, for $\tan\beta\gg 1$, one can never have $\cbma\tanb\sim\mathcal{O}(1)$ in the decoupling regime.   Thus, in the tree level Higgs sector of the MSSM, the phenomenon of delayed decoupling discussed below \eq{true} does not occur.  In light of \eq{hDD}, one cannot achieve the wrong-sign $h\overline{D}D$ Yukawa coupling in the region of the tree level MSSM Higgs sector parameter space where the $hVV$ coupling is SM-like.

It is well known that radiative corrections can significantly alter the properties of the MSSM Higgs sector (reviewed in, e.g., Refs.~\cite{Carena:2002es} and \cite{Djouadi:2005gj}).   In particular, the  MSSM prediction for $m_h^2$ is significantly shifted from its tree level value given in \eq{mssmhm} by radiative corrections~\cite{mssmhiggs}.   In addition, the radiative corrections can also generate significant shifts to the tree level values of the Higgs couplings.  
For example, consider the scenario in which the MSSM $\mu$ parameter and all supersymmetry-breaking mass parameters (excluding the $B$ parameter, which fixes the value of the mass $m_A$) are all of order of a common supersymmetry-breaking mass 
scale $M_{\rm SUSY}$.  If $m_A\ll M_{\rm SUSY}$, then one can integrate out all the supersymmetric states to obtain a low-energy effective theory below the scale $M_{\rm SUSY}$, which can be identified as a 2HDM extension of the Standard Model.  In this effective 2HDM, the tree level values of the $\lambda_i$ given in \eq{lamsusy} receive significant radiative corrections.
Moreover, nonzero values for $\lambda_5$, $\lambda_6$ and $\lambda_7$ are generated~\cite{Haber:1993an}, which can be complex if there are CP-violating phases associated with $\mu$, $A_t$ and the gluino mass parameter.  Likewise, nonzero values
for the so-called \textit{wrong-Higgs} Yukawa couplings~\cite{Haber:2007dj} that are absent in a type-II model
are also generated.    That is, the resulting effective 2HDM is no longer described by a softly broken $\mathbb{Z}_2$ symmetric 2HDM with type-II Higgs-fermion Yukawa couplings.
Thus, the results of this paper are not directly applicable to the radiatively corrected MSSM Higgs sector with $m_A\ll M_{\rm SUSY}$.
Nevertheless, using the approximations given in Ref.~\cite{Carena:2001bg}, one can check whether
it is possible to achieve a wrong-sign $h\overline{D}D$ coupling in a suitable region of the MSSM Higgs parameter space in which the radiative corrections to the Higgs couplings are potentially significant.  

There are two separate effects that must be taken into account.  First, the 
radiatively generated wrong-Higgs Yukawa couplings contribute an additional term to the $h\overline{D}D$ coupling that is enhanced in the limit of large $\tan\beta$.  Keeping only these $\tan\beta$ enhanced corrections
and neglecting any CP-violating phases of the MSSM parameters for simplicity, the following approximate expression (for $M_{\rm SUSY}\gg m_Z$ and $\tanb\gg 1$) is given in Ref.~\cite{Carena:2001bg} for the $hb\bar{b}$ coupling,\footnote{The factor $(1+\Delta_b)^{-1}$ in \eq{hbbloop} provides a resummation of the leading $\Delta_b$ corrections to all orders~\cite{Carena:1999py}.}
\beq \label{hbbloop}
g_{hb\bar{b}}=-\frac{m_b}{v}\,\frac{\sin\alpha}{\cos\beta}\left[1-\frac{\Delta_b}{1+\Delta_b}\left(1+\cot\alpha\cot\beta\right)\right]\,,
\eeq
where~\cite{db}
\beq  \label{Deltab}
       \Delta_b
        \simeq\left[
        \frac{2 \alpha_s}{3 \pi} \mu M_{\tilde g} \,
        I(M_{\tilde b_1}, M_{\tilde b_2}, M_{\tilde g})
        + \frac{h_t^2}{16 \pi^2} \mu A_t \,
        I(M_{\tilde t_1}, M_{\tilde t_2}, \mu)\right]\tan\beta\,.
\eeq
In \eq{Deltab}, $M_{\tilde g}$ is the gluino mass,
$M_{\tilde b_{1,2}}$ are the bottom squark masses, $h_t$ is the top-quark Yukawa coupling and 
the loop integral $I(a,b,c)$ is given by
\beq
I(a,b,c) = {a^2b^2\ln(a^2/b^2)+b^2c^2\ln(b^2/c^2)+c^2a^2\ln(c^2/a^2) \over
(a^2-b^2)(b^2-c^2)(a^2-c^2)}\,.
\eeq
Note that $I(a,a,a)=1/(2a^2)$.  Thus, if all supersymmetric parameters appearing in \eq{Deltab} are of $\mathcal{O}(M_{\rm SUSY})$, then $\Delta_b$ approaches a constant (nondecoupling) value in the limit of $M_{\rm SUSY}\gg m_Z$.  It is convenient to rewrite
\beq
1+\cot\alpha\cot\beta=\frac{\cbma}{\sinb\sin\alpha}\,.
\eeq
Inserting this result into \eq{hbbloop} and making use of \eq{hDD}, we end up with
\beq \label{ghbb}
g_{hb\bar{b}}=\frac{m_b}{v}\left[\sbma-\cbma\tan\beta\left(\frac{1-\Delta_b\cot^2\beta}{1+\Delta_b}\right)\right]\,.
\eeq

Second, after integrating out the supersymmetric particles to obtain the low-energy effective 2HDM,
one must take into consideration the renormalization of the CP-even mixing angle $\alpha$.  To include these effects, we diagonalize the radiatively corrected $2\times 2$ CP-even Higgs squared-mass matrix.  Denoting these loop corrections by $\delta\mathcal{M}_{ij}^2$, an approximate expression for $\cbma$ in the limit of $m_Z\ll m_A\ll M_{\rm SUSY}$ is given by~\cite{Carena:2001bg}:
\beq \label{capprox}
\cbma\simeq \left(1+\frac{\delta \mathcal{M}_{11}^2-\delta \mathcal{M}_{22}^2}{2m_Z^2\cos 2\beta}-
\frac{\delta \mathcal{M}_{12}^2}{m_Z^2\sin 2\beta}\right)\frac{m_Z^2\sin 2\beta\cos 2\beta}{m_A^2}\,.
\eeq
In the limit of $\tanb\gg 1$, the term proportional to $\delta \mathcal{M}_{12}^2$
in \eq{capprox} can
dominate over the tree level contribution.
Using the approximate one-loop expression given in Ref.~\cite{Carena:2001bg},
\beq \label{csim}
\cbma\sim \frac{\delta\mathcal{M}_{12}^2}{m_A^2} \simeq
-\frac{g^2 m_t^4}{32\pi^2 m_W^2 m_A^2 \sin^2\beta}\,\frac{\mu X_t}{M^2_{\rm SUSY}}
\left(6-\frac{X_t A_t}{M^2_{\rm SUSY}}\right)\,,
\eeq
where $X_t\equiv A_t-\mu\cot\beta$ (note that $X_t\simeq A_t$ for $\tanb\gg 1$).  

A quick back-of-the-envelope numerical analysis can reveal whether it is possible to achieve a value of $vg_{hb\bar{b}}/m_b$ close to $-1$.  We shall assume that $\sbma\sim 1$, corresponding to a SM-like $hVV$ coupling.  To maximize the effect of the radiative corrections, we shall also assume that $\tanb\gg 1$.   If we further assume that all supersymmetric particle masses are of $\mathcal{O}(M_{\rm SUSY})$, then \eq{Deltab} yields $\Delta_b\sim \pm 0.01\tanb$, where the sign is determined by the overall sign of $\mu M_{\tilde g}$ [since the first term in \eq{Deltab} typically dominates].
In light of \eq{range}, we conclude that $|\Delta_b|\lsim 0.5$, so at best the inclusion of $\Delta_b$ enhances the second term on the right-hand side of \eq{ghbb} by a factor of 2.  Thus, we examine whether it is plausible that $\cbma\tanb\sim\mathcal{O}(1)$.

In evaluating \eq{csim}, we must also ensure that the observed Higgs mass is correctly reproduced by the choice of supersymmetric parameters which govern the radiative corrections.   In the so-called maximal mixing scenario where $A_t^2=6M^2_{\rm SUSY}$, the approximate expression for $\delta\mathcal{M}_{12}^2$ vanishes.   For large values of $\tan\beta$, the measured Higgs mass,
$m_h\sim 125$~GeV is not compatible with the maximal mixing scenario as defined in Ref.~\cite{Carena:2013qia}, so it is reasonable to take $6-X_t A_t/M_{\rm SUSY}^2\sim\mathcal{O}(1)$.   As an example,
for $\tanb\gg 1$, $A_t\sim 2M_{\rm SUSY}$ and $\mu\sim -2M_{\rm SUSY}$,  one 
finds numerically that 
\beq
\cbma\sim \left({28\gev\over\mha}\right)^2\,.
\eeq
Choosing extreme parameters, $\tanb=50$ and $\Delta_b=-0.5$, we see that it is just possible to achieve 
a value of $vg_{hb\bar{b}}/m_b$ close to $-1$ if $m_A=200$~GeV.    However, this value of $m_A$ is uncomfortably close to $m_Z$ and $m_h$, in which case one must check that terms of $\mathcal{O}(m_Z^2/m_A^2)$, which have been neglected in the above analysis, do not spoil the estimate.   
Increasing the magnitude of $\mu$ or taking $A_t$ slightly above its maximal mixing value  would allow for a wrong-sign $hb\bar{b}$ coupling together with a somewhat higher value of $m_A$.

Similar considerations also apply to the $h\tau^+\tau^-$ coupling.  However, the expression for 
$\Delta_\tau$  [analogous to \eq{Deltab} for $\Delta_b$] involves only terms proportional to 
electroweak gauge couplings.  Hence, the effects of $\Delta_\tau$ only have a small impact on
$g_{h\tau^+\tau^-}$.  Thus, it is even harder to find a sensible parameter regime in which
$vg_{h\tau^+\tau^-}/m_\tau$ is close to~$-1$.
We conclude that in the MSSM, the wrong-sign $hb\bar{b}$ and $h\tau^+\tau^-$ couplings are not possible for generic choices of the MSSM parameters.  Nevertheless, based on an approximate treatment of the leading radiative corrections, it seems that some extreme regions of the parameter space do exist in which a value of $vg_{hb\bar{b}}/m_b$ close to $-1$ can be achieved due to large radiative correction effects in the large $\tan\beta$ regime.   
A more detailed study of the MSSM Higgs parameter space based on a more complete analysis of the radiative corrections lies beyond the scope of this paper.

\section{Nondecoupling of the $\boldsymbol{H^\pm}$ loop contribution
  to the  $\boldsymbol{h\to\gam\gam}$ amplitude and the $\boldsymbol{\kd<0}$ scenario}
\label{apjfg}

In this appendix, we give a detailed treatment of the nondecoupling 
of the  $H^\pm$ loop contribution
to the  $h\to\gam\gam$ amplitude discussed at the end of section \ref{sec:decoup}, focusing on its impact on the wrong-sign Yukawa coupling scenario, i.e.~$\kd<0$. 
In particular, we demonstrate
that the charged-Higgs contribution to the $h\gam\gam$ coupling in the $\kd<0$ case is approximately constant and {\it always} sufficiently significant as to eventually be observable at the LHC.
In addition, we display explicitly the constraints coming from tree level unitarity, which imply that the $\kd<0$ scenario is only perturbatively reliable for $\mhpm\lsim 650\gev$.  We also remark on nondecoupling of the charged-Higgs loop for some $\kd>0$ scenarios.\footnote{The phenomenological effects of the nondecoupling charged-Higgs loop contribution to the $h\to\gamma\gamma$ amplitude and other 2HDM observables have also been considered in Refs.~\cite{Arhrib:2003ph} and \cite{Bhattacharyya:2013rya}.}

To begin, let us first recall the basic formulae from Ref.~\cite{decoupling} in the case of $\lam_6=\lam_7=0$ considered in this paper, as summarized in section \ref{sec:decoup}.  The crucial ingredients are the mass-squared relation of \eq{mdiff} and the expression \eq{alt} for the $h\hp\hm$ coupling, $G_{h\hp\hm}$ [cf.~\eq{Hlags}].
For the purposes of this appendix, it is  useful to rearrange some of the angular factors and to define the dimensionless coupling
\beqa
g_{\hl\hp\hm}&\equiv&\frac{G_{\hl\hp\hm}}{v} \nonumber \\[8pt]
  &=&\frac{(2\mha^2-2\mhpm^2-\mhl^2)\sbma\sinb\cosb+(\mha^2-\mhl^2)\cos
    2\beta\cbma+\lam_5 v^2 \cbpa}{v^2\sinb\cosb}\,.
\label{gnodim}
\eeqa

In the decoupling limit  described in Section~\ref{sec:decoup}, 
we have $\sbma\to 1$, $\cbma\to 0$, and $\mha^2\sim\mhh^2\sim\mhpm^2\gg v^2$.
The first term inside the brackets of \eq{gnodim} is of order $v^2$
because of the mass relations (keeping the $\lam_i$ perturbative) and
the second term is of order $v^2$  because $\cbma\propto
v^2/\mha^2$. To discuss the third term we need to note that for
$\sbma\to 1$ we have $\alpha\to\beta-\pi/2$.  Then, the third term
approaches $2v^2\lam_5$ since $\cos(\beta+\alpha)\to \sin 2\beta=2\sinb\cosb$.  The net result is that $g_{h\hp\hm}$ is not growing with the Higgs mass squared and so the charged-Higgs loop contribution to the $h\to\gam\gam$  amplitude is suppressed by a factor of $m_W^2/\mhpm^2$ relative to the $W$ and $t$ and $b$ loops.  This is in correspondence with the idea that any heavy particle that does not acquire mass from the Higgs vacuum expectation value should decouple.

\begin{figure}[t!]
\centering
\includegraphics[width=3.5in,angle=0]{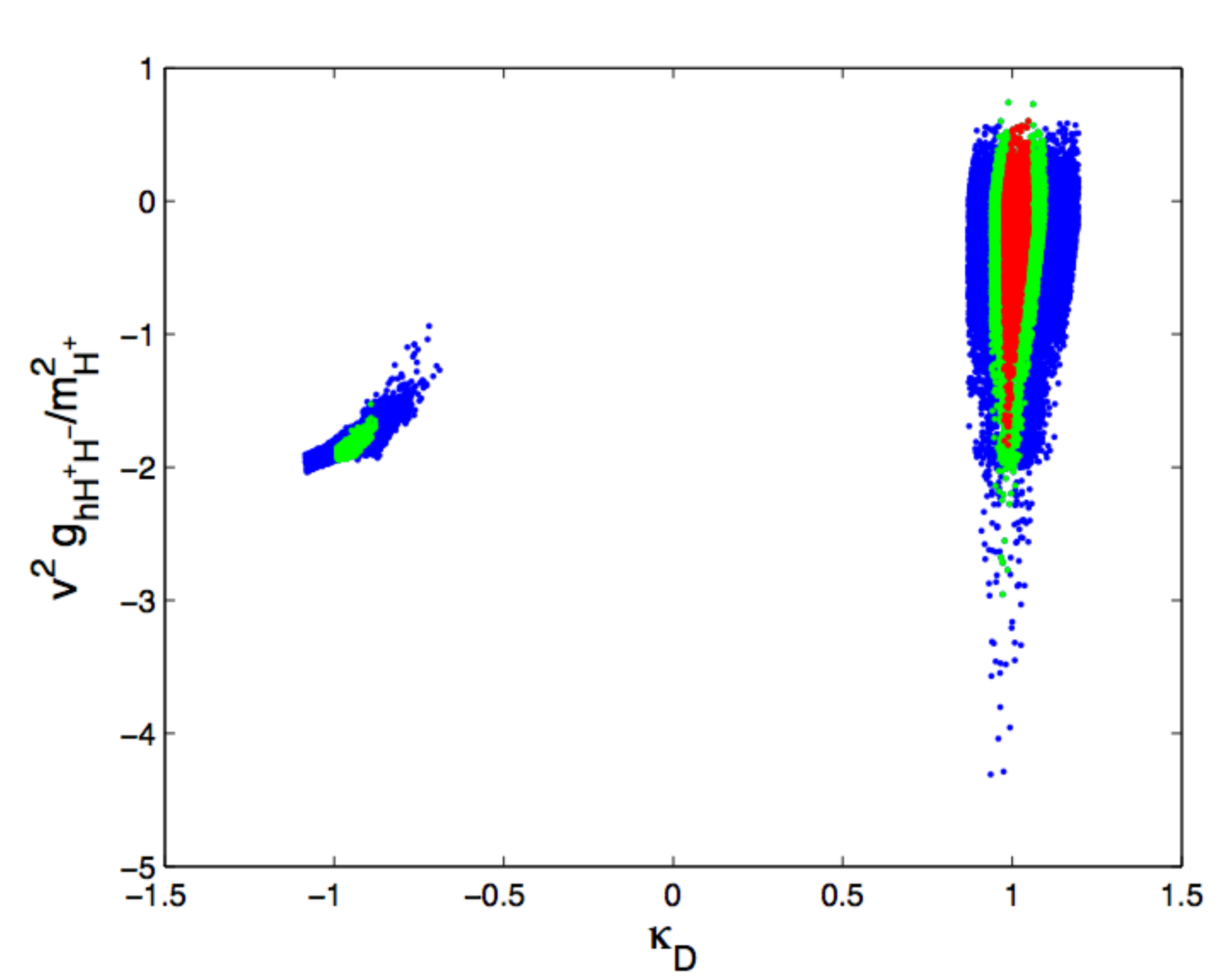}
\caption{We show points in the $v^2 g_{\hl\hp\hm}/\mhpm^2$ vs. $\kd$ plane with the standard color scheme of Fig.~\ref{fig:T2}. }
\label{constant}
\end{figure}

However, the situation is {\it necessarily} quite different in the case of $\kd<0$, where $\sbpa\to 1$, implying $\alpha\to  \pi/2-\beta$.  In this limit, $\cos(\beta-\alpha)\to \sin 2\beta$ so that the second term in the numerator of \eq{gnodim} is approximated by $2(\mha^2-\mhl^2)\cos{2\beta}$ which approaches $\sim 2\mhpm^2\cos{2\beta}$ as $\mha^2\sim \mhh^2\sim \mhpm^2
\to \infty$ (at fixed $\mhl\sim 125\gev$).  Of course, if $\tanb$ is large then $\cos{2\beta}\to -1$.  Thus, we see from \eq{gnodim} that for $\kd<0$ we have 
\beq \label{simtwo}
\frac{v^2 g_{\hl\hp\hm}}{\mhpm^2}\sim -2\,, 
\eeq
implying that the $H^\pm$ loop contribution to the $h\to\gam\gam$ amplitude will never decouple.  In practice, \eq{simtwo} implies that the modification cannot be detected if the $\muflhc$ values are only measured to be within 20\% or 10\% of unity, whereas no $\kappa_D<0$ points survive if the  $\muflhc$ values are found to be within $5\%$ of unity, as illustrated in Fig.~\ref{constant}.  In contrast, the range of
allowed values of $v^2 g_{\hl\hp\hm}/\mhpm^2$ in the case of $\kd>0$ is much larger, from nondecoupling values of $\mathcal{O}(1)$ (both positive and negative)  to decoupling values significantly less than~1.  Note that the results of Fig.~\ref{constant} indicate that, as in the $\kd<0$ scenario, the points in the case of $\kd>0$ with $v^2 g_{\hl\hp\hm}/\mhpm^2\lsim -2$ will not survive
if all the $\muflhc$ are measured to be within $5\%$ of the SM value of unity.

We have already noted that  in the $\kd<0$ scenario there will be a limitation on $\mhpm^2$ coming from perturbativity and unitarity. The relevant constraints are incorporated in all of our plots. Once $\mhpm$ becomes too large, the theory becomes perturbatively unreliable and insisting on tree level unitarity will then imply that only the $\kd>0$ possibility is allowed.  So, in this sense, nondecoupling is only possible temporarily for an intermediate range of heavy $H^\pm$ masses if we insist that  $\mhpm$ not be so large that the tree level unitarity bound is violated.
In order to illustrate the nature of the unitarity limits, we present some plots.

\begin{figure}[t!]
\centering
\includegraphics[width=3.5in,angle=0]{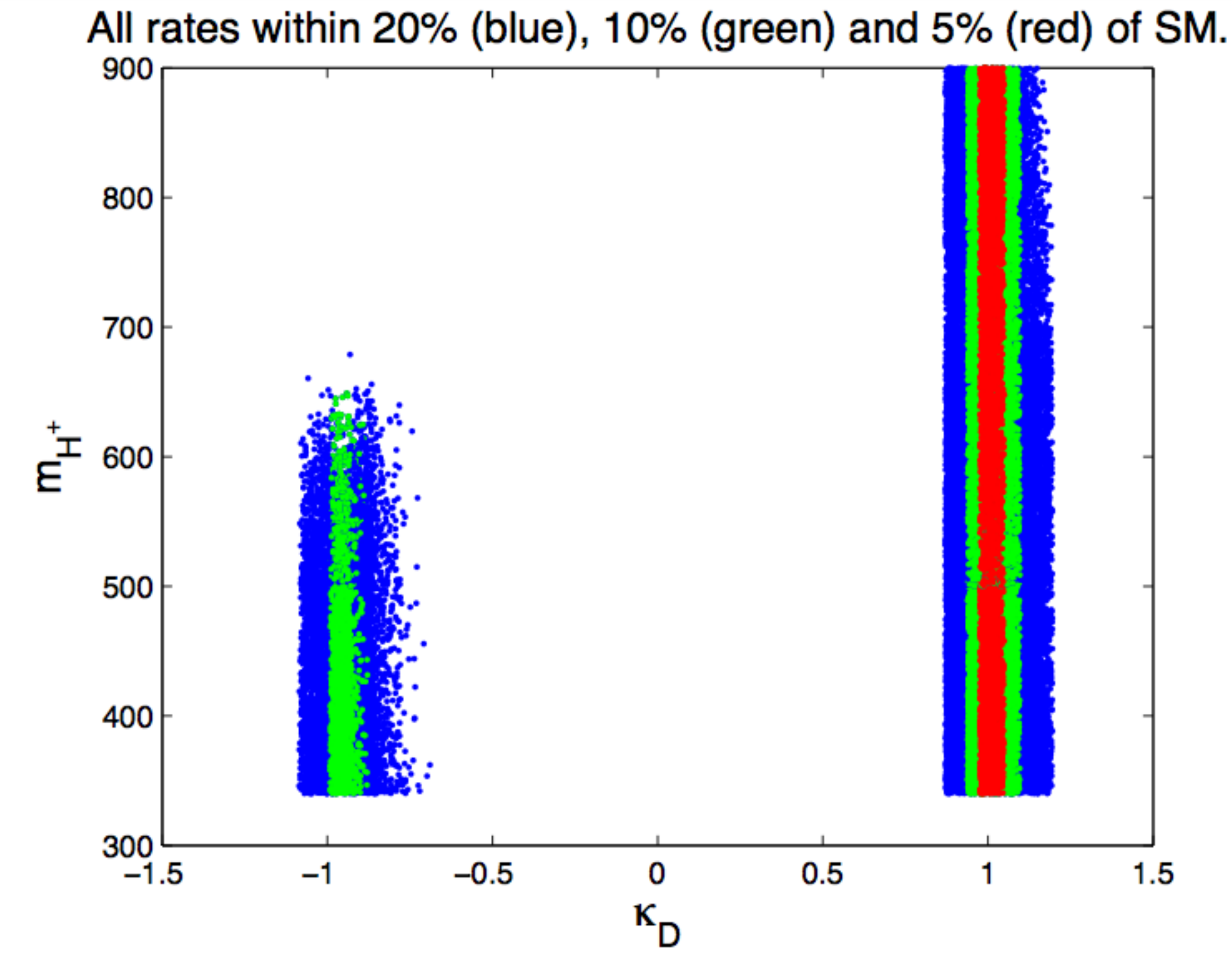}
\caption{We show points in the $\mhpm$ vs. $\kd$ plane with the standard color scheme of Fig.~\ref{fig:T2}. Note: $900\gev$ is the largest $\mhpm$ considered in the scans---the $\kd\sim +1$ region  would extend to arbitrarily large $\mhpm$ corresponding to the decoupling limit.}
\label{mchvskd}
\end{figure}
\begin{figure}[t!]
\bec
\includegraphics[width=3.5in,angle=0]{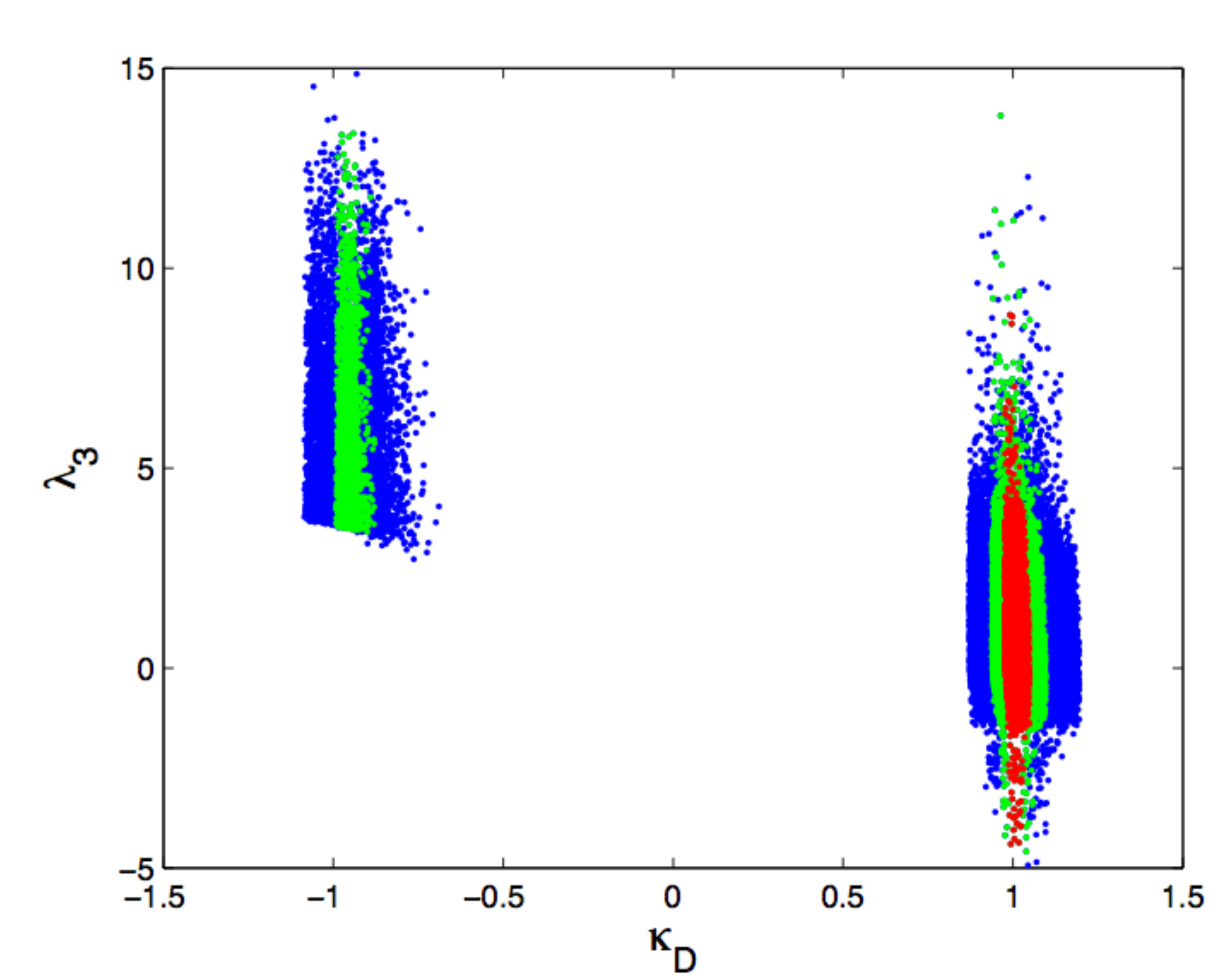}
\hspace{-.3cm}
\includegraphics[width=3.5in,angle=0]{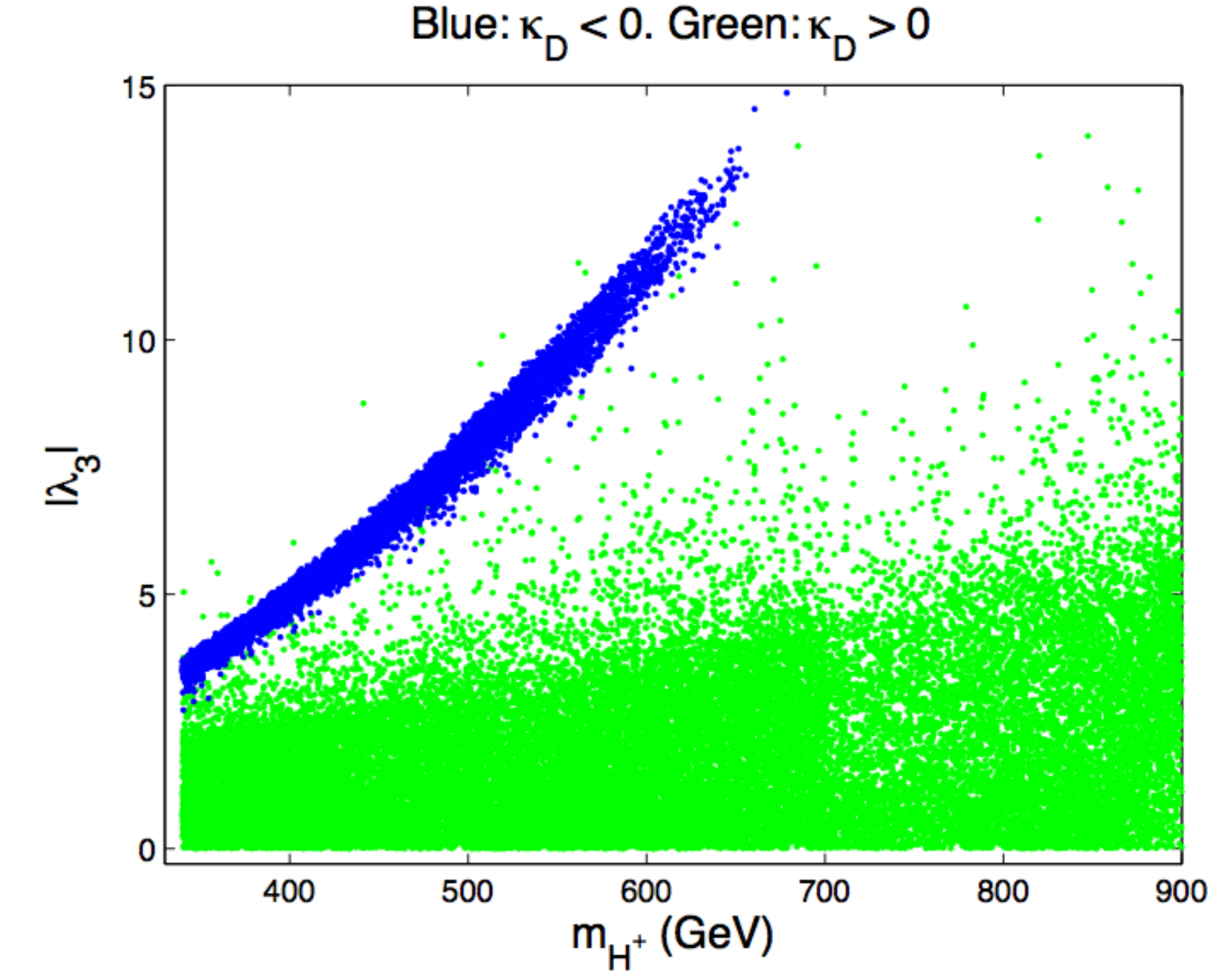}
\eec
\caption{In the left panel we plot $\lam_3$ vs. $\kd$ using the color scheme of Fig.~\ref{fig:T2}; in the right panel we plot $|\lambda_3|$ vs. $\mhpm$ for $\kd<0$ (blue/black) and $\kd>0$ (green/light grey) points with all $\muflhc$ within 20\% of unity. }
\label{uplimit}
\end{figure}

In Fig.~\ref{mchvskd}, we show points in the $\mhpm$ vs. $\kd$ plane allowed when all the $\muflhc$s are within 20\%, 10\% or 5\% of unity.  We see clearly that $\mhpm$ is limited to lie below about $650\gev$ in the $\kd<0$ case while it can be arbitrarily large (we only scan up to $900\gev$) for the standard $\kd>0$ scenario that allows for true decoupling.  We have found that the maximum $\mhpm$ value is limited by the tree level unitarity limits of the $\lam_i$, in particular $\lam_3$. 
In Fig.~\ref{uplimit}, we display in the left panel $\lam_3$ as a function of $\kd$ for both the $\kd>0$ and $\kd<0$ scenarios; and in the right panel we show $|\lam_3|$ as a function of  $\mhpm$ for the $\kd<0$ and $\kd>0$ scenarios requiring only that all $\muflhc$s be within 20\% of unity.  Given that the tree level unitarity bounds on the $\lam_i$ are of order $|\lam_i|\lsim 15$, we see that it is $\lam_3$ that encounters this upper limit at large $\mhpm$ in the $\kd<0$ case, whereas it is clear that in the $\kd>0$ case arbitrarily large $\mhpm$ is possible without violating tree level unitarity bounds, consistent with the decoupling limit. However, one should also note the significant number of $\kd>0$ points that hit the tree level unitarity bound for which nondecoupling is again possible.

\begin{figure}[t!]
\centering
\includegraphics[width=3.5in,angle=0]{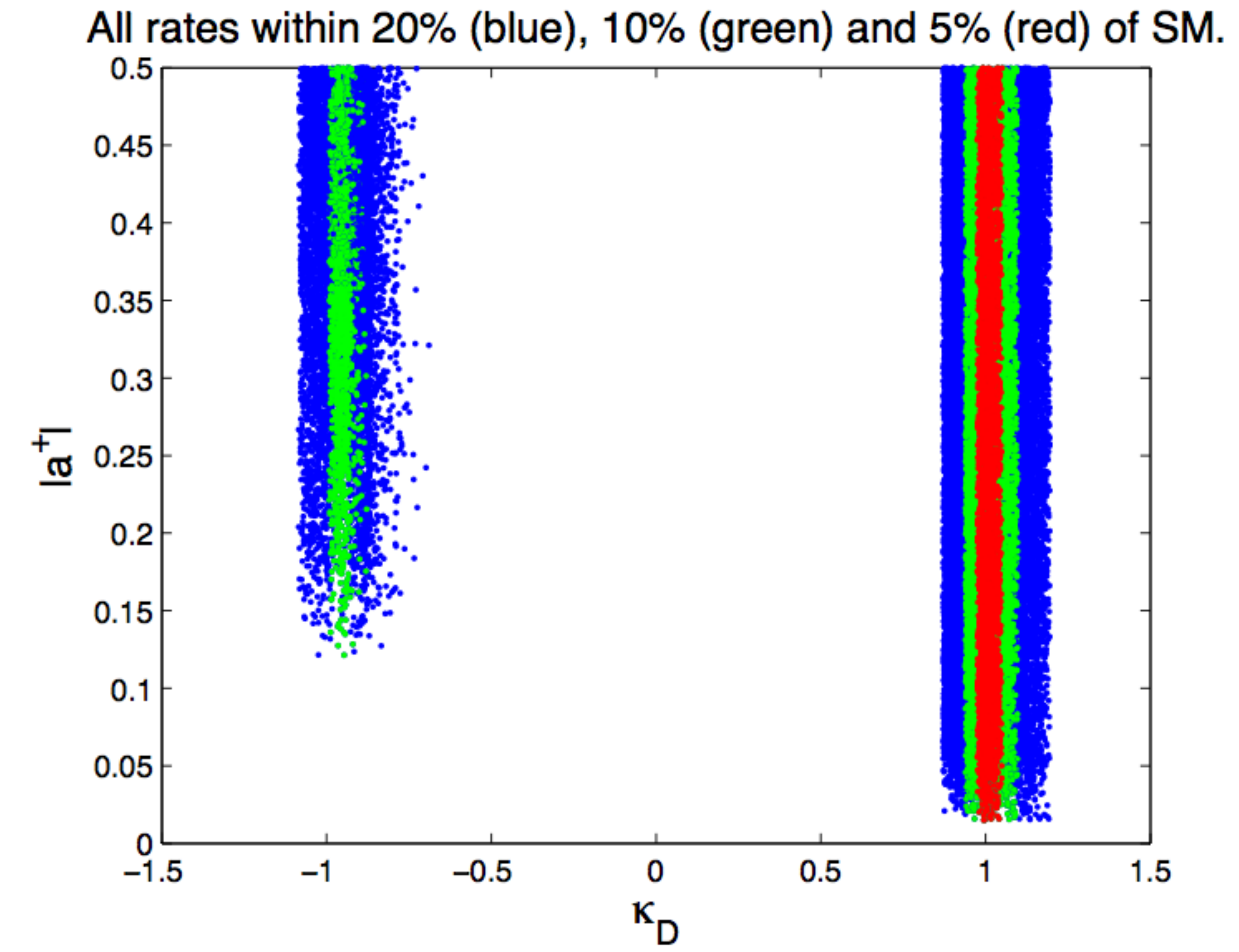}
\includegraphics[width=3.5in,angle=0]{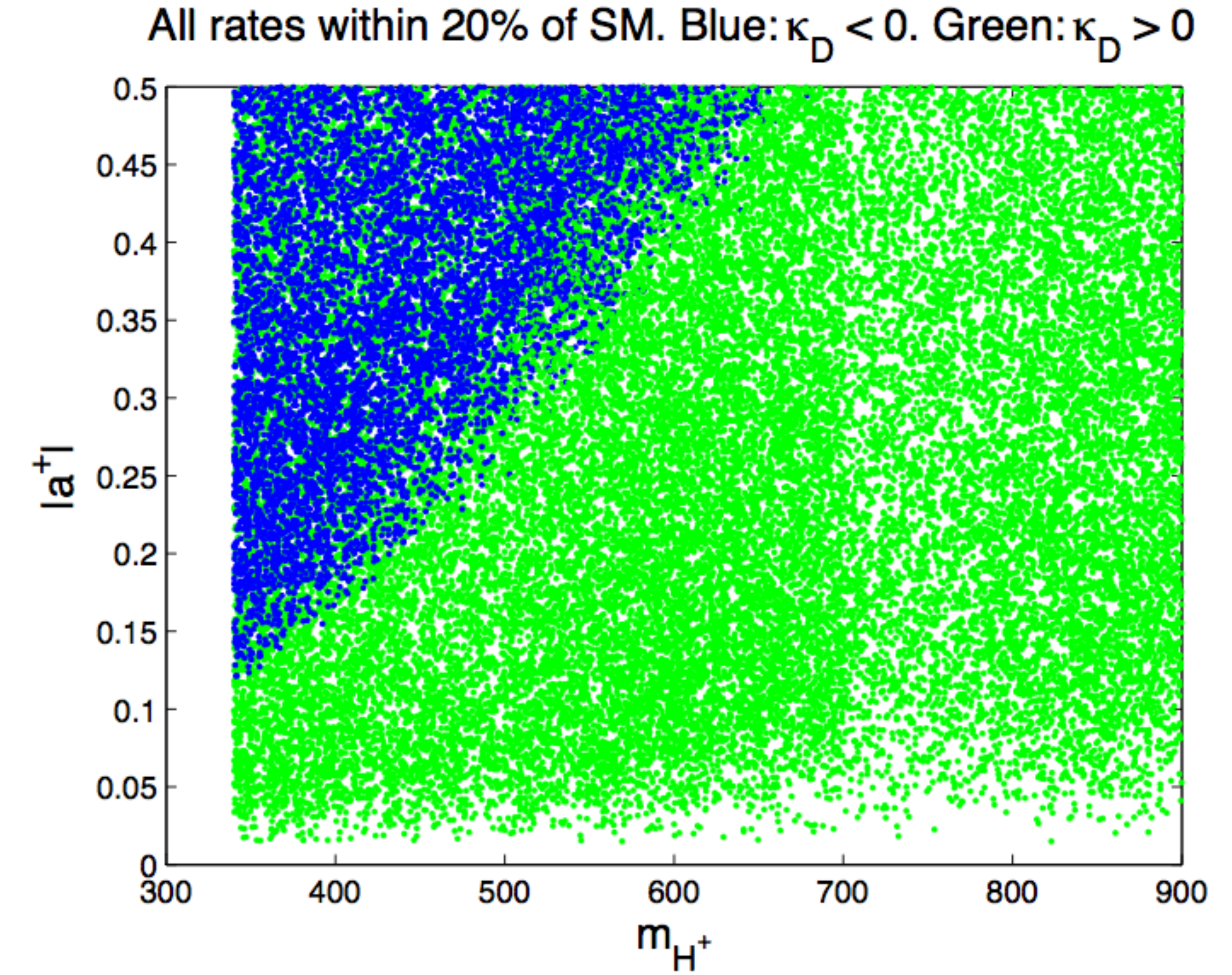}
\caption{In the left panel we plot $|a^+|$ vs. $\kd$ using the color scheme of Fig.~\ref{fig:T2}; in the right panel we plot $|a^+|$ vs. $\mhpm$ for $\kd<0$ (blue/black) and $\kd>0$ (green/light grey) points with all $\muflhc$ within 20\% of unity.}
\label{uplimitaplus}
\end{figure}

The actual limits based on tree level unitarity bounds are imposed in terms of various $\lam_i$ amplitude combinations, of which it is
\beq
a^+=\frac{1}{16 \, \pi}  \left[ {3\over 2}(\lam_1+\lam_3)+\sqrt{{9\over 4}(\lam_1-\lam_2)^2+(2\lam_3+
\lam_4)^2} \right]
\eeq
that is most constraining.  In Fig.~\ref{uplimitaplus} we plot $|a^+|$ as a function of $\kd$ and of $\mhpm$ using the same format as in Fig.~\ref{uplimit}.
Note that $|a^+|$ is hitting the tree level unitarity bound of $0.5$ for both the $\kd<0$ and $\kd>0$ scenarios.  However, there is no limit on the associated $\mhpm$ value in the latter case, whereas there is the already quoted limit of $\sim 650\gev$ in the former case.

We now show that for the type-II 2HDM with $\kd<0$, where $v^2 g_{\hl\hp\hm}/\mhpm\sim -2$, and with $\kd>0$,  where $v^2 g_{\hl\hp\hm}/\mhpm\lsim -2$  (cf.~Fig.~\ref{constant}), the loop functions are such that the charged-Higgs loop contributes with the same sign as the top-quark loop and thus will reduce the $h\to\gam\gam$ width, both canceling part of the $W$-loop contribution of the opposite sign. As we have seen earlier, and will show numerically below, we find that this reduction is sufficient to prevent the $\gam\gam$ channel from ever approaching the SM prediction and by an amount that will be seen at the LHC with high luminosity.

Let us now give more details.  We will employ a simplified version of the the notation of \texttt{CPsuperH} \cite{Lee:2003nta}.  One finds:
\beq
\Gamma(\hl\to\gam\gam)={\alpha^2 g^2\over 256 \pi^3}{\mhl^3\over v^2}\left\vert S \right\vert^2\,,
\eeq
where
\beqa
S&=&2\sum_{f=b,t}\left[ N_c q_f^2 g_f g_{\hl f\anti f}{v\over m_f}F_{1/2}(\tau_f) - g_{\hl WW} F_1(\tau_W)-g_{\hl\hp\hm}{v^2\over 2\mhpm^2}F_0(\tau_{\hpm})\right] \nonumber \\
&\equiv& 2\sum_f \left[ S^\hl_f F_{1/2}(\tau_f)+S^\hl_WF_1(\tau_W)+S^\hl_{\hpm}F_0(\tau_{\hpm})\right] \nonumber \\
&\equiv& \sum_f \left(I^\hl_f +I^\hl_W+I^\hl_{\hpm}\right)\,,
\label{sform}
\eeqa
with $\tau_i=\mhl^2/(4 m_i^2)$, $N_c=3$, and the various $F$'s given by\footnote{Relative to Ref.~\cite{hhg}, the $F_{1/2}$ defined here is one-half as large and $F_0$ has the opposite sign.}
\beq
F_{1/2}(\tau)=\taui[1+(1-\taui)f(\tau)]\,,\quad F_1(\tau)=2+3\taui+3\taui(2-\taui)f(\tau)\,,\quad F_0(\tau)=\taui[-1+\taui f(\tau)]\,.
\eeq
An explicit form for the function $f(\tau)$ is defined in Eq.~(40) of Ref.~\cite{Lee:2003nta}.  In the $\tau\to 0$ limit, $F_{1/2}\to 2/3$, $F_1\to 7$ and $F_0\to 1/3$.
In \eq{sform}, $g_f=gm_f/(2m_W)$ and the other $g$'s are defined by the interaction Lagrangians,
\beq \label{Hlags}
\call_{\hl WW}=(g\mw )g_{\hl WW} {\wp}_\mu{\wm}^\mu \hl\,,\quad \call_{\hl f\anti f}=-{gm_f\over 2\mw}g_{\hl f\anti f}\hl \anti f f\,,\quad \call_{\hl\hp\hm}=v g_{\hl \hp\hm}\hl\hp\hm\,,
\eeq
where $g_{\hl\hp\hm}\equiv G_{\hl\hp\hm}/v$ as defined in \eq{gnodim}.

In the $\kd>0$ case with $\sbma\to 1$ we have $\alpha\to \beta-\pi/2$, for which $\cbma\propto v^2/\mha^2$,  $\sin\alpha\to -\cos\beta$  and $\cos\alpha\to \sin\beta$ with the result
\beq
S^\hl_{u,c,t} \to 1\,,\quad S^\hl_{d,s,b,e,\mu,\tau}\to +1\,,\quad S_W^\hl=-\sbma\to -1\,,\quad S^\hl_{\hpm}\propto {v^2 \over \mw^2}\,.
\eeq

In the $\kd<0$ case we have $\alpha\to \pi/2-\beta$, for which $\sin\alpha\to \cos\beta$, $\cos\alpha\to \sin\beta$,  $\cbma\to \sin(2\beta)$, and $\sbma\to -\cos(2\beta)=(\tanbsq-1)/(\tanbsq+1)$ .  For simplicity, consider $\tanb\to \infty$ and the limit of large $\mha^2\sim \mhpm^2$.  We then have,
\beq
S_{u,c,t}^\hl\to 1\,,\quad S^\hl_{d,s,b,e,\mu,\tau}\to -1\,,\quad S_W^\hl=-\sbma\to -1\,,\quad S^\hl_{\hpm}\to 1\,.
\eeq
The important thing to note here is that the $\hpm$ loop contributes with the same sign as the top loop, {\it i.e.} it too will cancel against the negative $W$-loop and decrease the $\hl\to\gam\gam$ width.

In more detail, we have the following.  For both $\kd>0$ and $\kd<0$,
the relative contributions of the top-quark loop and the $W$ loop  to $S$ are $I^\hl_W\simeq -8.3233$ and $I^\hl_t=+1.8351$.
As regards the charged-Higgs loop, for $\kd<0$  and  large $\mhpm$ one gets $I^\hl_{\hpm}=+0.33333$.
As regards the $b$-quark loop, for the case of $\kd>0$ we have  $I^\hl_b=-0.0279+0.04 i$.  Of course, this changes sign for  $\kd<0$. We will neglect other quarks and leptons for simplicity since their contributions are quite small.

Then, in the SM $\kd>0$ case, neglecting the decoupled charged-Higgs loops, we find $\sum_{i=W,t,b} I^\hl_i=-6.5161+0.04 i$.  If we consider the $\kd<0$ case without including the charged-Higgs loop one finds  $\sum_{i=W,t,b} I^\hl_i=-6.4603-0.04 i$.   The ratio of the absolute values is $0.99$, a less than $1\%$ decrease in $\kgam$ and certainly not measurable at the LHC.  However, after including the  charged-Higgs loop we obtain $\sum_{i=W,t,b,\hpm} I^\hl_i=-6.127-0.04i$ with the charged-Higgs loop evaluated at large $\mhpm$, which translates to  $\kgam\sim 0.94$ corresponding to a 12\% decrease in $\Gamma(\hl\to \gam\gam)$.  In fact, this level of decrease is very characteristic of the full scan as shown in Fig.~\ref{constant} and is measurable at the LHC with $\rts=14\tev$ and $L\geq 300\fbi$.  As already noted, this same level of decrease also occurs for those $\kd>0$ scenarios for which the charged-Higgs loop does not decouple, i.e. roughly if $v^2 g_{\hl\hp\hm}/\mhpm\lsim -2$ (see Fig.~\ref{constant}).

Of course, in the computations presented in the main text,  the full set of quarks and leptons is included, the charged-Higgs mass is varied as part of the scan (with the lower bound of  $340\gev$) and current LHC Higgs constraints are imposed as well as constraints from perturbativity, unitarity and precision electroweak measurements.  As we have said above, all this leads to  only small numerical changes relative to the $\kgam$ decrease for $\kd<0$ quoted above; thus, the nondecoupling of the $\hpm$ loop for $\kd<0$ leads to a decrease in $\kgam$ that is at least as large as 5\%.

%
%%%%%%%%%%%%%%%%%%%%%%%%%%%%%%%%%%%%%%%%%%%%%%%%%%%%%%%%%%%%%%%%%%%%%%%
%
%
% REFERENCES
%


\begin{thebibliography}{99}

%%%%%%%%%%%%%%%%%% discovery

%\cite{Gunion:1989we}
\bibitem{ATLASHiggs}
G.~Aad {\it et al.}  [ATLAS Collaboration],
  %``Observation of a new particle in the search for the Standard Model Higgs boson with the ATLAS detector at the LHC,''
  Phys.\ Lett.\ B {\bf 716}, 1 (2012)
  [arXiv:1207.7214 [hep-ex]].
  %%CITATION = ARXIV:1207.7214;%%

\bibitem{CMSHiggs}
S.~Chatrchyan {\it et al.}  [CMS Collaboration],
  %``Observation of a new boson at a mass of 125 GeV with the CMS experiment at the LHC,''
  Phys.\ Lett.\ B {\bf 716}, 30 (2012)
  [arXiv:1207.7235 [hep-ex]].
  %%CITATION = ARXIV:1207.7235;%%

%\cite{Aad:2013wqa}
\bibitem{Aad:2013wqa}
  G.~Aad {\it et al.}  [ATLAS Collaboration],
  %``Measurements of Higgs boson production and couplings in diboson final states with the ATLAS detector at the LHC,''
  Phys.\ Lett.\ B {\bf 726} (2013) 88
  [arXiv:1307.1427 [hep-ex]].
  %%CITATION = ARXIV:1307.1427;%%
  %94 citations counted in INSPIRE as of 05 Jan 2014

  %\cite{Chatrchyan:2013mxa}
\bibitem{Chatrchyan:2013mxa}
  S.~Chatrchyan {\it et al.}  [ CMS Collaboration],
  %``Measurement of the properties of a Higgs boson in the four-lepton final state,''
  Phys.\ Rev.\ D {\bf 89}, 092007 (2014)
  [arXiv:1312.5353 [hep-ex]].
  %%CITATION = ARXIV:1312.5353;%%
  %3 citations counted in INSPIRE as of 05 Jan 2014

\bibitem{pdg}
M.~Carena, C.~Grojean, M.~Kado and V.~Sharma, in the
2013 partial update for the 2014 edition of
J. Beringer et al. (Particle Data Group), Phys. Rev. D {\bf 86}, 010001 (2012)
[\texttt{http://pdg.lbl.gov/2013/reviews/rpp2013-rev-higgs-boson.pdf}].
%%CITATION = PHRVA,D86,010001;%%

%%%%%%%%%%%%%%%%%%% effective

%\cite{Espinosa:2012im}
\bibitem{Espinosa:2012im}
  J.R.~Espinosa, C.~Grojean, M.~M{\"u}hlleitner and M.~Trott,
  %``First Glimpses at Higgs' face,''
  JHEP {\bf 1212} (2012) 045
  [arXiv:1207.1717 [hep-ph]].
  %%CITATION = ARXIV:1207.1717;%%
  %156 citations counted in INSPIRE as of 17 Jan 2014

%\cite{Falkowski:2013dza}
\bibitem{Falkowski:2013dza}
  A.~Falkowski, F.~Riva and A.~Urbano,
  %``Higgs at last,''
  JHEP {\bf 1311} (2013) 111
  [arXiv:1303.1812 [hep-ph]].
  %%CITATION = ARXIV:1303.1812;%%
  %86 citations counted in INSPIRE as of 17 Jan 2014

%\cite{Belanger:2013xza}
\bibitem{Belanger:2013xza}
  G.~Belanger, B.~Dumont, U.~Ellwanger, J.F.~Gunion and S.~Kraml,
  %``Global fit to Higgs signal strengths and couplings and implications for extended Higgs sectors,''
  Phys.\ Rev.\ D {\bf 88}, 075008 (2013)
  [arXiv:1306.2941 [hep-ph]].
  %%CITATION = ARXIV:1306.2941;%%
  %62 citations counted in INSPIRE as of 14 Feb 2014
 
 \bibitem{Celis:2013ixa} 
  A.~Celis, V.~Ilisie and A.~Pich,
  %``Towards a general analysis of LHC data within two-Higgs-doublet models,''
  JHEP {\bf 1312}, 095 (2013)
  [arXiv:1310.7941 [hep-ph]].
  %%CITATION = ARXIV:1310.7941;%%

%%%%%%%%%%%%%%%%%%%%%%%%%  2HDM refs

\bibitem{manyrefs}
%\cite{Ferreira:2011aa}
%\bibitem{Ferreira:2011aa}
   P.M.~Ferreira, R.~Santos, M.~Sher and J.P.~Silva,
  %``Implications of the LHC two-photon signal for two-Higgs-doublet models,''
  Phys.\ Rev.\ D {\bf 85}, 077703 (2012)
  [arXiv:1112.3277 [hep-ph]];
  %%CITATION = ARXIV:1112.3277;%%
  %
%\cite{Carmi:2012yp}
%\bibitem{Carmi:2012yp}
  D.~Carmi, A.~Falkowski, E.~Kuflik and T.~Volansky,
  %``Interpreting LHC Higgs Results from Natural New Physics Perspective,''
  JHEP {\bf 1207} (2012) 136
  [arXiv:1202.3144 [hep-ph]];
  %%CITATION = ARXIV:1202.3144;%%
   %
  H.S.~Cheon and S.K.~Kang,
  %``Constraining parameter space in type-II two-Higgs doublet model in light of a 125 GeV Higgs boson,''
  JHEP {\bf 1309}, 085 (2013)
  [arXiv:1207.1083 [hep-ph]];
  %%CITATION = ARXIV:1207.1083;%% 
%
W.~Altmannshofer, S.~Gori and G.D.~Kribs,
%``A Minimal Flavor Violating 2HDM at the LHC,''
Phys.\ Rev.\ D {\bf 86}, 115009 (2012)
[arXiv:1210.2465 [hep-ph]];
%%CITATION = ARXIV:1210.2465;%%
%
Y.~Bai, V.~Barger, L.L.~Everett and G.~Shaughnessy,
%``General two Higgs doublet model (2HDM-G) and Large Hadron Collider data,''
  Phys.\ Rev.\ D {\bf 87}, 115013 (2013)
  [arXiv:1210.4922 [hep-ph]];
  %%CITATION = ARXIV:1210.4922;%%
%
  C.-Y.~Chen and S.~Dawson,
  %``Exploring Two Higgs Doublet Models Through Higgs Production,''
Phys.\ Rev.\ D {\bf 87}, 055016 (2013)
 [arXiv:1301.0309 [hep-ph]];
  %%CITATION = ARXIV:1301.0309;%%
 %
%\cite{Celis:2013rcs}
%\bibitem{Celis:2013rcs}
  A.~Celis, V.~Ilisie and A.~Pich,
  %``LHC constraints on two-Higgs doublet models,''
   JHEP {\bf 1307}, 053 (2013)
  [arXiv:1302.4022 [hep-ph]];
  %%CITATION = ARXIV:1302.4022;%%
  %10 citations counted in INSPIRE as of 05 May 2013%
%\cite{Chiang:2013ixa}
%\bibitem{Chiang:2013ixa}
  C-W.~Chiang and K.~Yagyu,
  %``Implications of Higgs boson search data on the two-Higgs doublet models with a softly broken $Z_2$ symmetry,''
  JHEP {\bf 1307}, 160 (2013)
  [arXiv:1303.0168 [hep-ph]];
  %%CITATION = ARXIV:1303.0168;%%
 %
  %\cite{Krawczyk:2013gia}
%\bibitem{Krawczyk:2013gia}
  M.~Krawczyk, D.~Sokolowska and B.~Swiezewska,
  %``2HDM with Z_2 symmetry in light of new LHC data,''
  J.\ Phys.\ Conf.\ Ser.\  {\bf 447}, 012050 (2013)
  [arXiv:1303.7102 [hep-ph]];
  %%CITATION = ARXIV:1303.7102;%%
%
%\cite{Grinstein:2013npa}
%\bibitem{Grinstein:2013npa}
  B.~Grinstein and P.~Uttayarat,
  %``Carving Out Parameter Space in Type-II Two Higgs Doublets Model,''
   JHEP {\bf 1306}, 094 (2013)
  [Erratum-ibid.\  {\bf 1309}, 110 (2013)]
  [arXiv:1304.0028 [hep-ph]];
  %%CITATION = ARXIV:1304.0028;%%
 %
%\cite{Barroso:2013zxa}
%\bibitem{Barroso:2013zxa}
  A.~Barroso, P.M.~Ferreira, R.~Santos, M.~Sher and J.P.~Silva,
  %``2HDM at the LHC - the story so far,''
  arXiv:1304.5225 [hep-ph];
  %%CITATION = ARXIV:1304.5225;%%
 %
%\cite{Coleppa:2013dya}
%\bibitem{Coleppa:2013dya}
  B.~Coleppa, F.~Kling and S.~Su,
  %``Constraining Type II 2HDM in Light of LHC Higgs Searches,''
   JHEP {\bf 1401}, 161 (2014)
  [arXiv:1305.0002 [hep-ph]];
  %%CITATION = ARXIV:1305.0002;%%
%
%\cite{Ferreira:2013qua}
  P.M.~Ferreira, R.~Santos, M.~Sher and J.P.~Silva,
  %``2HDM confronting LHC data,''
  arXiv:1305.4587 [hep-ph];
  %%CITATION = ARXIV:1305.4587;%%
%
%\cite{Eberhardt:2013uba}
%\bibitem{Eberhardt:2013uba}
  O.~Eberhardt, U.~Nierste and M.~Wiebusch,
  %``Status of the two-Higgs-doublet model of type II,''
   JHEP {\bf 1307}, 118 (2013)
  [arXiv:1305.1649 [hep-ph]];
  %%CITATION = ARXIV:1305.1649;%%
%
%\cite{Chpoi:2013wga}
%\bibitem{Chpoi:2013wga}
  S.~Choi, S.~Jung and P.~Ko,
  %``Implications of LHC data on 125 GeV Higgs-like boson for the Standard Model and its various extensions,''
  JHEP {\bf 1310} (2013) 225
  [arXiv:1307.3948 [hep-ph]].
  %%CITATION = ARXIV:1307.3948;%%
%\cite{Barger:2013ofa}
%\bibitem{Barger:2013ofa}
  V.~Barger, L.L.~Everett, H.E.~Logan and G.~Shaughnessy,
  %``Scrutinizing h(125) in Two Higgs Doublet Models at the LHC, ILC, and Muon Collider,''
  Phys.\ Rev.\ D {\bf 88} (2013) 115003
  [arXiv:1308.0052 [hep-ph]];
  %%CITATION = ARXIV:1308.0052;%%
 %
%\cite{Lopez-Val:2013yba}
%\bibitem{Lopez-Val:2013yba}
  D.~L\a'{o}pez-Val, T.~Plehn and M.~Rauch,
  %``Measuring extended Higgs sectors as a consistent free couplings model,''
  JHEP {\bf 1310} (2013) 134
  [arXiv:1308.1979 [hep-ph]];
  %%CITATION = ARXIV:1308.1979;%%
  %
%\cite{Chang:2013ona}
%\bibitem{Chang:2013ona}
  S.~Chang, S.K.~Kang, J.-P.~Lee, K.Y.~Lee, S.C.~Park and J.~Song,
  %``Two Higgs doublet models for the LHC Higgs boson data at $\sqrt{s}=$ 7 and 8 TeV,''
  arXiv:1310.3374 [hep-ph];
  %%CITATION = ARXIV:1310.3374;%%
  %
%\cite{Cacciapaglia:2013ora}
%\bibitem{Cacciapaglia:2013ora}
  G.~Cacciapaglia, A.~Deandrea, G.D.~La Rochelle and J.-B.~Flament,
  %``Searching for a lighter Higgs: parametrisation and sample tests,''
  arXiv:1311.5132 [hep-ph];
  %%CITATION = ARXIV:1311.5132;%%
%
%\cite{Cranmer:2013hia}
%\bibitem{Cranmer:2013hia}
  K.~Cranmer, S.~Kreiss, D.~L\a'{o}pez-Val and T.~Plehn,
  %``A Novel Approach to Higgs Coupling Measurements,''
  arXiv:1401.0080 [hep-ph];
  %%CITATION = ARXIV:1401.0080;%%
  
 %%%%%%%%%%%%%%%%%%%%%
 
 \bibitem{mssm}
 %\bibitem{Inoue:1982ej} 
  K.~Inoue, A.~Kakuto, H.~Komatsu and S.~Takeshita,
  %``Low-Energy Parameters and Particle Masses in a Supersymmetric Grand Unified Model,''
  Prog.\ Theor.\ Phys.\  {\bf 67}, 1889 (1982);
  %%CITATION = PTPKA,67,1889;%% 
 % \bibitem{Flores:1982pr} 
  R.A.~Flores and M.~Sher,
  %``Higgs Masses in the Standard, Multi-Higgs and Supersymmetric Models,''
  Annals Phys.\  {\bf 148}, 95 (1983);
  %%CITATION = APNYA,148,95;%%
  %\bibitem{Gunion:1984yn}
 J.F.~Gunion and H.E.~Haber,
  %``Higgs Bosons in Supersymmetric Models. 1.,''
  Nucl.\ Phys.\ B {\bf 272}, 1 (1986)  [Erratum-ibid.\ B {\bf 402}, 567 (1993)];
  %%CITATION = NUPHA,B272,1;%%
  Nucl.\ Phys.\ B {\bf 278}, 449 (1986)
  [Erratum-ibid.\ B {\bf 402}, 569 (1993)].
  %\bibitem{Gunion:1986nh} 
  %J.~F.~Gunion and H.~E.~Haber,
  %``Higgs Bosons in Supersymmetric Models. 2. Implications for Phenomenology,''
 % Nucl.\ Phys.\ B {\bf 278}, 449 (1986).
 %[Erratum-ibid.\ B {\bf 402}, 568 (1993)].
  %%CITATION = NUPHA,B278,449;%%

\bibitem{Lee:1973iz}
 T.D.~Lee,
 %``A Theory of Spontaneous T Violation,''
 Phys.\ Rev.\  D {\bf 8} (1973) 1226.
 %%CITATION = PHRVA,D8,1226;%%

\bibitem{hhg}
  J.F.~Gunion, H.E.~Haber, G.L.~Kane and S.~Dawson,
  \textit{The Higgs Hunter's Guide}
  \mbox{(Westview Press, Boulder, CO, 2000)}.
  %%CITATION = BNL-41644;%%

\bibitem{Branco:2011iw}
  G.C.~Branco, P.M.~Ferreira, L.~Lavoura, M.N.~Rebelo, M.~Sher and J.P.~Silva,
  %``Theory and phenomenology of two-Higgs-doublet models,''
  Phys.\ Rept.\  {\bf 516}, 1 (2012)
  [arXiv:1106.0034 [hep-ph]].

 \bibitem{GWP}
S.L.~Glashow and S.~Weinberg,
%``Natural Conservation Laws For Neutral Currents,''
Phys.\ Rev.\ D {\bf 15}, 1958 (1977);
%%CITATION = PHRVA,D15,1958;%%
E.A.~Paschos,
%``Diagonal Neutral Currents,''
Phys.\ Rev.\ D {\bf 15}, 1966 (1977).
%%CITATION = PHRVA,D15,1966;%%

\bibitem{decoupling}
 J.F.~Gunion and H.E.~Haber,
  %``The CP conserving two Higgs doublet model: The Approach to the decoupling limit,''
  Phys.\ Rev.\ D {\bf 67}, 075019 (2003)
  [hep-ph/0207010].
  %%CITATION = HEP-PH/0207010;%%

\bibitem{vacstab}
P.M.~Ferreira, R.~Santos and A.~Barroso,
  %``Stability of the tree-level vacuum in two Higgs doublet models against charge or CP spontaneous violation,''
  Phys.\ Lett.\ B {\bf 603} (2004) 219
   [Erratum-ibid.\ B {\bf 629} (2005) 114]
  [hep-ph/0406231].

\bibitem{type1}
H.E.~Haber, G.L.~Kane and T.~Sterling,
%``The Fermion Mass Scale And Possible Effects Of Higgs Bosons On Experimental
%Observables,''
Nucl.\ Phys.\  B {\bf 161}, 493 (1979).
%%CITATION = NUPHA,B161,493;%%


\bibitem{hallwise}
L.J.~Hall and M.B.~Wise,
Nucl.\ Phys.\ B {\bf 187}, 397 (1981).
%%CITATION = NUPHA,B187,397;%%


\bibitem{type2}
J.F.~Donoghue and L.F.~Li,
%``Properties Of Charged Higgs Bosons,''
Phys.\ Rev.\ D {\bf 19}, 945 (1979).
%%CITATION = PHRVA,D19,945;%%


%%%%%%%%%%%%%%%%%%%%%%%%%%%%%%%%%%%%%%%%%%%%%%%%%%%%%%%%
% Constraints
%%%%%%%%%%%%%%%%%%%%%%%%%%%%%%%%%%%%%%%%%%%%%%%%%%%%%%%%

%\cite{Arhrib:2013oia}
\bibitem{Arhrib:2013oia}
  A.~Arhrib, P.M.~Ferreira and R.~Santos,
  %``Are There Hidden Scalars in LHC Higgs Results?,''
  JHEP {\bf 1403}, 053 (2014)
  [arXiv:1311.1520 [hep-ph]].
  %%CITATION = ARXIV:1311.1520;%%
  %2 citations counted in INSPIRE as of 23 Jan 2014

\bibitem{vac1}
%\cite{Deshpande:1977rw}
  N.G.~Deshpande and E.~Ma,
  %``Pattern Of Symmetry Breaking With Two Higgs Doublets,''
  Phys.\ Rev.\  D {\bf 18} (1978) 2574.


\bibitem{unitarity}
S.~Kanemura, T.~Kubota and E.~Takasugi,
%``Lee-Quigg-Thacker bounds for Higgs boson masses in a two doublet model,''
Phys.\ Lett.\  B {\bf 313} (1993)  155; %[arXiv:hep-ph/9303263].
%%CITATION = PHLTA,B313,155;%%
%\bibitem{abdesunit}
A.G.~Akeroyd, A.~Arhrib and E.M.~Naimi,
  %``Note on tree-level unitarity in the general two Higgs doublet model,''
  Phys.\ Lett.\  B {\bf 490} (2000)  119.
%  %%CITATION = HEP-PH/0012353;%%

%\cite{Peskin:1991sw}
\bibitem{Peskin:1991sw}
  M.E.~Peskin and T.~Takeuchi,
  %``Estimation of oblique electroweak corrections,''
  Phys.\ Rev.\ D {\bf 46}, 381 (1992).
  %%CITATION = PHRVA,D46,381;%%

\bibitem{STHiggs}
 %%%%%%%\bibitem{Froggatt}
  C.D.~Froggatt, R.G.~Moorhouse and I.G.~Knowles,
  %``Leading Radiative Corrections In Two Scalar Doublet Models,''
  Phys.\ Rev.\  D {\bf 45}, 2471 (1992);
  %%CITATION = PHRVA,D45,2471;%%
 W.~Grimus, L.~Lavoura, O.M.~Ogreid and P.~Osland,
  %``The Oblique parameters in multi-Higgs-doublet models,''
  Nucl.\ Phys.\ B {\bf 801}, 81 (2008)
  [arXiv:0802.4353 [hep-ph]];
  %%CITATION = ARXIV:0802.4353;%%
  H.E.~Haber and D.~O'Neil,
  %``Basis-independent methods for the two-Higgs-doublet model III: The CP-conserving limit, custodial symmetry, and the oblique parameters S, T, U,''
  Phys.\ Rev.\ D {\bf 83}, 055017 (2011)
  [arXiv:1011.6188 [hep-ph]].
  %%CITATION = ARXIV:1011.6188;%%

\bibitem{lepewwg}
The ALEPH, CDF,  D0, DELPHI, L3, OPAL, SLD Collaborations, the LEP Electroweak Working Group, the Tevatron Electroweak Working Group, and the SLD electroweak and heavy flavour Groups,
  %``Precision Electroweak Measurements and Constraints on the Standard Model,''
  arXiv:1012.2367 [hep-ex].
  %%CITATION = ARXIV:1012.2367;%%

\bibitem{gfitter1}
  M.~Baak, M.~Goebel, J.~Haller, A.~Hoecker, D.~Ludwig, K.~Moenig, M.~Schott and J.~Stelzer,
  %``Updated Status of the Global Electroweak Fit and Constraints on New Physics,''
  Eur.\ Phys.\ J.\ C {\bf 72}, 2003 (2012)
  [arXiv:1107.0975 [hep-ph]].
  %%CITATION = ARXIV:1107.0975;%%

\bibitem{gfitter2}
M.~Baak, M.~Goebel, J.~Haller, A.~Hoecker, D.~Kennedy, R.~Kogler, K.~Moenig, M.~Schott and J.~Stelzer,
  %``The Electroweak Fit of the Standard Model after the Discovery of a New Boson at the LHC,''
  Eur.\ Phys.\ J.\ C {\bf 72}, 2205 (2012)
  [arXiv:1209.2716 [hep-ph]].
  %%CITATION = ARXIV:1209.2716;%%


 \bibitem{Barroso:2013awa}
  A.~Barroso, P.M.~Ferreira, I.P.~Ivanov and R.~Santos,
  %``Metastability bounds on the two Higgs doublet model,''
  JHEP {\bf 1306} (2013) 045
  [arXiv:1303.5098 [hep-ph]].



\bibitem{BB}
%\cite{Hermann:2012fc}
%\bibitem{Hermann:2012fc}
  T.~Hermann, M.~Misiak and M.~Steinhauser,
  %``$\bar{B}\to X_s \gamma$ in the Two Higgs Doublet Model up to Next-to-Next-to-Leading Order in QCD,''
  JHEP {\bf 1211} (2012) 036
  [arXiv:1208.2788 [hep-ph]];
  %%CITATION = ARXIV:1208.2788;%%
  %41 citations counted in INSPIRE as of 06 Jan 2014
%
  M.~Misiak, H.M.~Asatrian, K.~Bieri, M.~Czakon, A.~Czarnecki, T.~Ewerth, A.~Ferroglia and P.~Gambino {\it et al.},
  %``Estimate of B(anti-B ---> X(s) gamma) at O(alpha(s)**2),''
  Phys.\ Rev.\ Lett.\  {\bf 98}, 022002 (2007)
  [hep-ph/0609232];
  %%CITATION = HEP-PH/0609232;%%
%\cite{Asner:2010qj}
%\bibitem{Asner:2010qj}
  D.~Asner {\it et al.}  [Heavy Flavor Averaging Group Collaboration],
  %``Averages of b-hadron, c-hadron, and $\tau-lepton Properties,''
  arXiv:1010.1589 [hep-ex];
  %%CITATION = ARXIV:1010.1589;%%

  \bibitem{BB2}
%\cite{Mahmoudi:2009zx}
%\bibitem{Mahmoudi:2009zx}
  F.~Mahmoudi and O.~Stal,
  %``Flavor constraints on the two-Higgs-doublet model with general Yukawa couplings,''
  Phys.\ Rev.\ D {\bf 81}, 035016 (2010)
  [arXiv:0907.1791 [hep-ph]];
  %%CITATION = ARXIV:0907.1791;%%
%\bibitem{Su:2009fz}
  S.~Su and B.~Thomas,
  %``The LHC Discovery Potential of a Leptophilic Higgs,''
  Phys.\ Rev.\ D {\bf 79}, 095014 (2009)
  [arXiv:0903.0667 [hep-ph]];
  %%CITATION = ARXIV:0903.0667;%%
%\bibitem{Aoki:2009ha}
  M.~Aoki, S.~Kanemura, K.~Tsumura and K.~Yagyu,
  %``Models of Yukawa interaction in the two Higgs doublet model,
  %and their collider phenomenology,''
  Phys.\ Rev.\ D {\bf 80}, 015017 (2009)
  [arXiv:0902.4665 [hep-ph]];
  %%CITATION = ARXIV:0902.4665;%%
 P.~Posch,
  %``Precision constraints and photonic Higgs decays in the Two Higgs Doublet Model,''
University of Vienna Ph.D. dissertation (2009).
  %%CITATION = INSPIRE-1087424;%%

%\cite{Freitas:2012sy}
\bibitem{Freitas:2012sy}
  A.~Freitas and Y.-C.~Huang,
  %``Electroweak two-loop corrections to sin^2{\theta}(eff,bb) and R(b) using numerical Mellin-Barnes integrals,''
  JHEP {\bf 1208}, 050 (2012)
  [arXiv:1205.0299 [hep-ph]].
  %%CITATION = ARXIV:1205.0299;%%

%\cite{Denner:1991ie}
\bibitem{Denner:1991ie}
  A.~Denner, R.J.~Guth, W.~Hollik and J.H.~Kuhn,
  %``The Z width in the two Higgs doublet model,''
  Z.\ Phys.\ C {\bf 51}, 695 (1991).
  %%CITATION = ZEPYA,C51,695;%%

%\cite{Boulware:1991vp}
\bibitem{Boulware:1991vp}
  M.~Boulware and D.~Finnell,
  %``Radiative corrections to BR (Z ---> b anti-b) in the minimal supersymmetric standard model,''
  Phys.\ Rev.\ D {\bf 44}, 2054 (1991).
  %%CITATION = PHRVA,D44,2054;%%

\bibitem{Grant:1994ak}
  A.K.~Grant,
  %``The Heavy top quark in the two Higgs doublet model,''
  Phys.\ Rev.\ D {\bf 51}, 207 (1995)
  [hep-ph/9410267].
  %%CITATION = HEP-PH/9410267;%%

%\cite{Haber:1999zh}
\bibitem{Haber:1999zh}
  H.E.~Haber and H.E.~Logan,
  %``Radiative corrections to the Z b anti-b vertex and constraints on extended Higgs sectors,''
  Phys.\ Rev.\ D {\bf 62}, 015011 (2000)
  [hep-ph/9909335].
  %%CITATION = HEP-PH/9909335;%%


%\cite{Abbiendi:2013hk}
\bibitem{Abbiendi:2013hk}
  G.~Abbiendi {\it et al.}  [ALEPH and DELPHI and L3 and OPAL and LEP Collaborations],
  %``Search for Charged Higgs bosons: Combined Results Using LEP Data,''
  Eur.\ Phys.\ J.\ C {\bf 73} (2013) 2463
  [arXiv:1301.6065 [hep-ex]].
  %%CITATION = ARXIV:1301.6065;%%
  %28 citations counted in INSPIRE as of 06 Jan 2014

\bibitem{ATLASICHEP}
  ATLAS collaboration, ATLAS-CONF-2013-090;
  G.~Aad {\it et al.}  [ATLAS Collaboration],
  %``Search for charged Higgs bosons decaying via H+ -> tau nu in top quark pair events using pp collision data at sqrt(s) = 7 TeV with the ATLAS detector,''
  JHEP {\bf 1206} (2012) 039
  [arXiv:1204.2760 [hep-ex]].

\bibitem{CMSICHEP}
  S.~Chatrchyan {\it et al.}  [CMS Collaboration],
  %``Search for a light charged Higgs boson in top quark decays in pp collisions at sqrt(s) = 7 TeV,''
  JHEP {\bf 1207} (2012) 143
  [arXiv:1205.5736 [hep-ex]].
  %%CITATION = ARXIV:1204.2760;%%

\bibitem{Lees:2012xj}
  J.P.~Lees {\it et al.}  [BaBar Collaboration],
  %``Evidence for an excess of $\bar{B} \to D^{(*)} \tau^-\bar{\nu}_\tau$ decays,''
  Phys.\ Rev.\ Lett.\  {\bf 109}, 101802 (2012)
  [arXiv:1205.5442 [hep-ex]].
  %%CITATION = ARXIV:1205.5442;%%


%%%%%%%%%%%%%%%%%%%%%%%%%%%%%%%%%%%%%%%%%%%%%%%%%%%%%%%%
% End of constraints
%%%%%%%%%%%%%%%%%%%%%%%%%%%%%%%%%%%%%%%%%%%%%%%%%%%%%%%%



\bibitem{Donoghue:1978cj}
 J.F.~Donoghue and L.F.~Li,
%``Properties Of Charged Higgs Bosons,''
Phys.\ Rev.\ D {\bf 19}, 945 (1979).
%%CITATION = PHRVA,D19,945;%%

\bibitem{Georgi}
H.~Georgi and D.V.~Nanopoulos,
%``Suppression Of Flavor Changing Effects From Neutral Spinless Meson Exchange
%In Gauge Theories,''
Phys.\ Lett. {\bf 82B}, 95 (1979).
%%CITATION = PHLTA,B82,95;%%

\bibitem{silva}
F.J.~Botella and J.P.~Silva,
Phys.\ Rev.\ D {\bf 51}, 3870 (1995).
%%CITATION = HEP-PH 9411288;%%

\bibitem{lavoura}
L.~Lavoura and J.P.~Silva,
Phys.\ Rev.\  D {\bf 50}, 4619 (1994);
%%CITATION = HEP-PH 9404276;%%

\bibitem{lavoura2}
L.~Lavoura,
%``Signatures of discrete symmetries in the scalar sector,''
Phys.\ Rev.\  D {\bf 50}, 7089 (1994)
[arXiv:hep-ph/9405307].
%%CITATION = HEP-PH 9405307;%%

\bibitem{branco}
G.C.~Branco, L.~Lavoura and J.P.~Silva, {\it CP Violation}
(Oxford University Press, Oxford, England, 1999),
Chapter 22.
%%CITATION = IMPHA,103,1;%%

\bibitem{Davidson:2005cw}
  S.~Davidson and H.E.~Haber,
  %``Basis-independent methods for the two-Higgs-doublet model,''
  Phys.\ Rev.\ D {\bf 72}, 035004 (2005)
  [Erratum-ibid.\ D {\bf 72}, 099902 (2005)]
  [hep-ph/0504050].
  %%CITATION = HEP-PH/0504050;%%



  \bibitem{Haber:1989xc}
  H.E.~Haber and Y.~Nir,
  %``Multiscalar Models With a High-energy Scale,''
  Nucl.\ Phys.\ B {\bf 335}, 363 (1990).
  %%CITATION = NUPHA,B335,363;%%

\bibitem{Carena:2001bg} 
  M.~Carena, H.E.~Haber, H.E.~Logan and S.~Mrenna,
  %``Distinguishing a MSSM Higgs boson from the SM Higgs boson at a linear collider,''
  Phys.\ Rev.\ D {\bf 65}, 055005 (2002)
  [Erratum-ibid.\ D {\bf 65}, 099902 (2002)]
  [hep-ph/0106116].

\bibitem{Craig:2013hca}
  N.~Craig, J.~Galloway and S.~Thomas,
  %``Searching for Signs of the Second Higgs Doublet,''
  arXiv:1305.2424 [hep-ph].
  %%CITATION = ARXIV:1305.2424;%%

%\cite{Asner:2013psa}
\bibitem{Asner:2013psa}
  D.M.~Asner, T.~Barklow, C.~Calancha, K.~Fujii, N.~Graf, H.E.~Haber, A.~Ishikawa, S.~Kanemura {\it et al.},
  %``ILC Higgs White Paper,''
  arXiv:1310.0763 [hep-ph].
  %%CITATION = ARXIV:1310.0763;%%
  %17 citations counted in INSPIRE as of 04 Jan 2014

\bibitem{Carena:2013ooa}
  M.~Carena, I.~Low, N.R.~Shah and C.E.M.~Wagner,
  %``Impersonating the Standard Model Higgs Boson: Alignment without Decoupling,''
  JHEP {\bf 1404}, 015 (2014)
  [arXiv:1310.2248 [hep-ph]].
  %%CITATION = ARXIV:1310.2248;%%

  \bibitem{Haberinprep}
 H.E.~Haber, 
 %``The Higgs data and the Decoupling Limit,''
  arXiv:1401.0152 [hep-ph];
  %%CITATION = ARXIV:1401.0152;%%
  and preprint in preparation.

\bibitem{Haber:2000kq} 
  H.E.~Haber, M.J.~Herrero, H.E.~Logan, S.~Penaranda, S.~Rigolin and D.~Temes,
  %``SUSY QCD corrections to the MSSM h0 $b \bar{b}$ vertex in the decoupling limit,''
  Phys.\ Rev.\ D {\bf 63}, 055004 (2001)
  [hep-ph/0007006].
  %%CITATION = HEP-PH/0007006;%%

 \bibitem{GKO}
  I.F.~Ginzburg, M.~Krawczyk and P.~Osland,
  %``Resolving SM like scenarios via Higgs boson production at a photon collider. 1. 2HDM versus SM,''
  LC Note LC-TH-2001-026,  [hep-ph/0101208];
  %%CITATION = HEP-PH/0101208;%%
  %``Potential of photon collider in resolving SM like scenarios,''
  Nucl.\ Instrum.\ Meth.\ A {\bf 472}, 149 (2001)
  [hep-ph/0101229];
  %%CITATION = HEP-PH/0101229;%%
 in \textit{Physics and Experiments
with Future Linear $e^+e^-$ Colliders}, Batavia, Illinois,
2000, edited by A. Para and H. E. Fisk, AIP Conf. Proc. No.
578 ~(AIP, Melville, NY, 2001), pp.~304-311 [hep-ph/0101331].
%%CITATION = HEP-PH/0101331;%%

\bibitem{Ginzburg}
 I.F.~Ginzburg and M.~Krawczyk,
  %``Symmetries of two Higgs doublet model and CP violation,''
  Phys.\ Rev.\ D {\bf 72}, 115013 (2005)
  [hep-ph/0408011].
  %%CITATION = HEP-PH/0408011;%%

\bibitem{Arhrib}
 A.~Arhrib, R.~Benbrik and C.-W.~Chiang,
  %``Probing triple Higgs couplings of the Two Higgs Doublet Model at Linear Collider,''
  Phys.\ Rev.\ D {\bf 77}, 115013 (2008)
  [arXiv:0802.0319 [hep-ph]].
  %%CITATION = ARXIV:0802.0319;%%

\bibitem{Arhrib:2003ph} 
  A.~Arhrib, M.~Capdequi Peyranere, W.~Hollik and S.~Penaranda,
  %``Higgs decays in the two Higgs doublet model: Large quantum effects in the decoupling regime,''
  Phys.\ Lett.\ B {\bf 579}, 361 (2004)
  [hep-ph/0307391].
  %%CITATION = HEP-PH/0307391;%%
  
\bibitem{Bhattacharyya:2013rya} 
  G.~Bhattacharyya, D.~Das, P.~B.~Pal and M.~N.~Rebelo,
  %``Scalar sector properties of two-Higgs-doublet models with a global U(1) symmetry,''
  JHEP {\bf 1310}, 081 (2013)
  [arXiv:1308.4297 [hep-ph]].
  %%CITATION = ARXIV:1308.4297;%%

\bibitem{rec}
A.~David {\it et al.}  [LHC Higgs Cross Section Working Group Collaboration],
  %``LHC HXSWG interim recommendations to explore the coupling structure of a Higgs-like particle,''
  arXiv:1209.0040 [hep-ph].


 %%%%%%%%%%%%%%%%%%  ILC

%\cite{Dawson:2013bba}
\bibitem{Dawson:2013bba}
  S.~Dawson, A.~Gritsan, H.~Logan, J.~Qian, C.~Tully, R.~Van Kooten {\it et al.},
  %``Higgs Working Group Report of the Snowmass 2013 Community Planning Study,''
  arXiv:1310.8361 [hep-ex].
  %%CITATION = ARXIV:1310.8361;%%
  %21 citations counted in INSPIRE as of 14 Feb 2014


%\cite{Ono:2012ah}
\bibitem{Ono:2012ah}
  H.~Ono and A.~Miyamoto,
  %``A study of measurement precision of the Higgs boson branching ratios at the International Linear Collider,''
  Eur.\ Phys.\ J.\ C {\bf 73} (2013) 2343
  [arXiv:1207.0300 [hep-ex]].
  %%CITATION = ARXIV:1207.0300;%%
  %9 citations counted in INSPIRE as of 04 Jan 2014


\bibitem{LHCHiggs}
\texttt{https://twiki.cern.ch/twiki/bin/view/LHCPhysics/CrossSectionsFigures\#Higgs\_production\_cross\_sections}

\bibitem{Harlander:2003ai}
R.V.~Harlander and W.B.~Kilgore,
Phys.\ Rev.\ D {\bf 68}, 013001 (2003)
[hep-ph/0304035].

\bibitem{Spira:1995mt}
M.~Spira,
arXiv:hep-ph/9510347.

%%%%%%%%%%%%%%%%%%%%%%%%%%%% decays

%\cite{dgjk2hdm}
\bibitem{dgjk2hdm}
B. Dumont, J.F.~ Gunion, Y. Jiang and S. Kraml, 
%``Constraints on and future prospects for Two-Higgs-Doublet Models in light of the LHC Higgs signal,''
  arXiv:1405.3584 [hep-ph].
%%CITATION = ARXIV:1405.3584;%%

%\cite{Djouadi:1997yw}
\bibitem{Djouadi:1997yw}
  A.~Djouadi, J.~Kalinowski and M.~Spira,
  %``HDECAY: A Program for Higgs boson decays in the standard model and its supersymmetric extension,''
  Comput.\ Phys.\ Commun.\  {\bf 108} (1998) 56
  [hep-ph/9704448].
  %%CITATION = HEP-PH/9704448;%%
  %1094 citations counted in INSPIRE as of 14 Jan 2014

%\cite{Harlander:2013qxa}
\bibitem{Harlander:2013qxa}
  R.~Harlander, M.~M{\"u}hlleitner, J.~Rathsman, M.~Spira and O.~St{\aa}l,
  %``Recommendations for the evaluation of Higgs production cross sections and branching ratios at the LHC in the Two-Higgs-Doublet Model,''
  arXiv:1312.5571 [hep-ph].
  %%CITATION = ARXIV:1312.5571;%%

%\cite{Eriksson:2009ws}
\bibitem{Eriksson:2009ws}
  D.~Eriksson, J.~Rathsman and O.~St{\aa}l,
  %``2HDMC: Two-Higgs-Doublet Model Calculator Physics and Manual,''
  Comput.\ Phys.\ Commun.\  {\bf 181} (2010) 189
  [arXiv:0902.0851 [hep-ph]].
  %%CITATION = ARXIV:0902.0851;%%
  %52 citations counted in INSPIRE as of 14 Jan 2014

\bibitem{Haber:2006ue} 
  H.E.~Haber and D.~O'Neil,
  %``Basis-independent methods for the two-Higgs-doublet model. II. The Significance of tan beta,''
  Phys.\ Rev.\ D {\bf 74}, 015018 (2006) [Erratum-ibid.\ D {\bf 74}, 059905 (2006)]
  [hep-ph/0602242].
  %%CITATION = HEP-PH/0602242;%%
    
 \bibitem{Carena:2002es} 
  M.~Carena and H.E.~Haber,
  %``Higgs boson theory and phenomenology,''
  Prog.\ Part.\ Nucl.\ Phys.\  {\bf 50}, 63 (2003)
  [hep-ph/0208209].
  %%CITATION = HEP-PH/0208209;%%
  
  \bibitem{Djouadi:2005gj} 
  A.~Djouadi,
  %``The Anatomy of electro-weak symmetry breaking. II. The Higgs bosons in the minimal supersymmetric model,''
  Phys.\ Rept.\  {\bf 459}, 1 (2008)
  [hep-ph/0503173].
  %%CITATION = HEP-PH/0503173;%% 
  
  \bibitem{mssmhiggs}
  %\bibitem{Haber:1990aw} 
  H.E.~Haber and R.~Hempfling,
  %``Can the mass of the lightest Higgs boson of the minimal supersymmetric model be larger than m(Z)?,''
  Phys.\ Rev.\ Lett.\  {\bf 66}, 1815 (1991);
  %%CITATION = PRLTA,66,1815;%%
  %\bibitem{Ellis:1990nz} 
  J.R.~Ellis, G.~Ridolfi and F.~Zwirner,
  %``Radiative corrections to the masses of supersymmetric Higgs bosons,''
  Phys.\ Lett.\ B {\bf 257}, 83 (1991);
  %%CITATION = PHLTA,B257,83;%%
  %\bibitem{Okada:1990vk} 
  Y.~Okada, M.~Yamaguchi and T.~Yanagida,
  %``Upper bound of the lightest Higgs boson mass in the minimal supersymmetric standard model,''
  Prog.\ Theor.\ Phys.\  {\bf 85}, 1 (1991).
  %%CITATION = PTPKA,85,1;%%

  \bibitem{Haber:1993an} 
  H.E.~Haber and R.~Hempfling,
  %``The Renormalization group improved Higgs sector of the minimal supersymmetric model,''
  Phys.\ Rev.\ D {\bf 48}, 4280 (1993)
  [hep-ph/9307201].
  %%CITATION = HEP-PH/9307201;%%
  
  \bibitem{Haber:2007dj} 
  H.E.~Haber and J.D.~Mason,
  %``Hard supersymmetry-breaking 'wrong-Higgs' couplings of the MSSM,''
  Phys.\ Rev.\ D {\bf 77}, 115011 (2008)
  [arXiv:0711.2890 [hep-ph]].
  %%CITATION = ARXIV:0711.2890;%%
    
  \bibitem{Carena:1999py} 
  M.~Carena, D.~Garcia, U.~Nierste and C.E.M.~Wagner,
  %``Effective Lagrangian for the $\bar{t} b H^{+}$ interaction in the MSSM and charged Higgs phenomenology,''
  Nucl.\ Phys.\ B {\bf 577}, 88 (2000)
  [hep-ph/9912516].
  %%CITATION = HEP-PH/9912516;%%
  
  \bibitem{db}
  %\bibitem{Hempfling:1993kv} 
  R.~Hempfling,
  %``Yukawa coupling unification with supersymmetric threshold corrections,''
  Phys.\ Rev.\ D {\bf 49}, 6168 (1994);
  %%CITATION = PHRVA,D49,6168;%%
  %\bibitem{Hall:1993gn} 
  L.J.~Hall, R.~Rattazzi and U.~Sarid,
  %``The Top quark mass in supersymmetric SO(10) unification,''
  Phys.\ Rev.\ D {\bf 50}, 7048 (1994)
  [hep-ph/9306309];
  %%CITATION = HEP-PH/9306309,;%%
  %\bibitem{Carena:1994bv} 
  M.~Carena, M.~Olechowski, S.~Pokorski and C.E.M.~Wagner,
  %``Electroweak symmetry breaking and bottom - top Yukawa unification,''
  Nucl.\ Phys.\ B {\bf 426}, 269 (1994)
  [hep-ph/9402253];
  %%CITATION = HEP-PH/9402253;%%
  %\bibitem{Pierce:1996zz} 
  D.M.~Pierce, J.A.~Bagger, K.T.~Matchev and R.-J.~Zhang,
  %``Precision corrections in the minimal supersymmetric standard model,''
  Nucl.\ Phys.\ B {\bf 491}, 3 (1997)
  [hep-ph/9606211].
  %%CITATION = HEP-PH/9606211;%%
    
  \bibitem{Carena:2013qia} 
  M.~Carena, S.~Heinemeyer, O.~St{\aa}l, C.E.M.~Wagner and G.~Weiglein,
  %``MSSM Higgs Boson Searches at the LHC: Benchmark Scenarios after the Discovery of a Higgs-like Particle,''
  Eur.\  Phys.\  J.\ C {\bf 73}, 2552 (2013)
  [arXiv:1302.7033 [hep-ph]].
  %%CITATION = ARXIV:1302.7033;%%
  
  %\cite{Lee:2003nta}
\bibitem{Lee:2003nta}
  J.S.~Lee, A.~Pilaftsis, M.~Carena, S.Y.~Choi, M.~Drees, J.R.~Ellis and C.E.M.~Wagner,
  %``CPsuperH: A Computational tool for Higgs phenomenology in the minimal supersymmetric standard model with explicit CP violation,''
  Comput.\ Phys.\ Commun.\  {\bf 156}, 283 (2004)
  [hep-ph/0307377].
  %%CITATION = HEP-PH/0307377;%%
  %227 citations counted in INSPIRE as of 28 Feb 2014



\end{thebibliography}
\end{document}